\documentclass[amsmath,amssymb,reprint,twocolumn,superscriptaddress,nobibnotes,prb]{revtex4-1}
\usepackage[utf8]{inputenc}
\usepackage[T1]{fontenc}

\usepackage[english]{babel}

\usepackage{amsmath,amssymb,amsfonts}
\usepackage{amstext, mathrsfs, textcomp}
\usepackage{mathptmx}
\usepackage{bigints}  
\usepackage{nicefrac}
\usepackage{multirow}
\usepackage{dcolumn}
\usepackage{bm}

\usepackage{subfigure}
\usepackage{graphicx}
\usepackage{xcolor}
\usepackage[export]{adjustbox}

\usepackage[colorlinks=true, citecolor=blue, urlcolor=blue, linkcolor=blue]{hyperref} 

\usepackage[final]{changes}

\usepackage[breakable, most]{tcolorbox}
\tcbset{colback=blue!5!white,colframe=blue!75!black,fonttitle=\bfseries}

\graphicspath{
	{Figures/}
	{Figures/data_generation/}
	{Figures/deep_learning_overview/}
	{Figures/deep_learning_workflow/}
	{Figures/deep_learning_when_to_use/}
	{Figures/inverse_design_ill_posed/}
	{Figures/methods/}
	{Figures/design_constraint/}
	{Figures/cVAE_multiple_solutions/}
}

\newcommand{\TITLE}{A newcomer's guide to deep learning for inverse design in nano-photonics}

\begin{document}

	\title{\TITLE}

	\author{\firstname{Abdourahman} \surname{Khaireh-Walieh}}
	\affiliation{LAAS, Universit\'e de Toulouse, CNRS, Toulouse, France}
	
	\author{\firstname{Denis} \surname{Langevin}}
	\affiliation{Universit\'e Clermont Auvergne, Clermont Auvergne INP, CNRS, Institut Pascal, F-63000 Clermont-Ferrand, France}
	
	\author{\firstname{Pauline} \surname{Bennet}}
	\affiliation{Universit\'e Clermont Auvergne, Clermont Auvergne INP, CNRS, Institut Pascal, F-63000 Clermont-Ferrand, France}
	
	\author{\firstname{Olivier} \surname{Teytaud}}
	\affiliation{Meta AI Research Paris, France}
	
	\author{\firstname{Antoine} \surname{Moreau}}
	\email[e-mail~: ]{antoine.moreau@uca.fr}
	\affiliation{Universit\'e Clermont Auvergne, Clermont Auvergne INP, CNRS, Institut Pascal, F-63000 Clermont-Ferrand, France}
	
	\author{\firstname{Peter R.} \surname{Wiecha}}
	\email[e-mail~: ]{pwiecha@laas.fr}
	\affiliation{LAAS, Universit\'e de Toulouse, CNRS, Toulouse, France}



	
	\begin{abstract}
		Nanophotonic devices manipulate light at sub-wavelength scales, enabling tasks such as light concentration, routing, and filtering. Designing these devices to achieve precise light-matter interactions using structural parameters and materials is a challenging task. Traditionally, solving this problem has relied on computationally expensive, iterative methods.
		In recent years, deep learning techniques have emerged as promising tools for tackling the inverse design of nanophotonic devices. While several review articles have provided an overview of the progress in this rapidly evolving field, there is a need for a comprehensive tutorial that specifically targets newcomers without prior experience in deep learning. Our goal is to address this gap and provide practical guidance for applying deep learning to individual scientific problems.
		We introduce the fundamental concepts of deep learning and critically discuss the potential benefits it offers for various inverse design problems in nanophotonics. We present a suggested workflow and detailed, practical design guidelines to help newcomers navigate the challenges they may encounter. By following our guide, newcomers can avoid frustrating roadblocks commonly experienced when venturing into deep learning for the first time.
		In a second part, we explore different iterative and direct deep learning-based techniques for inverse design, and evaluate their respective advantages and limitations. To enhance understanding and facilitate implementation, we supplement the manuscript with detailed Python notebook examples, illustrating each step of the discussed processes.
		While our tutorial primarily focuses on researchers in (nano-)photonics, it is also relevant for those working with deep learning in other research domains. We aim at providing a solid starting point to empower researchers to leverage the potential of deep learning in their scientific pursuits.
	\end{abstract}
	

	\maketitle


	\section{Introduction}
	
	The broad field of (nano-)photonics deals with the interaction of light with matter and with applications that arise from structuring materials at sub-wavelength scales in order to guide or concentrate light in a pre-defined manner.\cite{muhlschlegelResonantOpticalAntennas2005, girardFieldsNanostructures2005, novotnyPrinciplesNanooptics2006, kuznetsovOpticallyResonantDielectric2016, girardNearfieldOpticalProperties2006}
	Astonishing effects can be obtained in this way, such as unidirectional scattering, negative refraction, enhanced nonlinear optical effects, amplified quantum emitter yields or magnetic optical effects at visible frequencies.\cite{pendryNegativeRefractionMakes2000, wiechaStronglyDirectionalScattering2017, kauranenNonlinearPlasmonics2012, genevetRecentAdvancesPlanar2017, colasdesfrancsPlasmonicPurcellFactor2016, wangIntegratedPhotonicQuantum2020, wiechaEnhancementElectricMagnetic2019}
	Tailoring of such effects via the rational design of nanodevices is typically termed ``inverse design''.
	Unfortunately, like most inverse problems, nanophotonics inverse design is in general an ill-posed problem and cannot be solved directly.\cite{hadamardProblemesAuxDerives1902}
	Usually, iterative approaches like global optimization algorithms or high-dimensional gradient based adjoint methods are used, which however are computationally expensive and slow, especially if applied to repetitive design tasks.\cite{jensenTopologyOptimizationNanophotonics2011, elsawyNumericalOptimizationMethods2020}
	
	In the recent past it has been shown that deep learning models can be efficiently trained on predicting (nano-)optical phenomena. \cite{malkielPlasmonicNanostructureDesign2018, wiechaDeepLearningMeets2020, blanchard-dionneTeachingOpticsMachine2020, chenHighSpeedSimulation2022, maOptoGPTFoundationModel2023}
	This rapidly growing research interest stems from remarkable achievements that deep learning accomplished in computer science since around 2010, especially in the fields of computer vision,\cite{krizhevskyImageNetClassificationDeep2012, heDeepResidualLearning2015, guoDeepLearningVisual2016, kirillovSegmentAnything2023} and natural language processing.\cite{sundermeyerLSTMNeuralNetworks2012, brownLanguageModelsAre2020, otterSurveyUsagesDeep2021}
	The main underlying assumption is that neural networks are universal function approximators.\cite{hornikMultilayerFeedforwardNetworks1989}
	It has been shown that deep learning is capable to solve various inverse design problems in nano-photonics. 
	A non-exhaustive list of examples includes single nano-scatterers,\cite{peurifoyNanophotonicParticleSimulation2018, estrada-realInverseDesignFlexible2022} gratings,\cite{jiangGlobalOptimizationDielectric2019, jiangSimulatorbasedTrainingGenerative2019} Bragg mirrors,\cite{liuTrainingDeepNeural2018, unniDeepConvolutionalMixture2020, daiInverseDesignStructural2022} photonic crystals,\cite{asanoIterativeOptimizationPhotonic2019} waveguides,\cite{zhangEfficientSpectrumPrediction2019} or sophisticated light routers.\cite{tahersimaDeepNeuralNetwork2019, banerjiMachineLearningEnables2020, dinsdaleDeepLearningEnabled2021}
	For an extensive overview of the current state of research we refer the interested reader to recent review articles on the topic,\cite{zhouEmergingRoleMachine2019, soDeepLearningEnabled2020, jiangDeepNeuralNetworks2021, liuTacklingPhotonicInverse2021, wiechaDeepLearningNanophotonics2021, dengDeepInversePhotonic2022, kanyaoNanophotonicsMachineLearning2023, jiRecentAdvancesMetasurface2023a}
	or to more methodological reviews and comparative benchmarks.\cite{schneiderBenchmarkingFiveGlobal2019, elsawyNumericalOptimizationMethods2020, hegdeDeepLearningNew2020, renInverseDeepLearning2022}

	\begin{figure*}
		\includegraphics[width=\linewidth]{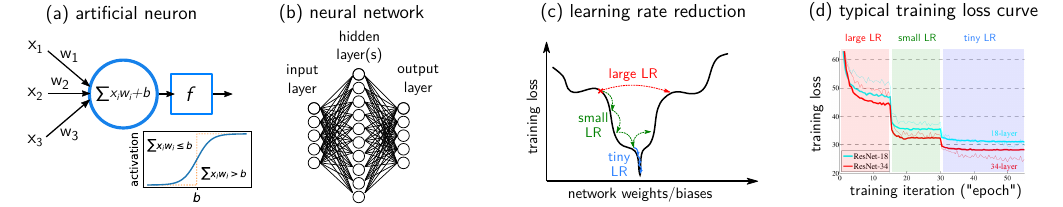}
		\caption{
			(a) A single neuron where $x$ is the input vector, $w$ and $b$ are respectively the weights and biases and $f$ is the activation function. The inset shows the output of the sigmoid activation function.
			(b) A set of neurons arranged in several, consecutively connected layers, forming a neural network. 
			(c) Learning rate size effect on the loss function optimization. 
			(d) A typical training loss or validation error plot. When the loss decrease starts to stagnates, reducing the learning rate leads to further convergence. Adapted from Ref.~\onlinecite{heDeepResidualLearning2015}.
		}\label{fig:neuron_and_neuralnet}
	\end{figure*}

	This work aims at providing a comprehensive tutorial for deep learning techniques in nano-photonics inverse design. Rather than assembling a complete review of the literature, we try to develop a pedagogical guide through the typical workflow, and focus in particular on the discussion of practical guidelines and good habits, that will hopefully help with the creation of robust models, avoiding frustration through typical pitfalls.
	We start with a concise introduction to the basic ideas behind deep learning and discuss their practical implications. We then study the question about which types of problem may benefit from solving them with deep learning, and which problems are probably better solved with other approaches.
	Subsequently we provide considerations on the typical deep learning workflow, including detailed advice for best practices. This ranges from the choice of the model architecture, data generation, parameterization and normalization, to the setup and running of the training procedure and tuning of the associated hyperparameters.
	
	In the second part, we introduce a selection of methods from the two most popular categories of inverse design using deep learning models. 
	The first group of methods is based on iterative optimization, using deep learning models as ultra-fast and differentiable surrogates for slow numerical simulations.
	The second approach aims at developing end-to-end network training for solving the inverse problem. The latter, so-called ``one-shot'' solvers, can be implemented in different ways, we specifically discuss the Tandem network, as well as conditional Variational Autoencoders and conditional Generative Adversarial Networks(cVAE, cGAN).
	Finally we provide a short overview of further techniques.
	The paper is accompanied by a set of extensively commented Python notebook tutorials,\cite{wiechaNewcomerGuideDeep2023} that demonstrate the practical details of the presented techniques on two specific examples from (nano-)photonics: The design of reflective multi-layer systems and tailoring of the scattering response of individual nano-scatterers.

	\section{Introduction to deep learning and the typical workflow}\label{sec:deep_learning_general}
	
	Before diving into the technical details of different approaches to use deep learning (DL) for inverse design, it is crucial to become familiar with some basic concepts and good practices.
	In the following section we therefore provide a very concise introduction to artificial neural networks, we discuss how to assess whether it could make sense to apply deep learning to a problem and in which cases one should rather stick to conventional methods.

	\subsection{Short introduction to artificial neural networks}

	\paragraph{Artificial neurons and neural networks}
	The basic building block of an artificial neural network (ANN) is the artificial neuron.
	An artificial neuron is simply a mathematical function taking several input values. A weight parameter is associated with each of the inputs.
	As depicted in figure~\ref{fig:neuron_and_neuralnet}a, first the sum of the weighted input values is calculated, then the value of an additional bias parameter is added. Subsequently a so-called activation function $f$ is applied to the resulting number. 
	The result of the activation function is the neuron's output value, also simply called its activation.
	A neural network is nothing else than several of these artificial neurons connected to each other in some way, for instance by feeding the output of one neuron into other neurons' input layers (c.f. figure~\ref{fig:neuron_and_neuralnet}b). 
	Please note that some network architectures also implement other types of connections, for example of neurons within a layer or to preceding layers, etc.
	
	\paragraph{Deep networks with nonlinear activations}
	A key hypothesis underlying deep learning (``deep'' means that a network has many layers) is that the ANN is learning a hierarchy of features from the input data, where each layer is extracting a deeper level of characteristics. In an image for instance, the first layer could be recognizing lines and edges, the second layer may ``understand'' how they form specific shapes like eyes or ears and a subsequent layer may then analyze the relative positions, orientations and sizes in the ensemble of these features, to identify complex objects like animals or human faces. 
	In consequence, using many layers is essential to ensure that a network has a high abstraction capacity.
	
	Now, it is technically possible to use linear activation functions throughout a network. However, it is trivial to show, that any neural network with multiple layers of linear activation can be identically represented by a single linear layer, as the chain of two linear functions is still linear.
	Hence, using nonlinear activation functions is crucial in any deep ANN in order to perform hierarchical feature extraction. Various activation functions can be used. The activation which is closest to a biological neuron's response is probably the sigmoid neuron, which implements a logistic activation function, depicted in the inset of figure~\ref{fig:neuron_and_neuralnet}a. 
	The sigmoid is however suffering from the fact that for inputs far away from the bias, the gradient is very small. For such inputs learning is very slow.
	Therefore, other activations such as Rectified Linear Units (ReLU) expressed as $\text{max}(0,x)$ often preferred in the hidden layers of an ANN, because of its constant gradient and cheap computation cost.
	In the ``body'' of the network, ReLU is usually a very good first choice and leads to robust networks. 
	
	The activation of the very last layer of a network, its \textit{output layer}, needs to be chosen depending on the task to solve, as well as on the numerical range of the output data values. 
	A softmax activation layer is a form of a generalized logistic function, where the sum of the neurons' activations of the entire layer is normalized. Hence it is adequate for outputs that correspond to a probability distribution like in classification tasks.
	In regression tasks -- inverse design typically falls in this category -- a linear output activation function is the simplest first choice, since it can take arbitrary values. 
	Be aware that in case of a linear output activation the ANN needs to learn the data range along with the actual problem to solve.
	Therefore one should consider normalizing also the output part of the dataset (see also below). In this case Sigmoid (data range $[0, 1]$) or tanh activation (data range $[-1, 1]$) can help the network to focus on the essential learning task (c.f. ``inductive bias'').

	\paragraph{Training}
	Deep learning is a statistical method, the goal is to adapt the network parameters (weights and biases of the artificial neurons) such that the ANN learns to solve a task that is implicitly defined by a large dataset. 
	The trick to learn from such data, is to define a loss function that quantifies the network's prediction error. 
	More precisely, the loss expresses how much the neural network's predictions of a set of samples is different from the expected network outputs. The notion of ``set'' is important, because the loss is defined in a stochastic approximation for \textit{batches} of samples.\cite{robbinsStochasticApproximationMethod1951} The expected outputs obviously need to be known and thus are also part of the training data. 
	
	Using an optimization algorithm, the loss function is minimized by modifying the weight and bias parameters of the network.
	During this optimization, the model repeatedly computes a batch of training samples in forward propagation to calculate the loss.
	Subsequently the gradients of the loss function with respect to all network weights is calculated in a backpropagation step, using automatic differentiation (autodiff).\cite{lukasheinrichPyHEP2020Autodiff2020}
	The network parameters are finally adapted towards the negative gradient direction, in order to minimize the loss function.
	Autodiff is the core of deep learning and hence all DL libraries are essentially autodiff libraries with tools for neural network optimization.
	
	In practice, the model parameters are randomly initialized, hence DL is typically not deterministic. Restarting a training process will not give the exact same network.
	During training, the size of the parameter modification steps is a crucial parameter.
	It is typically called ``learning rate'' and takes generally values smaller than $1$ ($10^{-3}$ or $10^{-4}$ are common start values). 
	For a given network parameter $ w_i $, the update is expressed as  $ w_i = w_i - LR \times \partial \text{loss} / \partial w_i $. 
	If the learning rate ($LR$) is very small, the algorithm requires many iterations to reach the loss function minimum, and if it is too large, it can miss the optimal solution which often is a steep minimum. Thus it is important to reduce the learning rate during training, as depicted in figure~\ref{fig:neuron_and_neuralnet}c-d.

	\paragraph{Which optimizer}
	The most popular optimization algorithm in deep learning is stochastic gradient descent (SGD),\cite{robbinsStochasticApproximationMethod1951, goodfellowDeepLearning2016} which performs the weight updates as described above.
	A popular alternative, that often offers a faster convergence are the SGD variants ``adam''\cite{kingmaAdamMethodStochastic2014} or adamW''.\cite{loshchilovDecoupledWeightDecay2019} 
	It is also more robust with respect to its hyperparameter configuration, and therefore an excellent first choice.

	\paragraph{Practical implementation of neural networks}
	The de-facto standard programming language in the deep learning community is Python, but frameworks exist for virtually all programming languages. 
	The most popular libraries to build and train neural networks are ``PyTorch''\cite{paszkePyTorchImperativeStyle2019}, ``TensorFlow'' with its high-level API ``Keras''\cite{abadiTensorFlowLargeScaleMachine2015, cholletDeepLearningPython2017}, ``Flax''/``JAX''\cite{heekFlaxNeuralNetwork2023}, or ``MXNet''\cite{chenMXNetFlexibleEfficient2015}, among others.

	\subsection{When is deep learning useful?}
	
	With the rapidly growing research interest around deep learning in the last years, a newcomer can easily get the impression that DL is the perfect solution to basically any problem.
	This is a dangerous fallacy. Assuming limitation to reasonable computational invests, deep learning will in many situations actually lead to inferior results compared to conventional methods, and due to data generation and network training, it will often come with a higher total computational cost.
	We therefore want to start with a survey of inverse design scenarios, and discuss situations in which deep learning may, or rather may not, be an adequate method.
	
	The question of whether deep learning may or may rather not be an interesting option stands and falls with the quantity of available data. 
	If huge amounts of data are available, with an appropriate model layout, deep learning likely works well on essentially any problem.\cite{brownLanguageModelsAre2020, kaplanScalingLawsNeural2020, yuScalingAutoregressiveModels2022}
	Unfortunately, data is often expensive. In photonics for instance, many simulation methods are relatively slow.
	In our considerations below, we will therefore assume the case of datasets with high computational cost and limited size, in the order of thousands to tens of thousands of samples.

	\subsubsection{When to use deep learning}
	
	\begin{figure}
		\includegraphics[width=\columnwidth]{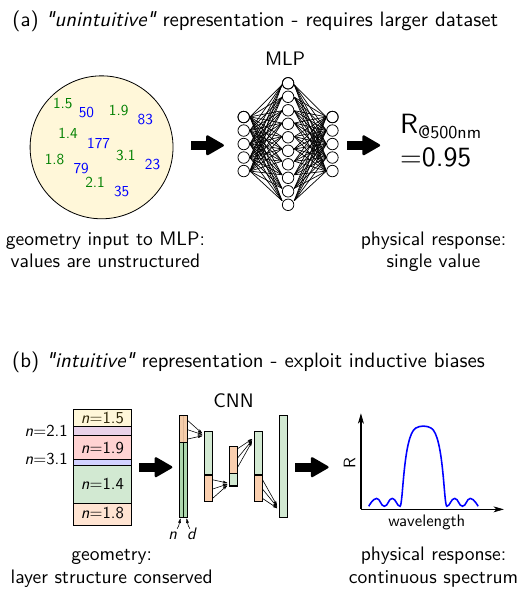}
		\caption{
			Comparison of two different parameterizations of the same problem.
			(a) Using a dense neural network, the order of the thicknesses (blue values) and refractive indices (green values) is lost, because every input value is fed into every neuron in the first layer. The network needs to learn during training that the order matters. Furthermore, the correlation between the many input values and the single output reflectivity are difficult to understand.
			(b) using a CNN, the layer order can be conserved and using two input channels, even the association of thickness and refractive index of a single layer can be directly passed to the network, hence during training these correlations do not need to be learned. Furthermore, predicting a whole reflectivity spectrum facilitates to identify correlations between changes of the geometry and for example resonance peak shifts. Returning this again via a convolutional layer conserves the order of the spectrum.
			On small to medium size datasets, exploiting these inductive biases of a CNN can significantly improve performance.
		}\label{fig:when_to_use_dl_intuition}
	\end{figure}

	\paragraph{Intuition}
	
	Deep learning is a data-driven approach: during training, the neural network is trying to figure out correlations in the dataset that allow to link the input to the output values, eventually converging to an empirical model describing the implicit rules behind the dataset (in our case the implicit physics).
	In that sense, neural network training can be somehow compared to human learning: By mere observations, humans figure out causal correlations in nature. In nano-photonics for instance, a person who is studying plasmonic nanostructures will sooner or later develop an intuition for the expected red-shift of a particle's localized surface plasmon resonance with increasing particle size.
	
	Therefore, a good question to ask is: With the given dataset, does it seem easily possible to develop an \textit{intuition} of the physical response? 
	If this is the case, training a deep learning model on a large enough dataset promises to be successful. 
	On the other hand, if a problem or its description is highly entangled and we can only hardly imagine that an intuitive understanding can be learned without further guidance, then an artificial neural network is likely going to have a hard time understanding the correlations in the dataset.
	
	An example is shown in figure~\ref{fig:when_to_use_dl_intuition}. 
	A Bragg mirror design problem is parameterized by its geometry of $N$ dielectric layers with arbitrary material and thickness, and the target physical observable is the reflectivity at a fixed wavelength. The learning problem is a mapping of $2N$ values to a single reflectivity value. Providing a human with a large set of such data samples would most likely result in confusion rather than in an intuitive understanding of the correlation between geometry and physical property.
	The problem becomes easier to grasp intuitively if the layer-order is given to the network so it does not need to learn that the order is important. In deep learning this can be done using a convolutional layer, which keeps the input structure and searches correlations only between ``neighbor'' values. We exploit the convolution's \textit{inductive bias} (see also below).
	We know that periods of the same pair of layers form the ideal Bragg mirror. With that prior knowledge we could further simplify the intuitive accessibility and parameterize the geometry with only two thicknesses and two materials values, this layer-pair being repeated for $N/2$ times.
	Finally, instead of predicting only a single reflectivity at one wavelength, we could train the network on predicting reflection spectra in a large spectral window. Now, the geometry can be understood easier and the impact of changes in the layer pair geometry is easier to interpret. A wavelength shift of the stop-band for instance can be easily quantified if a spectrum is given instead of just one reflectivity value. In the same way as the latter representation is easier to interpret for a human, a deep learning model will be capable to develop an empirical model much easier.
	Other examples where such additional tasks, also called auxiliary tasks, are useful for improving the performance on the machine learning principle task, have been investigated in reference \onlinecite{caruanaMultitaskLearning1997}, and more recently in the case of the game ``Go''.\cite{wuAcceleratingSelfPlayLearning2020} In conclusion, a richer information about the problem is often useful for making training faster and less data intensive.

	\paragraph{Repetitive and speed-critical design tasks}
	
	In case it appears reasonable to assume that an intuitive comprehension of the problem can be built, one should also think about the motivation of using deep learning.
	On often highlighted advantage of DL models is their high evaluation speed (leaving aside the training phase).
	Using deep learning with the goal to speed up inverse design therefore seems reasonable. The training phase however requires significant computational work for network training and often also for data generation. Hence deep learning makes most sense in cases of highly repetitive design tasks like metasurface meta-atom creation, or in speed-critical scenarios like real-time applications, such as spatial light modulator control.
	
	\paragraph{Differentiable surrogate models}
	
	Another tremendous strength of deep learning models is the fact that they are differentiable. While gradients in numerical simulations may be obtained with adjoint methods,\cite{jensenTopologyOptimizationNanophotonics2011} deep learning models can learn differentiable models even from empirical data, for example from experimental measurements.
	
	\paragraph{Latent descriptions of high-dimensional data}
	
	A key capability of deep learning is the possibility to learn latent representations of complex, high-dimensional data. The latent space is not only a crucial concept in deep generative models,\cite{karrasAnalyzingImprovingImage2020, rombachHighResolutionImageSynthesis2022} it can also be used to compress bulky data,\cite{kingmaIntroductionVariationalAutoencoders2019, khaireh-waliehMonitoringMBESubstrate2023} or to gain insight into hidden correlations.\cite{melatiMappingGlobalDesign2019, kiarashinejadKnowledgeDiscoveryNanophotonics2020, zandehshahvarManifoldLearningKnowledge2022}
	
	\paragraph{Empirical models from experimental data}
	
	On data that is very complex and/or high-dimensional, it can be difficult to fit a conventional physical model. 
	A deep learning neural network may be a promising alternative to obtain a differentiable description for the physics, based on experimental data. 
	A specific use-case application may be when a theoretical model fails to reproduce experimental observations. A so-called ``multi-modal''\cite{bachmannMultiMAEMultimodalMultitask2022} model could learn in parallel from experimental data and the simulated data. Both inputs are separately projected in a shared latent space. After successful training, the latent description creates a learned link between experiment and simulated data.

	\subsubsection{When to NOT use deep learning}
	
	Writing a good data generation routine to create a useful training set is at least as challenging, but often even more challenging than writing a good fitness function for a conventional optimization technique.
	Additionally, it has been demonstrated on several occasions, that simple conventional methods often outperform heavy GPU-based black-box optimization methods.\cite{liuSurveyEvolutionaryNeural2023, liRandomSearchReproducibility2020, phamEfficientNeuralArchitecture2018, realRegularizedEvolutionImage2019}
	Before rushing into data generation, we therefore urge the reader to consider whether conventional techniques may not be a sufficient alternative to deep learning on their specific problem.
	
	\begin{figure}
		\includegraphics[width=\columnwidth]{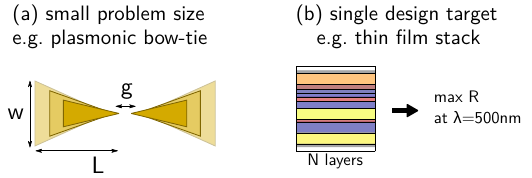}
		\caption{
			Examples of problem configurations, for which deep learning is probably \textbf{\textit{not}} adequate.
			(a) Problems with few parameters like the here depicted plasmonic bow-tie antenna design (3 free parameters) can easier be solved by conventional approaches, intuition or may even be systematically explored.
			(b) In problems with a single design target, the computational overhead of deep learning is not paying back. Conventional global optimization is the method of choice.
		}\label{fig:when_to_use_dl_problem_size}
	\end{figure}

	\paragraph{``Unintuitive'' problems or parameterizations}
	
	Deep learning may perform badly if the problem or its parameterization are not intuitive, hence where the correlations between input and output are highly abstract. This is the case in the above example of a sequence of $N$ random dielectric layers, where the goal is to map the $2N$ values of layer thicknesses and materials to a single output value (the reflectivity). 
	While the problem is very easy to solve with the right physical model at hand (e.g. with transfer matrix or s-matrix method), a human presented with examples of such $2N + 1$-value samples will have a very hard time to understand the correlation.
	This can typically be solved by proper pre-processing and network architecture design, but requires supplementary effort.

	\begin{figure*}
		\includegraphics[width=\linewidth]{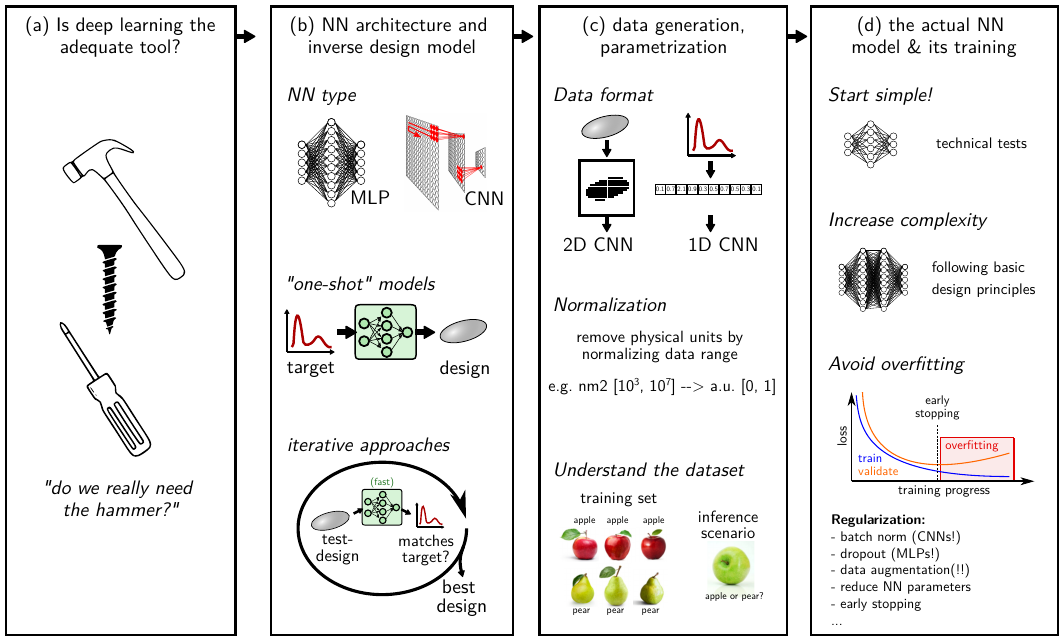}
		\caption{
			Workflow for the application of deep learning in inverse design.
			(a) The workflow should start with an unbiased assessment, whether other methods might not be more adequate than deep learning (``do we need the hammer, or is it rather a screwdriver?'').
			(b) Think about how the problem can be parameterized. Which network type will be necessary? Is the goal ultimate acceleration (use one-shot inverse design), or a less time critical, best possible optimization (use iterative inverse design).
			(c) prepare and understand the dataset.
			(d) implement the actual mode. Start simple, increase complexity, avoid overfitting (regularization!). Always keep in mind literature design guidelines.
		}\label{fig:deep_learning_workflow}
	\end{figure*}

	\paragraph{Single design tasks and simple problems}
	
	We argue that the main argument against deep learning is that it is inappropriately expensive in many situations. 
	If the geometric model used in the design problem is described by only a few free parameters (like 3 or 4), a systematic analysis is probably very promising and maybe even cheaper than generation of a large dataset. 
	In many cases this may furthermore be done around an intuitive, first guess.
	Similarly, if the geometry is more or less known and only small variations are to be optimized, conventional optimization or again a systematic exploration of all parameters are more appropriate.
	Such situations are depicted in figure~\ref{fig:when_to_use_dl_problem_size}.
	Even if a systematic analysis cannot be performed with sufficiently dense parameter steps, it may be worth considering conventional interpolation approaches such as Chebyshev expansion or also Bayesian optimization.\cite{hammerschmidtSolvingInverseProblems2018, garcia-santiagoBayesianOptimizationImproved2021, wuEfficientHybridMethod2021, elsawyMultiobjectiveStatisticalLearning2021}
	Finally, if the problem consists in solving a single design target, even for complex scenarios a conventional global optimization run is probably the more adequate approach.\cite{elsawyNumericalOptimizationMethods2020}

	In conclusion, finding problems that really benefit from deep learning based inverse design is not as obvious as it is often suggested in literature. 
	As an example to support this claim we recall that to date all production-scale metasurface design is being done with conventional lookup tables, while deep learning is still only used on toy-problems for testing.

	\subsection{General workflow and good habits}

	Let's suppose we have decided to go for deep learning as method of choice.
	Before diving into the details of methods for inverse design, we want to provide a loose guide of the typical workflow with suggestions for good habits that should be useful when applying deep learning to any problem, not limited to inverse design. In the following we aim at providing guidelines that implement findings of modern deep learning and that will be useful to get good results with little trial and error.
	The following section is structured in the way that we believe the workflow should be organized. A schematic overview is depicted in figure~\ref{fig:deep_learning_workflow}.

	\subsubsection{Overview of the major model architectures}

	Neurons can be connected in different ways to form an ANN. The most adequate neural network type will also depend on the dataset to be processed. Multilayer perceptrons (MLP), convolutional neural networks (CNN) and recurrent neural networks (RNNs) are the most widely used general architectures. Recently so-called ``attention''-layers have gained significant popularity as well. 
	In the following we will discuss the advantages and drawbacks of the different network architectures and give suggestions concerning their applications.

	\paragraph{Multilayer Perceptron (MLP)} The MLP is also called fully connected or dense network, and consists of layers of unstructured neurons, where each neuron of a layer is connected to every neuron of the preceding and of the following layer.\cite{goodfellowExplainingHarnessingAdversarial2015}
	Because they scale badly with dimension, MPLs should not be used on raw data, but on \textit{features} of the data. Exceptions could be natively relational data, or if the dataset consists already of high-level features. In geometric inverse design, raw data might be images of the structure, whereas size and shape parameters would be the features. 
	In consequence, MLPs are typically part of a larger neural network, where features are extracted in a first stage. Those features are then passed into the MLP, for example in various convolutional neural network architectures, or similarly also in the attention heads of transformers (see below).\cite{simonyanVeryDeepConvolutional2015, vaswaniAttentionAllYou2017}
	With large layers the number of free parameters in fully connected networks can quickly diverge, which makes this architecture particularly prone to overfitting.
	
	We argue that MLPs can be a valid option for inverse design, for example when a model is parameterized with only a low number of parameters and/or when the dataset comprises relatively high-level features of low dimensionality (e.g. a low number of size and/or position parameters). However, for more general problems and parameterizations (e.g. parameterization at the pixel/voxel level), we recommend CNNs (see below).

	\paragraph{Convolutional neural networks (CNN)} CNNs are used since the 1980ies,\cite{fukushimaNeocognitronSelforganizingNeural1980, atlasArtificialNeuralNetwork1987, lecunBackpropagationAppliedHandwritten1989, zhangParallelDistributedProcessing1990} and were inspired by how receptive fields in the visual cortex process signals in a hierarchical manner.\cite{hubelReceptiveFieldsSingle1959}
	They work similarly to popular computer vision algorithms of the 1990s that processed images via feature detection kernels,\cite{loweObjectRecognitionLocal1999} with the difference that the CNN kernels are composed of artificial neurons and automatically learn from data, instead of being manually defined. 
	Feature detection itself is performed by convolutions of the kernels with the structured input data (e.g. images).\cite{goodfellowDeepLearning2016}
	In the early 2010s, GPUs rendered calculation of discrete convolutions computationally cheap. This allowed to scale up CNNs, which lead to a breakthrough in computer vision performance. Ever since, CNNs are amongst the most popular artificial neural network architectures that scale well from small up to gigantic data set and problem sizes.\cite{krizhevskyImageNetClassificationDeep2012, liSurveyConvolutionalNeural2022}
	In particular the idea of residual blocks with identity connections allowed to scale the depth of CNNs by orders of magnitude (thousand layers and more), while maintaining efficient training performance.\cite{heDeepResidualLearning2015, heIdentityMappingsDeep2016, huangDeepNetworksStochastic2016}
	
	The popularity of CNN is a result of many favorable properties. For instance they offer inductive biases (see also below), that are useful for many applications like locality (assuming that neighbor input values are closely related e.g. in an image or a spectrum), or translation invariance of the feature detection (a feature will be detected regardless of its position in the input data).
	CNNs are easy to configure and very robust in training. They are applicable and perform very well over a large range of input dimensions, e.g. from tiny input arrays to megapixel images, and they scale well to large dataset sizes.\cite{karrasAnalyzingImprovingImage2020}
	They are computationally friendly since the convolutions can be very efficiently calculated on modern GPU accelerators. 
	Please note, that an often forgotten advantage of CNNs is also their ability to work on variable input sizes.\cite{linNetworkNetwork2014}
	And finally, due to their tremendous success in the past 10 years, an abundance of recipes and optimization guides are available in the literature, which lowers the entry barrier for newcomers to an absolute minimum.

	\paragraph{Recurrent neural networks (RNN)} Recurrent neural networks implement memory mechanisms in networks which process data sequentially. Thanks to this memory, long-term correlations in data sequences can be processed.\cite{hochreiterLongShortTermMemory1997, hochreiterGradientFlowRecurrent2001}
	RNNs have been the state of the art in natural language processing for some time, and due to their success in this area, researchers have applied RNN concepts to various other fields as well.\cite{sutskeverSequenceSequenceLearning2014}
	However, RNNs suffer from a main drawback, their training is largely sequential and can not easily be parallelized, hence training scales badly with increasing amounts of data.\cite{vaswaniAttentionAllYou2017}
	With the recent exploding availability of data and highly parallelized accelerator clusters, RNNs were loosing foot in terms of performence and have today been mostly replaced by other architectures like CNNs or attention based models like transformers.\cite{lakewComparisonTransformerRecurrent2018, wolfTransformersStateoftheArtNatural2020}
	
	We argue that RNNs should not be the first choice for typical inverse design tasks, because of their drawbacks such as non-parallel training and more complex hyperparameter tuning compared to CNNs.

	\paragraph{Graph neural networks (GNN)}
	Graph Neural Networks (GNNs) are an emerging class of network architectures that process data represented as graphs. Graphs are natural representations for various data structures, for example social networks or molecules in chemistry.\cite{scarselliGraphNeuralNetwork2009, bronsteinGeometricDeepLearning2021}
	Several variants of GNNs exist, such as convolutional GNNs,\cite{kipfSemiSupervisedClassificationGraph2017} Graph-attention networks,\cite{velickovicGraphAttentionNetworks2018} or recurrent GNNs. \cite{liGatedGraphSequence2017}
	Please note that also CNNs are strictly speaking a specific type of GNN, processing image-``graphs'' where connections exist only between neighbor pixels.
	One of their advantages is that they can operate on data of variable input size and are very flexible regarding the data format.
	In physics GNNs have been proposed for example to learn dynamic mesh representations.\cite{deshpandeMAgNETGraphUNet2023}
	In nano-photonics, GNNs have started to be used only recently. 
	GNNs have also been used for the decription of optically coupled systems like  metasurfaces, including non-local effects.\cite{khoramGraphNeuralNetworks2023}
	It has been also demonstrated recently that GNNs can learn domain-size agnostic computation schemes. A GNN learned the finite difference time domain (FDTD) time-step update scheme to calculate light propagation through complex environments.\cite{kuhnExploitingGraphNeural2023}

	In our opinion, GNNs can be a very interesting approach, i.e. for difficult parametrizations. But we argue that they should not be the first choice for a newcomer, because of the scarce literature in nano-photonics and challenges in its proper configuration.


	\paragraph{Transformer} 
	
	The transformer is a recent, very successful, attention-based model class.\cite{vaswaniAttentionAllYou2017} 
	The underlying attention mechanism\cite{bahdanauNeuralMachineTranslation2016} mimics cognitive focusing on important stimuli while omitting insignificant information. 
	The attention module gives the network the capacity to learn a hierarchy of correlations within an input sequence.
	In natural language processing (NLP), transformers largely outperform formerly used recurrent neural network architectures, which they entirely replaced in all NLP applications.\cite{lakewComparisonTransformerRecurrent2018, brownLanguageModelsAre2020} 
	In 2020 the concept was adopted to computer vision with so-called vision transformers (ViT), followed by important research efforts with some remarkable results.\cite{cordonnierRelationshipSelfAttentionConvolutional2020, dosovitskiyImageWorth16x162021, liuSwinTransformerHierarchical2021, naseerIntriguingPropertiesVision2021}
	However, tuning the hyperparameters of a transformer is significantly more difficult than conceiving a good CNN design, therefore also approaches that combine CNNs with ViTs were proposed.\cite{xiaoEarlyConvolutionsHelp2021, daiCoAtNetMarryingConvolution2021}
	Furthermore, while the transformer's main advantage is its excellent scaling behavior to huge datasets and model sizes, the need for gigantic datasets is probably also their biggest shortcoming. When the size of the datasets decrease, or for smaller model sizes, their advantages diminish.\cite{liuConvNet2020s2022}
	In fact, ``small'' datasets in the world of transformers still range in the order of hundreds of thousands samples.\cite{leeVisionTransformerSmallSize2021}
	
	We argue that transformers are in most cases not the most adequate architecture for inverse design, since their advantages unleash only for datasets of gigantic sizes, in the order of millions or even billions of samples.\cite{liuConvNet2020s2022, maOptoGPTFoundationModel2023}

	\subsubsection{Which network type to use}
	
	Comparing the arguments in favor and against the different network architecture that are presented in the preceding section, we come to the conclusion that, whenever applicable, CNNs are generally the best first choice for approaching an inverse design (or other) problem with deep learning. They are simple to design, robust in training and have excellent scaling behavior.
	If the data is not in an adequate format, often it is even advantageous to reformat the data in order to render it compatible with a CNN.
	There is no ``one'' recipe to do so, but if a meaningful structure exists in the data values, it should be conserved. In case of a sequence of thin film layers for instance, it makes sense to concatenate layer thicknesses and layer materials in two lists of values that can be fed to a one-dimensional CNN (c.f. figure~\ref{fig:when_to_use_dl_intuition}b).
	
	Besides the first technical tests, in general one should not use simple sequences of convolutions, typically called ``Visual Geometry Groups'' (VGG). 
	It is long known that such VGG-type CNNs have severe limitations and do not scale well with model size.\cite{heDeepResidualLearning2015}
	Instead, convolutional layers should be organized in residual blocks\cite{heIdentityMappingsDeep2016} or, even better and only slightly more complex, in residual ``ResNeXt'' blocks with inverted bottleneck layers and grouped convolutions.\cite{xieAggregatedResidualTransformations2017, liuConvNet2020s2022} 
	We refer the reader to the supplementary Python notebook tutorials for a detailed technical description and example implementation.\cite{wiechaNewcomerGuideDeep2023}

	\subsubsection{Choosing the inverse design method}

	\paragraph{Iterative DL based approaches}
	
	Following conventional inverse design by global optimization, a deep learning surrogate forward model can be used to accelerate iterative optimization for inverse design.\cite{hegdePhotonicsInverseDesign2020, renInverseDeepLearning2022}
	
	Drawbacks of iterative methods are: They are slower than one-shot approaches (due to multiple evaluations through the iterations) and coupled with deep learning, they may bear the risk of convergence to network singularities, since they actively search for extrema in the parameter space.\cite{huangAdversarialAttacksNeural2017} 
	Also convergence to parameters in the extrapolation zone needs to be avoided, since often the physical model collapses in the extrapolation regime, adding an additional technical challenge since the interactive solver needs proper regularization.\cite{blanchard-dionneTeachingOpticsMachine2020, dengNeuraladjointMethodInverse2021}
	Such dangerous extrapolations can also be mitigated by increasing ground truth data: a real physical simulator is sometimes used (typically at designs proposed by the surrogate model), and the learned model is updated to take into account these new data. This is a classical procedure in machine-learning enhanced global optimization.\cite{luksicMetaModelFrameworkSurrogateBased2019, khowajaSurrogateModelsOptimization2021, huSurrogateLocallyInterpretableModels2020, popovMultifidelityEnsembleKalman2021, daveDeepSurrogateModels2020}

	\paragraph{Direct inverse design (``one-shot'')}
	
	An alternative class of inverse design methods are one-shot methods. The goal is to create a neural network, which takes the design target as input and immediately returns a geometry candidate that implements the desired functionality.
	The main advantage of such methods is the ultimate speed. A single network call gives the response to the inverse problem.
	However, this also means that no optimization is performed in this case, and the solution is likely not the optimum. To obtain close-to-optimum solutions from one-shot approaches, considerable effort needs to be invested in dataset generation, network design, optimum training and proper testing.\cite{wenRobustFreeformMetasurface2020}
	Alternatively, a one-shot inverse model can be very useful to provide high-quality initial guesses for a subsequent (gradient based) optimization.

	\subsubsection{The dataset -- questions to ask about the data}
	
	The data are the most important resource in a deep learning model. 
	It is therefore essential to meticulously carry out data generation or, if data already exists, to understand the dataset. 
	In both cases it is often helpful to perform appropriate preprocessing.
	As a rough guide, we provide a few questions that one should ask about the data:
	
	\begin{figure}
		\includegraphics[width=\columnwidth]{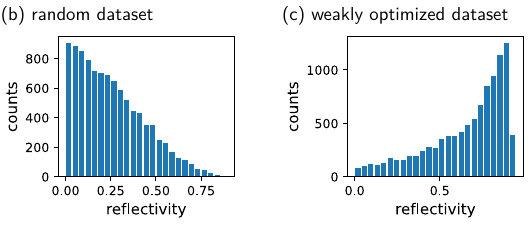}
		\caption{
			Histograms of two illustrative datasets of the reflectivity (at $\lambda=500\,$nm) of 10,000 dielectric thin-film sequences.
			(a) Randomly generated thin film sequences. Their reflectivity is in general low. A network for the design of high reflectivity solutions, that is trained on this dataset, will very likely fail.
			(b) Weakly optimized data generation, starting from random samples. The dataset offers a large portion of medium to high reflectivity samples. Care should be taken that still enough ``randomness' is present to avoid biases towards high $R$ solutions. 
			A network trained on the optimized set will perform better at the task of reflectivity maximization.
		}\label{fig:data_randomized_pre_optimization_stat}
	\end{figure}
	
	\begin{figure}
		\includegraphics[width=\columnwidth]{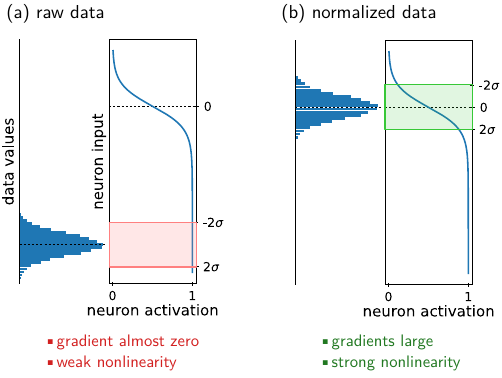}
		\caption{
			Normalizing the data is important. 
			(a) Large numerical values do not exploit the non-linearity of typical neuron activation functions. In case of specific activations like Sigmoid or tanh, furthermore the gradients of the neuron output are very small. Both situations are unfavorable for network training.
			(b) The numerical values of normalized data cover the full range of an artificial neuron's non-linearity. Also the neuron gradients are large for all typically used activation functions. This significantly helps the learning process.
			On non normalized data, during training first the neuron biases need to be optimized which consumes unnecessary computation.
			The same argument explains why batch normalization is so effective.
		}\label{fig:data_normalization}
	\end{figure}
	
	\begin{itemize}
		\item \textit{How much data is necessary?}\newline
		Most deep learning architectures require at the very least a few thousand samples, likely more. There are cases where less data can suffice, for instance in transfer learning / fine-tuning of an existing model. In some cases it's possible to learn on slices of large but few samples, for instance in segmentation tasks.\cite{zeilerVisualizingUnderstandingConvolutional2014, ronnebergerUNetConvolutionalNetworks2015}
		\item \textit{Is random data generation possible?}\newline
		The design targets may be sparse in the parameter space, in which case training using a random dataset may not work. 
		This is often the case in free-form geometries, where the number of design parameters is high. Using a weak optimization method for data generation can help in such cases, but special care must be taken to avoid bias of the dataset towards a specific type of pre-optimized solution, so enough randomness is important.\cite{dinsdaleDeepLearningEnabled2021}
		Also systematic data sampling approaches (e.g. Sobol sampling) may be useful.\cite{provostEfficientProgressiveSampling1999, bierkensZigZagProcessSuperefficient2019, renardySobolNotSobol2021}
		The potential difference between random and optimized data is illustrated in figure~\ref{fig:data_randomized_pre_optimization_stat}.
		
		\item \textit{Are there obvious biases in the data?}\newline
		Biased data is one of the most important challenges in deep learning. In short: ``what you put in is what you get out''.
		For instance if a dataset is built with resonant nanostructures, a network trained on this data may not correctly identify non-resonant cases.
		\item \textit{Is the data meaningful?}\newline 
		Samples carrying the essential information may be sparse in the dataset or the important features may be hidden in a large number of irrelevant attributes. A careful pre-processing of the dataset may help in such cases.
		\item \textit{Can I shuffle the data?}\newline
		Are subsets of the data partly redundant (e.g. after sequential data generation)?
		If yes, a splitting method is required that guarantees that training, validation, and test subsets are independent.
		\item \textit{Are the subsets (training data, validation data) representative for the problem?}\newline
		In order for the training loss and benchmark metrics to be meaningful, each subset needs to fully cover the entire problem to be solved. 
		If we apply the network on significantly different samples than it was trained on, we probe the extrapolation regime, where performance is generally weak. A non-representative validation or test dataset will therefore provide a bad benchmark.\cite{blanchard-dionneTeachingOpticsMachine2020}
		An illustration depicting non-representative subsets is shown in figure~\ref{fig:deep_learning_workflow}c.
		\item \textit{How frequent are outliers}\newline
		If necessary, can the impact of exceptions or anomalies in the data be reduced, e.g. by selective removal or by pre-processing with a non-linear scaling function?\cite{yeoNewFamilyPower2000, karvanenEstimationQuantileMixtures2006}
		\item \textit{How should the data be normalized?}\newline
		Input values should in general be standardized: Subtract the mean and divide by the standard deviation. Figure~\ref{fig:data_normalization} depicts why this is essential.
		The choice of the normalization for the network output is closely linked to the choice of the loss function, but also depends on the data representation and on the significance of outliers, among other factors. 
		If the data has multiple output values or channels, are those of similar numerical magnitude? If not, should they be normalized with the same scaling process, or each channel individually? Appropriate normalization is highly important for an accurate model.
		\item \textit{Can we exploit inductive biases?}\newline
		The term \textit{inductive bias} denotes any assumption about how to solve a task, that is implicitly included in the model. In a classification task for instance, the output layer is chosen as a probability distribution. In a regression task, the output layer is a continuous activation function. Such inductive biases improve the network training. In their absence, a network needs to learn them from the data in addition to the actual task. For example a classifier would need to learn that the output is a binary yes/no decision, and parts of the network would be used for this aspect.
		Often it is possible to include some properties of the problem implicitly in the network architecture. CNNs typically imply translation invariance. Likewise, symmetries may be included, or the valid data range, by choice of an adequate output activation. In physics, a possibility to exploit inductive bias is causality, which may be enforced through Kramers-Kronig relations or, more simply, using a Lorentzian output layer.\cite{blanchard-dionneTeachingOpticsMachine2020, khatibLearningPhysicsAllDielectric2022}
	\end{itemize}

	\subsubsection{Practical network implementation and training}
	
	Once the data is prepared, the full set needs to be split into training, validation and test subsets. With these subsets we are ready for the actual training of a network. The next step is to code and train the model. This section is a little technical, we ask the reader to look at the Python notebook tutorials that demonstrate explicit implementations of networks following the below guidelines.\cite{wiechaNewcomerGuideDeep2023}
	
	Since a sophisticated model may train very slowly, it is important to do a first technical test with a very simple network model. Even though you shouldn't use it as a final layout, the initial test network can be a sequence of a few stacked convolutional layers (``Visual Geometry Group'', VGG). With this we test the implementation of the input and output layer dimensions, if the data scales are correct, and if the network output format is adequate. Finally we get a first idea about the typical learning loss.
	As for the specific configuration of the convolutions, small 3x3 kernels, strides 1, and ``ReLU'' activations can be used as a relatively foolproof rule of thumb.\cite{xieAggregatedResidualTransformations2017}
	Kernels with larger receptive field are occasionally used in literature,\cite{liuConvNet2020s2022} but increasing the convolutional kernel size increases quadratically (in 2D) the computational cost as well as the number of network parameters. 
	It can be easily shown that stacking $N$ $3\times 3$ convolutions covers a receptive field equivalent to a single layer with kernel size $(2N+1) \times (2N+1)$, but with significantly less trainable parameters, reduced computational effort and potentially the additional benefit of hierarchical feature extraction capabilities.\cite{luoUnderstandingEffectiveReceptive2017}
	
	Once this technical test is passed, it is time to increase the model complexity. 
	Simple convolutions are converted into residual blocks\cite{heIdentityMappingsDeep2016} or, even better, ``ResNeXt'' blocks,\cite{xieAggregatedResidualTransformations2017, liuConvNet2020s2022} and more and more layers are added to the network, while paying attention to overfitting.
	Once overfitting occurs, the network complexity needs to be reduced again, or if the accuracy is not yet sufficient, other regularization strategies can be considered, such as data augmentation (for example, shifting or rotating the input data, if possible). 
	Following dense layers, dropout can be applied, which randomly deactivates some of the neurons during training, leading to a model less prone to overfitting.\cite{srivastavaDropoutSimpleWay2014}

	\paragraph*{A note on overfitting}
	
	Recent research showed that overfitting occurs mainly in medium size networks, at the so-called interpolation threshold, where the number of model parameters is comparable to the number of degrees of freedom of the dataset. 
	It appears that overfitting can not only be avoided by reducing the model size (or increasing the amount of data), but it can in fact be circumvented also by increasing the model size far into the overdetermined region.\cite{belkinReconcilingModernMachinelearning2019, loogBriefPrehistoryDouble2020, nakkiranDeepDoubleDescent2021, schaefferDoubleDescentDemystified2023}
	These findings are in direct link with the modern concept of so-called \textit{foundation models} like the large language models GPT3/GPT4.\cite{brownLanguageModelsAre2020, bubeckSparksArtificialGeneral2023} 
	Such foundation models are extremely large network models that are pre-trained on a gigantic corpus of generic data, and in a second step are fine-tuned on specific downstream tasks, which is usualyl successful even on very small datasets.\cite{bommasaniOpportunitiesRisksFoundation2022, maOptoGPTFoundationModel2023}
	
	In a typical scenario of deep learning for nano-photonics inverse design however, the model and dataset sizes are likely in a regime where the conventional considerations regarding overfitting hold and corresponding parameter tuning will be beneficial.
	
	\paragraph*{Dropout and batch normalization}
	
	We feel it is important to dedicate a paragraph on dropout and batch normalization. 
	Both methods are applied on training time. Dropout deactivates random subsets of neurons with the goal to avoid overfitting because the model cannot rely on exactly memorized information.\cite{srivastavaDropoutSimpleWay2014}
	Batch normalization normalizes the samples that it receives according to the statistics of the current batch of samples, with the goal to ideally exploit gradients and non-linearities of the activation functions (see also figure~\ref{fig:data_normalization}).\cite{ioffeBatchNormalizationAccelerating2015}
	
	Even though dropout is often found in internet tutorials, using it in convolutions should be avoided. Its effect in CNNs is entirely different than in dense layers, for which it was originally proposed. If it does help regularizing a CNN, this will be rather accidental, than an effect that can in general be expected.\cite{mianjyImplicitBiasDropout2018}
	
	Instead of dropout, batch normalization (BN) should be used in deep CNNs (in particular in deep res-nets), after the convolution and before the nonlinearity.\cite{ioffeBatchNormalizationAccelerating2015} 
	As a rule of thumb, BN and dropout should not be acting on the same layer of neurons.\cite{liUnderstandingDisharmonyDropout2018}
	BN is indispensable in very deep architectures as it counteracts internal covariate shift. Furthermore, BN typically stabilizes training in the sense that it allows to use larger learning rates.
	
	However, in smaller networks (in the order of a few tens of convolutions) and in particular with normalized data, BN is often not very useful, since the data normalization is more or less conserved through late layers in smaller networks. 
	On the other hand, BN slows down data processing and training due to its computational overhead.\cite{brockCharacterizingSignalPropagation2021} 
	Even worse though, in physics regression tasks, using batch normalization (and also dropout) can be an unexpected pitfall, often difficult to pinpoint.
	This is because BN and dropout are both dynamic regularization techniques that act differently on every training batch. This can in fact be problematic when the statistics of individual batches fluctuate significantly, which is the case if the training data has a large variance.  
	The network then re-normalizes each data batch very differently, introducing a new statistical error.
	In many applications, such as in segmentation tasks or in classification problems, the network output goes through a final normalization layer (typically softmax), then this problem is usually negligible.\cite{lianRevisitBatchNormalization2019, ozgurEffectDropoutLayer2020}
	For regression tasks on the other hand, this can be a source of a considerable error. 
	Large batch-sizes in the late training can counteract the problem to a certain extent. Still, in many regression problems, in particular with small to medium size network models, the best solution is often to avoid using batch normalization (or dropout).\cite{wuRethinkingBatchBatchNorm2021}
	When large models are required, BN should be used with an adequate batch-size increase schedule.
	
	To conclude, we want to remind again, that normalization of the input data using the whole data-set statistics is a small, yet crucial step in the pre-processing workflow (see figure~\ref{fig:data_normalization}).
	This by itself speeds up training due to the same mechanisms as used in BN, while avoiding some of the problems mentioned here.\cite{lecunEfficientBackProp1998}

	\subsubsection{Training loop good habits}
	
	A lot of frustration can be avoided if some good habits are respected concerning the training loop itself.

	\begin{itemize}
		\item \textit{Learning rate:} The learning rate is an important training hyperparameter. If it is too small, training gets stuck in a local minimum. If it is too large, the network parameters may ``overshoot'', resulting in a diverging training. 
		The choice of the initial learning rate (LR) is strongly dependent on the problem and on the network architecture. Typically it needs to be tested which LR works well. Once a good starting value is found, the LR should not be left constant through the entire training loop. Training rate schedules of varying complexity exist, but a good starting procedure is usually to gradually reduce the LR (see also figure~\ref{fig:neuron_and_neuralnet}c-d). 
		A simple learning rate decay schedule such as dividing the LR by a factor 10 every time the validation loss stagnates for a few epochs, is often already very efficient.
		\item \textit{Batch size:} Evaluating the training loss on small batches of random samples is one of the key mechanisms in deep learning. It is the pivotal procedure that impedes the optimizer to get stuck in local minima. As a result the batch size (BS) is also a crucial hyperparameter, that has a tremendous impact on the training convergence and model performance. 
		While it depends also on each specific problem, in general the BS should be small in early training. 16 or 32 are good initial values. Too large starting batch sizes usually converge to local minima and thus lead to less accurate models.\cite{keskarLargeBatchTrainingDeep2017, mastersRevisitingSmallBatch2018}
		However, once the loss stagnates, it is good practice to increase the batch size since it renders the gradients smoother, and helps further decreasing the loss function, similar to the reduction of the learning rate.\cite{smithDonDecayLearning2018}
		Please note that, as in any global optimization scheme, it is impossible to evaluate whether the actual global minimum was reached, because in general training will always end in a local minimum.
	\end{itemize}
	
	A few additional, rather technical tips are given in a dedicated section~\ref{app:techical_tips}.

	\subsection{Alternative data-based approaches}
	
	Developing, optimizing and training a deep learning model can require a considerable time investment and can be very resource consuming. 
	Having a look at other data based approaches can be worth a try. 
	Methods such as k-means or principle component analysis,\cite{fournierEmpiricalComparisonAutoencoders2019} clustering algorithms like DBscan\cite{schubertDBSCANRevisitedRevisited2017} or t-SNE,\cite{vandermaatenVisualizingHighDimensionalData2008} and machine learning methods like support vector machines\cite{cortesSupportvectorNetworks1995} or random forest\cite{breimanRandomForests2001} are typically straightforward in their application and often computationally more efficient than large deep learning models.
	Since this is not the scope of this work, we refer the reader to the above cited literature.
	The python package \href{https://scikit-learn.org}{scikit-learn} is also a very accessible opensource collection of machine learning algorithms and tools with an extensive and well-written online documentation.\cite{pedregosaScikitlearnMachineLearning2011}

	\section{Deep learning based inverse design}\label{sec:deep_learning_invdesign}

	\begin{figure}
		\includegraphics[width=\columnwidth]{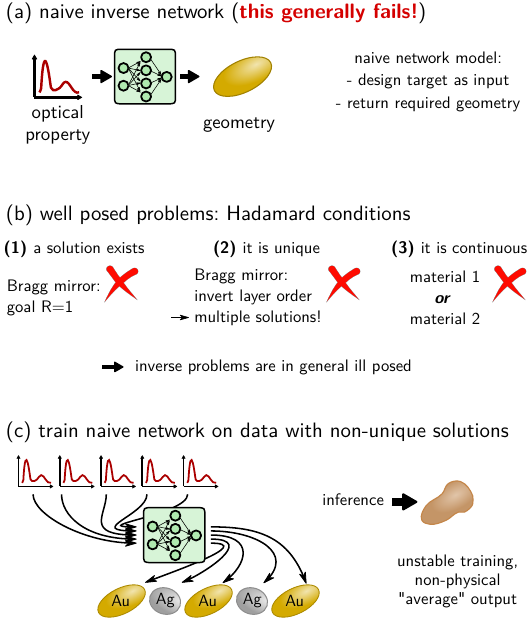}
		\caption{
			The crux of the ill posed problem.
			(a) A naive, not working implementation of a simple feed-forward inverse network would take as input the design target (e.g. an optical property) and returns the design that is required to obtain it.
			(b) Only well posed physical problems can be solved this way. Such problem obeys the three Hadamard conditions. 
			However, neither of these conditions is in general fulfilled in photonics inverse design, as illustrated by a selected example under each condition.
			(c) In case of multiple solutions, the training process would iterate of these several times, every time adapting the network parameters to return a different design. Training is unstable and eventually the network will learn some non-physical mix of the multiple solutions. If a non-continuous parameterization is used (here: two distinct materials), the naive network may also return non-allowed mixtures of those.
		}\label{fig:ill_posed_inv_problem}
	\end{figure}

	\begin{figure}
		\includegraphics[width=\columnwidth]{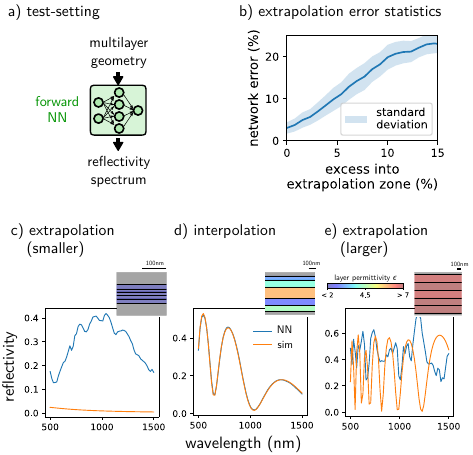}
		\caption{
			Illustration of failed extrapolation. 
			(a) A neural network was trained on predicting the reflectivity spectrum of a thin film layer stack made of ideal dielectrics with constant permittivity $\varepsilon$. 
			(b) average forward network error as function of the excess of the input parameters outside of the training data range.
			(c) Failed prediction example outside of the training data range (smaller permittivities and thinner layers).
			(d) Example inside the range of the training-set design parameters (interpolation). The network predictions are accurate.
			(e) Same as (b) but for larger permittivities and thicker layers than used in training.
			Orange lines: PyMoosh simulated reflectivity spectra. 
			Blue lines: neural network predicted spectra.
			Insets represent the used layer stack. Bar heights correspond to the layer thickness according to the scale bar, the color code indicates the layer dielectric's permittivity.
		}\label{fig:extrapolation_fail_example}
	\end{figure}

	After having discussed the general approach and workflow for deep learning, we will now explain different possible approaches to solve inverse design problems via deep artificial neural networks. 
	
	The naive approach to solve inverse design with deep learning would be to use a feed-forward neural network that takes the optical property as input and returns the geometry parameters as output. This would be trained on a large dataset.
	Unfortunately such approach, as depicted in figure~\ref{fig:ill_posed_inv_problem}a, does not work.
	
	The main challenge in nanophotonics inverse design is that the problem is in general ill posed, in consequence it is impossible to solve the problem directly.
	J. Hadamard described a so-called ``well posed problem'' as one for which a solution does exist, this solution is unique and continuously dependent on the parameterization (c.f. figure~\ref{fig:ill_posed_inv_problem}b).\cite{hadamardProblemesAuxDerives1902}
	The typical inverse design problem however has in general non-unique solutions (multiple geometries yield the same or very similar property). 
	Often design targets exist that cannot be optimally implemented, hence no exact solution exists (e.g. a mirror with unitary reflectivity).
	And finally, in many cases the physical property of a device is not continuously dependent on the geometry, but the parameter space is at least partially discrete (e.g. if a choice from a finite number of materials has to be made).
	Training of a naive network on a problem with multiple possible solutions will oscillate between the different possible outputs and finally learn some non-physical average between those different solutions, as illustrated in Fig.~\ref{fig:ill_posed_inv_problem}c.\cite{wiechaDeepLearningNanophotonics2021}
	
	Fortunately, methods exist to solve ill posed inverse problems with deep learning. 
	We discuss in the following two popular groups of approaches.
	The first contains iterative methods that use optimization algorithms to discover the best possible solution(s). 
	The second group consists of direct (``one-shot'') inverse design methods.

	\begin{figure}
		\includegraphics[width=\columnwidth]{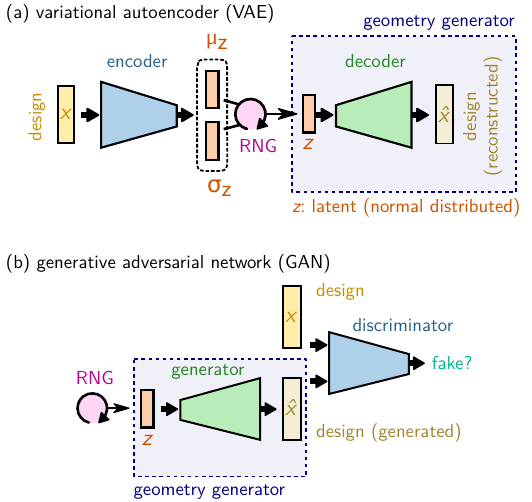}
		\caption{
			Avoid the extrapolation regime of the forward model with learned design parameterization.
			(a) Sketch of a variational autoencoder (VAE) trained to reconstruct the design. In a VAE, the encoder is trained to return the mean value (\textmu$_z$) and standard deviation ($\sigma_z$) of the latent variable $z$. A randomized, normal distributed latent vector is passed to the decoder for the reconstruction task (via random number generator ``RNG''). By further constraining $\sigma_z$ with a KL loss (c.f. text), one obtains a compact and smooth latent space that is normally distributed.
			(b) Sketch of a generative adversarial network (GAN). 
			As in the VAE, by using a normal distributed random number generator during training for the latent space input, the generator develops a smooth and compact latent space, essentially representing the interpolation regime of the dataset.
		}\label{fig:design_methods_design_parametrization}
	\end{figure}

	\subsection{Iterative approaches}
	
	Conventional inverse design generally uses iterative methods to search for an optimum solution to a given design problem.
	One way to speed up inverse design with deep learning is hence to accelerate the iterative process with a deep learning surrogate model.
	A surrogate model is a forward deep learning network that is trained on predicting the physical property of a structure, hence to solve the ``direct'' problem.\cite{hegdeDeepLearningNew2020, hegdePhotonicsInverseDesign2020, wiechaDeepLearningMeets2020, hegdeSampleefficientDeepLearning2021, majorelDeepLearningEnabled2022}
	This is usually straightforward, and consists in a feed-forward neural network that takes the design parameters as input (e.g. geometry, materials, ...) and returns the physical property of interest via its output layer (e.g. reflection or absorption spectrum, ...).
	Provided a large enough dataset is available, designing and training of such forward model is generally not difficult. 
	We advise to follow the design guidelines from section~\ref{sec:deep_learning_general} and to consult the Python tutorial notebooks.

	An accurate forward model can be used in various ways for rapid iterative inverse design.

	\begin{figure*}
		\includegraphics[width=\linewidth]{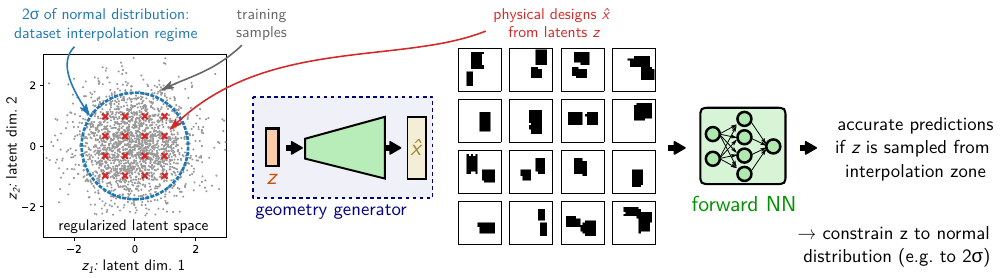}
		\caption{
			Re-parameterized the forward model using a learned latent representation as design input:
			The trained geometry generator (e.g. from a VAE or a GAN) is simply plugged before the input of the forward network. It converts a latent vector $z$ to a physical design $\hat{x}$.
			If the latent space was properly regularized, sampling from within the range of a normal distribution with unitary variance will generate geometries in the interpolation regime of the training data, where the forward network works accurately.
			Instead of optimizing the physical design parameters, we can now run the optimization on the latent variable of the geometry generator. 
			Constraining the numerical range of the optimization parameters accordingly, renders iterative optimization robust.
		}\label{fig:design_methods_reparametrized_fwd_model}
	\end{figure*}

	\subsubsection{Avoiding the extrapolation zone}
	
	The most imminent danger when using iterative optimization with deep learning surrogates is that the optimizer pushes the design parameters outside of the model's validity range.
	Most neural networks are strong at interpolation but bad at extrapolation tasks.
	This is illustrated in figure~\ref{fig:extrapolation_fail_example}, where a neural network has been trained on predicting the reflectivity of a dielectric thin film layer stack. This network totally fails in doing so outside of the range of the training data.
	For input values that are $10\%$ larger than the largest training data, the average error is already in the order of $20\%$ (Fig.~\ref{fig:extrapolation_fail_example}b).
	This illustrates that it is important to constrain the allowed designs to the parameter range that has been used for training, hence to the interpolation regime of the forward model.
	Please note that even if the model is efficiently constrained, a non-negligible risk of converging to network singularities remains in any case, since optimization algorithms seek extrema of the target function.\cite{goodfellowExplainingHarnessingAdversarial2015, huangAdversarialAttacksNeural2017}
	We insist therefore that careful verification of results obtained from deep learning is crucial in any case.
	A classical approach is also to retrain the surrogate model with the new data, after output data has been verified by a classical physical model.
	
	In order to constrain the designs, in specific cases dependent on the design parameterization, it is possible to formulate a penalty loss. For instance when the design is described by size values for which the limits are known.\cite{dengNeuraladjointMethodInverse2021}
	The penalty loss is added to the optimization fitness function, which then increases, if the optimizer leaves the allowed parameter range.
	A drawback of this method can be the increasing complexity of the fitness function, and the requirement of problem-specific weight tuning to balance the contributions to the total fitness function.

	In many situations this tuning can turn out to be complicated or the parameterization of the design is difficult to constrain, for example in free-form optimization, where the design may be given as a 2D image.
	In such cases, a separate deep learning model can provide an elegant solution.
	By training in a first step a generative model on the designs alone (without any physics knowledge), a design parameterization can be \textit{learned} from the dataset.\cite{melatiMappingGlobalDesign2019, liuHybridStrategyDiscovery2020, zandehshahvarManifoldLearningKnowledge2022}
	Adequate models are for instance variational autoencoders (VAEs) or generative adversarial networks (GANs).
	These two approaches are depicted in figure~\ref{fig:design_methods_design_parametrization}a, respectively~\ref{fig:design_methods_design_parametrization}b.
	For technical details and an example implementation, we refer the reader to the tutorial notebooks in the supplemental material.
	
	The latent space of such a VAE or GAN represents a learned parameterization of the designs. 
	Thanks to the properties of the regularized latent space of VAEs or GANs, the parameterization is compact and continuous, even if the original design description was not (e.g. if a list of discrete materials is used). 
	Please note that some techniques like variational regularization are not differentiable. Since backpropagation necessarily requires deterministic gradients, reparametrization tricks need to be applied to the backward path in such cases \cite{kingmaAutoEncodingVariationalBayes2022}.
	Provided the VAE or GAN training did converge, this means that every point in the latent space yields a physically meaningful solution.
	Most importantly, the latent space of a well trained VAE or GAN is regularized to a normal distribution with unitary standard deviation. 
	
	This means that in an iterative optimization loop, we can now constrain the latent variable that describes the designs to the numerical range of a normal distribution (e.g. to a $2\sigma$ confidence interval). 
	This is then equivalent to constraining the entire problem to the interpolation regime of the dataset.
	Practically we replace the original design parameters by the latent input variable of the trained geometry generator, as depicted in figure~\ref{fig:design_methods_reparametrized_fwd_model}.
	In the subsequent optimization, the latent input can be conveniently constrained to a normal distribution e.g. by simply penalizing large values as in inspirational generation,\cite{roziereInspirationalAdversarialImage2021} or by using a Kullback–Leibler divergence loss (KL loss), as used for variational autoencoders.\cite{kullbackInformationSufficiency1951, kingmaIntroductionVariationalAutoencoders2019}

	\paragraph*{Technical hint: GAN normalization.}
	To conclude, we like to put emphasis on an important technical detail in the generative adversarial network layout, especially in deep GAN architectures. 
	In fact, the activation function of the generator output and the normalization of the associated data is important for a robust model design. 
	As discussed above (c.f. sections on data normalization and on batch normalization), the statistical assumption behind deep learning is that the data follows a normal distribution with mean value of zero and unity variance (c.f. also Fig.~\ref{fig:data_normalization}). 
	Therefore, the design parameters ($x$ in Fig.~\ref{fig:design_methods_design_parametrization}b) should follow this assumption, since the output of the GAN generator is being fed back into the discriminator network. 
	The most simple mean to accomplish this is to normalize the data (e.g. the design images) between $[-1,1]$, and use a tanh activation function at the generator output.

	\subsubsection{Heuristics: Forward network with global optimization}
	
	A robust way to overcome local extrema and assure convergence to the global optimum is using gradient-free heuristics such as evolutionary optimization or genetic algorithms.\cite{wiechaEvolutionaryMultiobjectiveOptimization2017, wiechaDesignPlasmonicDirectional2019, elsawyNumericalOptimizationMethods2020, liuVersatileBlackboxOptimization2020, barryEvolutionaryAlgorithmsConverge2020, bruleMagneticElectricPurcell2022}
	
\begin{figure}
		\includegraphics[width=\columnwidth]{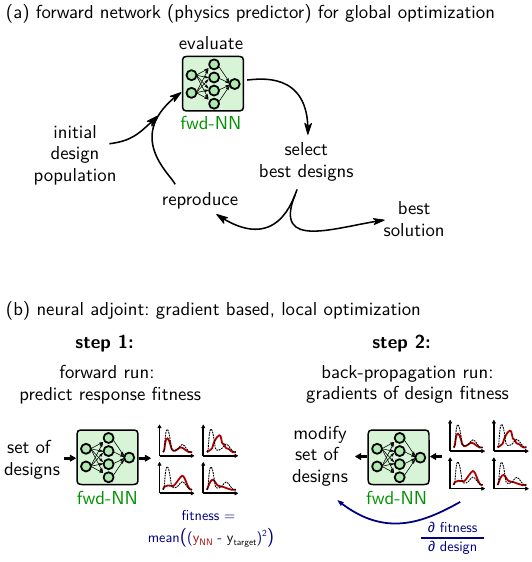}
		\caption{
			Inverse design via iterative optimization using a forward model as fast physics solver surrogate.
			(a) Use forward model to accelerate a global optimization loop. 
			(b) Neural adjoint method: neural networks are differentiable and allow gradient based optimization. Reduce risk of local minima by operating on a large set of random initial designs. The optimization tries to minimize the error between the predicted optical property (solid red line) and the design target (dashed black line).
		}\label{fig:design_methods_forward_model}
	\end{figure}

	Accelerating global optimization based inverse design with deep learning is in principle straightforward. 
	Instead of using numerical simulations, the evaluation step in the loop of the optimizer (e.g. evolutionary optimization, particle swarm, genetic algorithm...) is done with a deep learning surrogate model.
	This is depicted in figure~\ref{fig:design_methods_forward_model}a. 
	As discussed in the previous section, it is recommended to constrain the designs to the interpolation regime, for example using a generative model that precedes the physics predictor, as illustrated in figure~\ref{fig:design_methods_reparametrized_fwd_model}.
	The actual optimization can be done with any algorithm, since the operation of the deep learning model is restricted to the calculation of the fitness function.
	
	One of the major advantages of deep learning surrogates is their differentiability. We therefore encourage to not use gradient-free global optimization alone, but combine it with gradient-based optimization for faster convergence. This will be discussed in the following.

	\subsubsection{Gradient descent -- Neural adjoint method}
	
	While global optimization is robust and generally converges well towards the overall optimum, such methods are also inherently slow since they do not take advantage of gradients. 
	This is unfortunate because gradients are available ``for free'' in deep learning surrogate models. 
	On the other hand, gradient based approaches tend to get stuck in local minima. 
	This can be accounted for to a certain extent, but it usually depends strongly on the individual design problem if a gradient based method will work.
	
	The idea of gradient based optimization is the same as in the Newton-Raphson method.
	A fitness function is defined such that it is a measure of the error of a solution compared to the design target.
	Then, the derivatives of the fitness function with respect to the design parameters of a test solution are calculated and used to modify the test-design towards the negative gradient. By minimizing the fitness function in this way, the solution iteratively gets closer and closer to the ideal design target until a minimum is reached.
	
	Typical numerical simulation methods are not differentiable, and hence gradient based methods cannot be applied directly.
	While gradients can be calculated using adjont methods,\cite{jensenTopologyOptimizationNanophotonics2011} these still require multiple calls of the, generally, slow simulation, and hence are usually computationally expensive.
	Both problems can be solved to some extent by forward neural network models.
	A key advantage is, besides the evaluation speed, that gradients can be calculated ``for free'', because the network is an analytical mathematical function.
	For the same reason, the gradients of the surrogate model are also continuous, since this is a key requirement for the network training alogrithms.
	As stated in the beginning, the training procedure of a neural network is in fact a gradient based optimization by itself, therefore the main functionality of all deep learning toolkits is automatic differentiation.
	A forward neural network model can thus always be used for gradient based inverse design, which consists of two steps that are illustrated in figure~\ref{fig:design_methods_forward_model}b.
	In a first step, a set of test-designs (typically random initial values) is evaluated with the forward model. 
	Their predicted physical behavior is compared to the design target, for which a fitness function evaluates the error between target and prediction. 
	Now the deep learning toolkit is used to calculate the gradients of this fitness with respect to the input design parameters via backpropagation and the chain rule. 
	Finally, the designs are modified by a small step towards the negative gradients. Repeating this procedure minimizes the fitness.\cite{peurifoyNanophotonicParticleSimulation2018, asanoIterativeOptimizationPhotonic2019, dengNeuraladjointMethodInverse2021, renInverseDeepLearning2022, jingDeepNeuralNetwork2023, augensteinNeuralOperatorbasedSurrogate2023}
	
	Note that, if the gradients of the underlying physics data source are known, they can be added in the training step. This often leads to better convergence of the model, since the training can use deeper correlations to build its model.\cite{sitzmannImplicitNeuralRepresentations2020}
	
	\begin{figure}
		\includegraphics[width=\columnwidth]{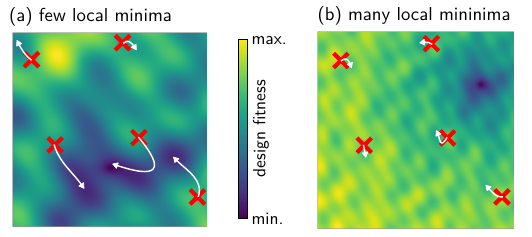}
		\caption{
			Schematic fitness landscapes of (a) a friendly problem with relatively few local extrema. (b) a complicate problem with many local fitness minima. 
			Using gradient based methods with a large number of initial test-sets, problem (a) will likely converge to the global optimum. 
			In problem (b) on the contrary, the chance is high that none of the initial designs is close enough to the global optimum, and optimization will converge to a local solution. The paths taken by a gradient-based method path are indicated by white arrows.
		}\label{fig:design_methods_many_global_minima}
	\end{figure}

	As mentioned before, the main difficulty in this approach is to avoid getting stuck in local minima of the fitness function. To a certain extent this can be accounted for by optimizing a large number of random designs in parallel (see Fig.~\ref{fig:design_methods_many_global_minima}a). 
	While such strategy would be prohibitively expensive using numerical simulations, with a machine learning surrogate model it is typically possible to optimize several hundreds or even thousands of designs in parallel.
	However, depending on the problem, the number of local extrema may be too large for successful convergence (see Fig.~\ref{fig:design_methods_many_global_minima}b).
	This can be tested by running the optimization several times. 
	If multiple runs do not converge to a similar solution, the parameter landscape of the problem is probably too ``bumpy'' for gradient based inverse design.
	
	As explained above, it is crucial also in gradient based optimization to remain in the forward model's interpolation regime since extrapolation bears a high risk of converging towards non-physical minima of the deep learning model.\cite{dengNeuraladjointMethodInverse2021}
	Also, if the dimensionality of a problem is high, the risk of strongly varying gradients further increases and optimization may always converge to unsatisfying local minima. 
	As discussed above, in such cases it is helpful to train a separate generator network that maps the design parameters onto a regularized latent space (e.g. VAEs or GANs, c.f. also Fig~\ref{fig:design_methods_design_parametrization}). 
	Instead of optimizing the physical design parameters, the optimizer then acts on this design latent space. 
	Because the latent space is regularized, it is possible to constrain the designs to the neural network's interpolation regime e.g. by using a KL loss term in the fitness function.
	
	Finally we want to recall once again, that unlike conventional numerical simulation methods, deep learning surrogates can possess singularities, also called failure modes or adversarial examples.\cite{huangAdversarialAttacksNeural2017} 
	Gradient based optimization, especially, comes with the risk of converging to such singularities of the surrogate network.
	In fact, the Neural Adjoint method is very similar to the ``Fast Gradient Signed Method'' that is specifically used to find network failure modes.\cite{goodfellowExplainingHarnessingAdversarial2015}
	To avoid convergence to a network singularity, it has been proposed to alternate the evaluation in the optimizer loop between the surrogate network model and exact numerical simulations. 
	Such occasional verification of the optimized solutions effectively eliminates non-physical designs.\cite{hegdePhotonicsInverseDesign2020}

	\begin{figure}
		\includegraphics[width=\columnwidth]{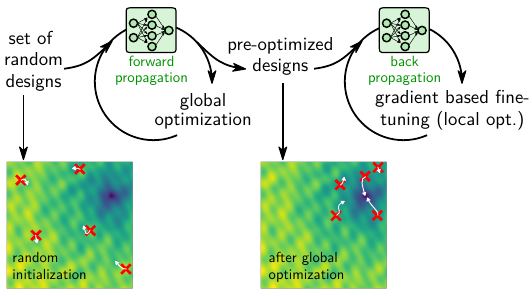}
		\caption{
			Global pre-optimization.
			If too many local minima exist, a promising solution is to start by pre-optimizing a set of random designs with global optimization.
			The positions of the random initial samples in an illustrative fitness landscape are depicted in the bottom left.
			After a few iterations of a global optimizer, the solutions are closer to the global optimum, as illustrated in the bottom right. Using this set as initial population for gradient based neural adjoint is likely to converge.
		}\label{fig:design_methods_pre_optimization}
	\end{figure}

	\subsubsection{Hybrid approach: Global optimization followed by neural adjoint}

	As mentioned above, inverse design tasks often possess of a large number of local extrema. As illustrated in figure~\ref{fig:design_methods_many_global_minima}b, in such case gradient based algorithms may get stuck in those local extrema, even if a large number of designs is optimized in parallel.
	A possible solution can be a combination of iterative global optimization and local gradient based neural adjoint.
	In such a scenario, a global optimizer runs for a few iterations with a rather large population of solutions. During the first generations global solvers usually converge the most rapidly towards the global optimum. However, they can be expensive in the final convergence towards the exact extremum. 
	Using the population obtained from a few iterations of a global optimization run can be very helpful as the initial set of designs for the neural adjoint method. 
	Those designs are then relatively close to the global optimum and the chance that at least a few manage to avoid local minima is considerably increased.
	The approach is depicted schematically in figure~\ref{fig:design_methods_pre_optimization}c.

	\subsection{Direct inverse design networks}
	
	\begin{figure}
		\includegraphics[width=\columnwidth]{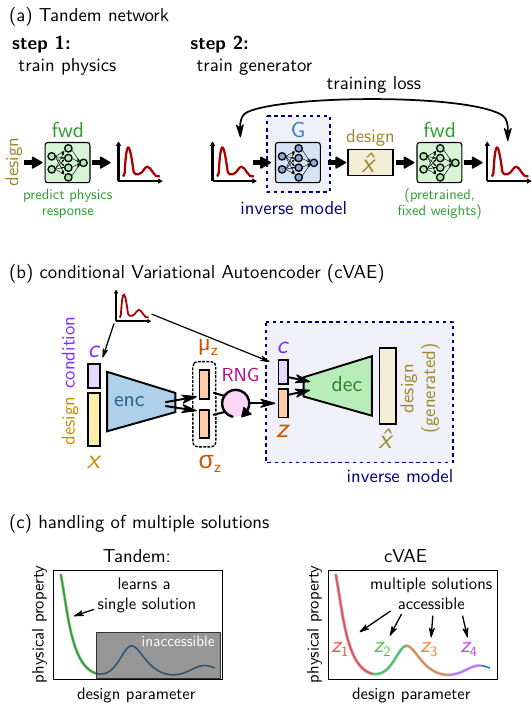}
		\caption{
			Direct inverse design models.
			(a) Tandem model. The training is divided in two steps. At first a forward predictor is trained on the direct problem. Subsequently, the forward network is fixed and used to train the generator,
			(b) The conditional Variational Autoencoder (cVAE) is trained end-to-end in a single run. A latent space $z$ is used to provide additional degrees of freedom to handle ambiguities in the design problem.
			(c) inverse problems typically can be solved by multiple solutions. A Tandem model will learn only one of possibly multiple solutions, the other remain inaccessible. The cVAE on the other hand typically learns the set of possible solutions which can be retrieved via the latent vector $z$.
		}\label{fig:design_methods_direct_models}
	\end{figure}
	
	\begin{figure*}
		\includegraphics[width=\linewidth]{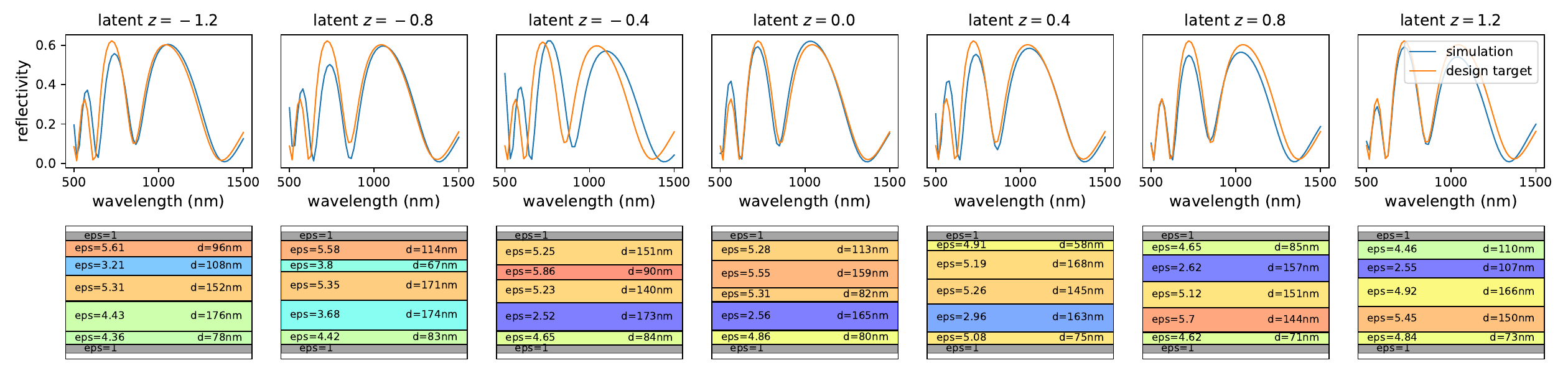}
		\caption{
			Dielectric layer stacks implementing an arbitrary reflectivity design spectrum. Inverse designed by a cVAE.
			By sweeping through the latent space of the cVAE generator with fixed target spectrum, multiple possible design solutions can be identified.
			Note that the cVAE discovered that mirrored structures yield the same reflectivity spectra (c.f. for example latents $z=0$ and $z=0.8$).
			A systematic latent inspection can also be done for further optimizing the solution, for example by a search for the best possible spectral match, or by identification of the most robust design, etc.
		}\label{fig:multiple_solutions_via_cVAE}
	\end{figure*}

	In the above discussed iterative approaches, the optimality of the solution is the highest priority. 
	Deep learning methods for direct inverse design on the other hand, put the design speed over all other criteria.
	The goal is to solve the inverse problem with a single network call. These approaches typically yield the ultimate acceleration, but results are generally not optimum, since no optimization algorithm pushes the solution to the extremum.
	These ``one-shot'' techniques rather perform a kind of similarity matching and typically yield solutions that resemble the design target.
	
	For the sake of accessibility we focus first on two popular variants, the Tandem network and conditional Variational Autoencoders, for which also detailed jupyter notebook tutorials are provided as supplemental documents.\cite{wiechaNewcomerGuideDeep2023}
	Subsequently we give a brief overview of other direct inverse design techniques.

	\subsubsection{Tandem network}
	
	One of the most simple configurations for a one-shot inverse design network is the so-called ``tandem network''.\cite{liuTrainingDeepNeural2018}
	The tandem network is a variation on an autoencoder acting on the physics domain. It takes as input the desired physical property and returns a reconstruction of these physics (for instance a target reflectivity spectrum and its reconstruction).
	The difference to a conventional autoencoder is, that the decoder is trained in a first step on predicting the physical properties using the design parameters as input.
	This means, the decoder is simply a ``forward'' physics predictor, solving the direct problem (``fwd'' in figure~\ref{fig:design_methods_direct_models}a). 
	Subsequently, a second training step is performed, in which the forward model weights are fixed, and the encoder, which is actually trained on generating the designs, is added to the model (generator ``G'' in Fig.~\ref{fig:design_methods_direct_models}a).
	In this second step, the full model is trained, but now only the physical responses from the training set are used. 
	The physical property (e.g. reflectivity spectrum, etc...) is fed into the encoder, which predicts a design. However, instead of comparing this design to the known one from the dataset, the generated design is fed into the forward model, that predicts the physical property of the suggestion. This predicted response is finally compared with the input response, the error between both being minimized as training loss.
	This means, that even if multiple possible design solutions exist, the training remains unambiguous since only the physical response of the design is evaluated, regardless of how it is achieved.
	The full model is then essentially an autoencoder of which the latent space is being forced to correspond to the design parameters by using the fixed, pre-trained forward network as decoder.
	
	A practical advantage of the Tandem is, that the inverse problem is split in two sub-problems, that are individually easier to fit, compared to end-to-end training of the full inverse problem.
	In a first step the forward problem is learned, which is usually a relatively straightforward task. This physics knowledge is then used in the second step to guide training of the generator network.

	\subsubsection{Conditional variational autoencoder (cVAE)}
	
	A drawback of the Tandem network is that only a single solution is learned, even if multiple designs are possible to reach the design target. Several network architectures have been developed to learn mappings to the set of multiple solutions in ambiguous inverse problems. We discuss in the following a very efficient and robust model, the conditional Variational Autoencoder (cVAE).
	
	The reason why a ``vanilla'' variational autoencoder (VAE) can not be directly trained on an inverse design task is the correlation between geometry and physics domain. 
	The latent space of the autoencoder forms during training and represents the most efficient and compact, reduced representation of the inputs. This is generally \textit{not} the design parametrization.
	In consequence it is necessary to force the latent space to correspond to the design space, which is achieved in the ``Tandem'' architecture via a two-step training procedure (see above).
	The tandem is hence an autoencoder with design-regularized latent space.
	
	By \textit{conditioning} the designs on their physical properties (here: optical), a modified variant of a VAE, a so-called \textit{conditional} variational autoencoder (cVAE), can however be trained as an inverse design network. 
	To this end, the classical VAE, that reproduces a design through an encoder-decoder architecture, is extended by an additional input, the design condition. Here this is a physical property (e.g. a reflectivity spectrum) -- the design target.
	As depicted in figure~\ref{fig:design_methods_direct_models}b, this additional condition is added as input to both, the encoder (blue) as well as the decoder (green).
	During training, multiple possible solutions are associated with different values of the latent vector $z$, i.e. can be treated without training ambiguities, as illustrated in figure~\ref{fig:design_methods_direct_models}c.
	After successful network training, only the decoder is used for the inverse design. 
	The possibility to identify multiple possible solutions with a cVAE is illustrated in figure~\ref{fig:multiple_solutions_via_cVAE} by the example of multi-layer designs for a fixed reflectivity target. 
	
	An advantage of the cVAE is its generally robust training. It often also works well with low-dimensional latent spaces, so that the latent space can be explored systematically, to identify different possible solutions.\cite{daiInverseDesignStructural2022}
	A recent comparison indicates, that cVAEs are among the most effective methods for direct inverse design tasks.\cite{maBenchmarkingDeepLearningbased2022}
	
	\paragraph*{Technical hint: Regularization.} 
	As mentioned before, (c)VAEs require a latent regularization scheme in their training. The goal is for the latent space to become continuous and smooth (to allow meaningful interpolation).
	This is achieved using perturbative random latent sampling in the forward path, so the network learns that similar latent values correspond to similar solutions.
	In order to additionally achieve compactness (no blank regions in latent space), a weighted (``$\beta$''-coefficient) KL-loss is added to the training, which pushes the latent variables to a normal distribution around zero with unitary variance. 
	If the KL loss weight is too large, the latent space will be normally distributed around zero, but reconstruction will fail. 
	If the KL loss weight is too small, it has no effect. Then reconstruction will be good, but blank spaces in latent space may occur that do not carry useful information and impede to perform meaningful interpolation between solutions.
	In consequence, the weight of the KL loss with respect to the reconstruction loss needs to be carefully chosen (``$\beta$-VAE''\cite{i.higginsVVAELearningBasic2017}).
	
	Unfortunately this value needs to be adapted for each problem / network model, so some trial and error is required to find the adequate value. Good starting values are typically $\beta=0.01$ or $\beta=0.001$. Blank latent spaces may be difficult to spot in training, so the easier approach to find a good weight is to increase the $\beta$ value until the reconstruction loss starts suffering notably.
	
	\paragraph*{Technical hint: Condition.} 
	The condition is the design target and may be a high-dimensional construct such as a reflectivity spectrum. It is possible to process the condition with a sub-network (e.g. a 1D-CNN), that can be a common, shared network, before the two input branches (encoder and decoder).

	\subsection{Further deep inverse design approaches}
	
	The scope of the present work is to provide an entry-level guide to deep learning for inverse design, specifically targeting a newcomer audience. 
	We therefore focused on a few popular examples and provided a detailed discussion of practical challenges.
	As an outlook, here we want to briefly summarize a few further deep-learning based methods that can be used for inverse design.

	\subsubsection{Global optimization by deep reinforcement learning}
	
	Some key successes of recent deep learning are based on combining forward modeling and direct policy learning (such as Alpha-Zero\cite{silverMasteringChessShogi2017}).
	In this regard, conventional global optimization can be replaced by deep reinforcement learning (RL)~\cite{zophNeuralArchitectureSearch2017}, or \cite{banerjiMachineLearningEnables2020, wangAutomatedMultilayerOptical2020} a RL policy can be applied directly for choosing the design parameters. 
	However, comparison with classical evolutionary methods is in most cases not clearly in favor of deep reinforcement learning, so we believe it is questionable whether the extra effort of using a less studied method is advisable at the moment.\cite{realLargeScaleEvolutionImage2017, chengAssessmentReinforcementLearning2023}
	Please note that there are debates whether RL can indeed outperform conventional global optimization in a general picture.\cite{markovFalseDawnReevaluating2023}

	\subsubsection{Conditional generative adversarial network (cGAN)}
	
	Generative adversarial networks (GANs) can be seen as a variation of VAEs. The position of the decoder and the encoder are interchanged, and rather than compressing information, the encoder acts as a dynamically trained loss-function, which has as objective to distinguish generated samples from real data.
	Analogously to the case of conditional VAEs, in order to solve an ill posed inverse design problem, the GAN inputs need to be conditioned on the physics design. The GAN becomes a \textit{conditional} GAN (cGAN).
	
	Note that, the original GAN uses a min-max loss function between generator and discriminator, which is in practice very difficult to handle and often suffers from severe convergence problems.\cite{goodfellowGenerativeAdversarialNetworks2014} 
	The convergence of the shortly later proposed Wasserstein-distance loss for GAN training (WGAN) is significantly more robust.\cite{arjovskyWassersteinGAN2017}
	Typically it is combined with gradient penalty regularization (WGAN-GP), which leads to even better robustness, and is today used in the majority of large GAN models.\cite{gulrajaniImprovedTrainingWasserstein2017}
	
	Note also that very well performing inverse design GANs have been proposed that take inspiration from NVIDIA's ``style GAN''. By progressively growing the network model as well as the design resolution during training, very good design results and high accuracy can be obtained, however at the cost of a highly increased computational training budget.\cite{karrasStyleBasedGeneratorArchitecture2019, karrasAnalyzingImprovingImage2020, wenRobustFreeformMetasurface2020}

	\subsubsection{Diffusion models}
	
	Recently, stochastic generative modeling was shown to be capable to solve inverse problems. 
	These so-called ``diffusion models'' were described by Y. Song et al. by: ``Creating noise from data is easy; creating data from noise is generative modeling.'' \cite{songScoreBasedGenerativeModeling2021}
	The key idea is inspired by thermodynamic processes and is based on performing a sequence of denoising steps by a deep learning model. The model is trained on removing a small amount of noise from input data, which was perturbed by different amounts of noise.\cite{sohl-dicksteinDeepUnsupervisedLearning2015}
	After several denoising steps, a large amount of initial noise can be entirely removed.
	Starting this iterative process on random noise, and combining it with some guidance using latent information about the original content, a denoising diffusion model can generate samples that match the target latent description. 
	Such networks are very popular in computer vision and image generation (``text to image''). \cite{rombachHighResolutionImageSynthesis2022, croitoruDiffusionModelsVision2023, changDesignFundamentalsDiffusion2023}
	Very recently, a first application of a diffusion model for metasurface inverse design has been demonstrated.\cite{zhangDiffusionProbabilisticModel2023}

	\subsubsection{Invertible neural networks}
	
	A possibility to solve the inverse problem in a one-shot manner are invertible neural networks. Such networks are constructed using exclusively mathematical operations that guarantee that the full network remains an invertible mathematical expression, so that it is bijective. Every point in the target space matches exactly one point in the input space.\cite{behrmannInvertibleResidualNetworks2019}
	Obviously, this approach runs into troubles with the ill-posed character of typical inverse design problems, that prohibits to find a bijective projection between physics and design space. 
	To allow the network to learn such a bijective projection, latent dimensions need to be added to the design space, which during training are fitted to distinguish between multiple solutions or to identify cases with no solution at all.\cite{ardizzoneAnalyzingInverseProblems2018}

	\subsubsection{Physics informed neural networks}
	
	As discussed above, a very interesting aspect of deep learning models is that they are differentiable. Furthermore, the basis of every deep learning application is the universal approximator theorem, stating that \textit{any} function can be approximated with arbitrary accuracy with a sufficiently large neural network.\cite{hornikMultilayerFeedforwardNetworks1989}
	The idea behind physics informed neural networks (PINNs)\cite{raissiPhysicsinformedNeuralNetworks2019, grossmannCanPhysicsInformedNeural2023} is to learn an approximation to a solution for a partial differential equation (PDE), by testing arbitrary values and minimizing a physics-based loss, that tests the validity of the predicted solution in the PDE. 
	PINNs are known for their extremely high accuracy and have been used for inverse design as well, where design parameters are typically included in the model and iteratively fitted during the physics-based training.\cite{chenPhysicsinformedNeuralNetworks2020, fangDeepPhysicalInformed2020, luPhysicsinformedNeuralNetworks2021}
	
	In the context of PINNs, we want to note that periodic activation functions can be very powerful in the context of differential equation solving, since their derivatives naturally represent many typical solutions.\cite{klocekHypernetworkFunctionalImage2019, sitzmannImplicitNeuralRepresentations2020}

	\subsection{Improving inverse design performance}
	
	There are several ways to improve the inverse design performance. 
	This includes neural network model optimization, tuning of the training hyper-parameters, application of regularization techniques or multi-step training methods.
	
	Eventually, the accuracy of a deep learning model stands and falls with the quality of the data (and its quantity).
	As discussed before, it is crucial that the dataset is representative for the problem and that it includes as little bias as possible. However, merely generating more and more random samples has often only a very limited effect on the quality of the design predictions.
	A very efficient way of improving the dataset in a more purposeful manner is iterative or interactive data generation, also called active learning.
	The idea behind active learning is to reduce the effort of data generation by letting a neural network ``learn from its own mistakes''. To this end, first a network is trained on an imperfect, but cheap dataset. 
	Subsequently, the model predictions are evaluated and data is added to the training set with a particular focus on cases where the network model performs weakly.
	This evaluation can be done by ensemble statistics, in which evaluations of multiple neural networks are gathered and the statistics of the predictions are used to assess the quality of the results.\cite{pestourieActiveLearningDeep2020}
	In inverse design it is even easier, since the generated designs can be simply simulated and the results appended to the dataset.\cite{wenRobustFreeformMetasurface2020, dinsdaleDeepLearningEnabled2021, blanchard-dionneSuccessiveTrainingGenerative2021}

	\section{Additional Technical Tips}\label{app:techical_tips}
	\textit{Mixed precision training:} If supported by the accelerator (GPU, TPU), mixed precision (bf16) training should be used. With this option, the computation device uses half precision (16 bit per number) for most calculations, which requires half the memory and runs very fast on modern hardware. 
		On large models a factor 2 can be easily achieved both in runtime acceleration and memory reduction, allowing larger models to fit in the GPU RAM. It also allows for larger batch sizes in the late training, which typically comes with additional acceleration.

	\textit{Checkpoint saving:} A callback to automatically save the best validation model can be very useful, in this way the best non-overfitted model is saved automatically even if severe overfitting occurs later in the training (``early stopping'').\cite{yaoEarlyStoppingGradient2007}

	\textit{Ensemble averaging:} A typical procedure is to train a network several times, saving the best model of every run. In inference, all of these models are then used and the average solution is taken. The variance of the prediction statistics provides further information about the prediction certainty.\cite{wangMassiveComputationalAcceleration2019, wiechaDeepLearningNanophotonics2021}

	\textit{Mixture of experts:} For very large models, it is common practice to essentially split the model into smaller sub-networks. 
	The details are beyond the scope of this newcomer's guide, but the technique can significantly reduce computational cost for very large architectures.\cite{eigenLearningFactoredRepresentations2014, shazeerOutrageouslyLargeNeural2017}

	\textit{Make the problem more specific:} Often it is possible to exploit the fact that deep learning models usually perform better, the more specifically a problem is defined. It may be possible to re-formulate a problem, or to use multiple networks that predict different sub-problems, which may then be combined by a further neural network.
	In nano-photonics, for example in a metasurface problem, at the network prediction stage one may separate local light-matter interaction and far-field propagation, instead of directly predicting the far-field response.\cite{wiechaDeepLearningMeets2020, chenHighSpeedSimulation2022}

	\section{Tutorial notebooks}
	
	The typical workflow described above is demonstrated in a series of python notebooks as supplemental material, accessible online.\cite{wiechaNewcomerGuideDeep2023}
	We demonstrate the full workflow from data generation, and data-processing, over network architecture design and hyperparameter tuning, to an implementation of the above discussed different inverse design approaches.
	We use two specific problems for the tutorial notebooks:

	\subsection{Problem 1: Reflectivity of a layer stack with \textit{PyMoosh}}
	
	The first problem used to demonstrate typical deep learning workflow is a dielectric multi-layer stack with the goal to tailor the reflectivity spectrum.
	For the physics calculations these tutorials use \textit{PyMoosh}\cite{moreauPyMoosh2023}, the python version of \textit{Moosh}, an s-matrix based solver for multilayer optics problems.\cite{defranceMooshNumericalSwiss2016}
	As deep learning framework, we use \textit{Keras},\cite{abadiTensorFlowLargeScaleMachine2015} a \textit{PyTorch} implementation of the notebooks is planned for the near future.
	For global optimization we use the package \textit{Nevergrad}.\cite{bennetNevergradBlackboxOptimization2021}
	This covers following tutorials:
	
	\begin{itemize}
		\item Data generation: Fully random designs. \href{https://gitlab.com/wiechapeter/newcomer_guide_dl_inversedesign/-/blob/main/keras/01a_data_generation_random.ipynb}{link}
		\item Data generation: Weakly optimized designs. \href{https://gitlab.com/wiechapeter/newcomer_guide_dl_inversedesign/-/blob/main/keras/01b_data_generation_with_optimization.ipynb}{link}
		\item Forward network training of increasingly complex models. \href{https://gitlab.com/wiechapeter/newcomer_guide_dl_inversedesign/-/blob/main/keras/02_forward_models.ipynb}{link}
		\begin{itemize}
			\item Forward problem: Using batch normalization for regression tasks. \href{https://gitlab.com/wiechapeter/newcomer_guide_dl_inversedesign/-/blob/main/keras/02x_batchnorm_dropout_regression_task.ipynb}{link}
			\item Forward problem: Danger of extrapolation. \href{https://gitlab.com/wiechapeter/newcomer_guide_dl_inversedesign/-/blob/main/keras/02y_extrapolation_regression_task.ipynb}{link}
		\end{itemize}
		\item Direct inverse design: The Tandem model. \href{https://gitlab.com/wiechapeter/newcomer_guide_dl_inversedesign/-/blob/main/keras/03a_inverse_direct_tandem.ipynb}{link}
		\item Direct inverse design: The conditional Variational Autoencoder. \href{https://gitlab.com/wiechapeter/newcomer_guide_dl_inversedesign/-/blob/main/keras/03b_inverse_direct_cVAE.ipynb}{link}
		\item Iterative optimization: Gradient based (Neural Adjoint). \href{https://gitlab.com/wiechapeter/newcomer_guide_dl_inversedesign/-/blob/main/keras/04a_neural_adjoint.ipynb}{link}
		\item Iterative optimization: Global pre-optimizaiton, then gradient descent. \href{https://gitlab.com/wiechapeter/newcomer_guide_dl_inversedesign/-/blob/main/keras/04b_global_pre_opt_before_neural_adjoint.ipynb}{link}
		\item Iterative / active learning: Iterative fine-tuning of the inverse design model on the actual downstream problem. \href{https://gitlab.com/wiechapeter/newcomer_guide_dl_inversedesign/-/blob/main/keras/04c_active_learning_finetune_on_design_task.ipynb}{link}
	\end{itemize}

	\subsection{Problem 2: Scattering of dielectric nanostructures}
	
	A second problem is used to illustrate the case of structure parameterization via images (here of the geometry top-view). 
	This is a typical scenario for many top-down fabricated nano-photonic devices like metasurfaces. 
	We demonstrate how a Wasserstein GAN with gradient penalty\cite{gulrajaniImprovedTrainingWasserstein2017} can be trained on learning a regularized latent description for 2D geometry top-view images.
	Using this in combination with a forward predictor model is then demonstrated using global and gradient based optimization of scattering spectrum inverse design, simultaneously for two incident polarizations.
	The nano-scattering dataset is created using simulations with the \textit{pyGDM} toolkit.\cite{wiechaPyGDMPythonToolkit2018, wiechaPyGDMNewFunctionalities2022}
	This problem contains following tutorials:

	\begin{itemize}
		\item Data generation: Random nano-scatterer 2D geometries. \href{https://gitlab.com/wiechapeter/newcomer_guide_dl_inversedesign/-/blob/main/keras/05a_nanoscattering_data_generation.ipynb}{link}
		\item Learned design parameterization: Train a WGAN on the geometries. \href{https://gitlab.com/wiechapeter/newcomer_guide_dl_inversedesign/-/blob/main/keras/05b_nanoscattering_structure_WGAN_GP.ipynb}{link}
		\item Forward Model: ResNet for nano-scattering prediction. \href{https://gitlab.com/wiechapeter/newcomer_guide_dl_inversedesign/-/blob/main/keras/05c_nanoscattering_forward_model.ipynb}{link}
		\item Constrained inverse design: Gradient based iterative optimization of nano-scatterer geometries in the WGAN latent space. \href{https://gitlab.com/wiechapeter/newcomer_guide_dl_inversedesign/-/blob/main/keras/05d_nanoscattering_neural_adjoint_via_GAN.ipynb}{link}
	\end{itemize}

	\section{Conclusions and perspectives}

	In conclusion, we provided a practical newcomer's guide to approach inverse design problems with deep learning.
	We gave an introduction to the key concepts in deep learning, and critically discussed how should be assessed whether deep learning is a promising strategy for solving a specific problem.
	We discussed guidelines for the evaluation of a dataset in a first step and for the subsequent, practical implementation and training of a deep neural network architecture.
	We then discussed deep learning based inverse design approaches, which fall into in two main groups: iterative techniques and direct (one-shot) inverse networks. 
	After specifically discussing the possibility to train an auxiliary network on learning a new, regularized design parameterization, we concluded with referring to a set of python tutorial notebooks, which demonstrates practically all above discussed steps and methods.
	
	Specialists working on topics around nano-photonics, who want to apply statistical deep learning techniques to their problems, often do not have the time to learn all the subtleties of deep learning model implementation the hard way.
	We believe that our tutorial will be particularly useful for such deep learning newcomers, since we try to discuss various possible pitfalls and provide hints and guidelines for robust architectures that may "just work" without too much painful parameter tuning. 
	
	To conclude, we want to recall once again, that deep learning is not the solution to all problems.\cite{wiechaDeepLearningNanophotonic2023}
	For many applications of inverse design there exist highly optimized, specific algorithms and other solutions, which will often outperform deep learning. Considering whether a deep neural network is indeed the way to go, is an extremely important first question that one should always ask before starting.
	However there are many situations where deep learning can offer unique assets in inverse design. For instance if ultimate speed is required, or when a large number of design tasks need to be solved for the same problem, deep learning can be a literal game changer. 
	If, for instance, a large dataset for a photonic platform exists and struggles understanding the underlying physical mechanisms, deep learning offers an ideal platform with latent space methods. 
	Finally, a key property of deep learning models is automatic differentiation, which allows to build analytical, differentiable models from empirical data, for instance from experimental measurements.
	We believe that, used with the necessary caution and upon proper verification of the predictions, deep learning offers a very powerful platform not only for inverse design problems but far beyond.

	\subsection*{Acknowledgments}
	
	The authors thank Arnaud Arbouet and Otto Muskens for fruitful discussions.
	A.M. is an Academy CAP 20-25 chair holder. He acknowledges the support received from the Agence Nationale de la Recherche of the French government through the program Investissements d'Avenir (16-IDEX-0001 CAP 20-25). This work was supported by the International Research Center "Innovation Transportation and Production Systems" of the Clermont-Ferrand I-SITE CAP 20-25.
	P.R.W. acknowledges the support of the French Agence Nationale de la Recherche (ANR) under grant ANR-22-CE24-0002 (project NAINOS), and from the Toulouse high performance computing facility CALMIP (grant p20010).

	\subsection*{Disclosures}
	The authors declare no conflicts of interest.

	\bibliography{2023_tutorial_invdesign_main.bbl}

\begin{thebibliography}{211}%
\makeatletter
\providecommand \@ifxundefined [1]{%
 \@ifx{#1\undefined}
}%
\providecommand \@ifnum [1]{%
 \ifnum #1\expandafter \@firstoftwo
 \else \expandafter \@secondoftwo
 \fi
}%
\providecommand \@ifx [1]{%
 \ifx #1\expandafter \@firstoftwo
 \else \expandafter \@secondoftwo
 \fi
}%
\providecommand \natexlab [1]{#1}%
\providecommand \enquote  [1]{``#1''}%
\providecommand \bibnamefont  [1]{#1}%
\providecommand \bibfnamefont [1]{#1}%
\providecommand \citenamefont [1]{#1}%
\providecommand \href@noop [0]{\@secondoftwo}%
\providecommand \href [0]{\begingroup \@sanitize@url \@href}%
\providecommand \@href[1]{\@@startlink{#1}\@@href}%
\providecommand \@@href[1]{\endgroup#1\@@endlink}%
\providecommand \@sanitize@url [0]{\catcode `\\12\catcode `\$12\catcode
  `\&12\catcode `\#12\catcode `\^12\catcode `\_12\catcode `\%12\relax}%
\providecommand \@@startlink[1]{}%
\providecommand \@@endlink[0]{}%
\providecommand \url  [0]{\begingroup\@sanitize@url \@url }%
\providecommand \@url [1]{\endgroup\@href {#1}{\urlprefix }}%
\providecommand \urlprefix  [0]{URL }%
\providecommand \Eprint [0]{\href }%
\providecommand \doibase [0]{http://dx.doi.org/}%
\providecommand \selectlanguage [0]{\@gobble}%
\providecommand \bibinfo  [0]{\@secondoftwo}%
\providecommand \bibfield  [0]{\@secondoftwo}%
\providecommand \translation [1]{[#1]}%
\providecommand \BibitemOpen [0]{}%
\providecommand \bibitemStop [0]{}%
\providecommand \bibitemNoStop [0]{.\EOS\space}%
\providecommand \EOS [0]{\spacefactor3000\relax}%
\providecommand \BibitemShut  [1]{\csname bibitem#1\endcsname}%
\let\auto@bib@innerbib\@empty
\bibitem [{\citenamefont {M{\"u}hlschlegel}\ \emph {et~al.}(2005)\citenamefont
  {M{\"u}hlschlegel}, \citenamefont {Eisler}, \citenamefont {Martin},
  \citenamefont {Hecht},\ and\ \citenamefont
  {Pohl}}]{muhlschlegelResonantOpticalAntennas2005}%
  \BibitemOpen
  \bibfield  {author} {\bibinfo {author} {\bibfnamefont {P.}~\bibnamefont
  {M{\"u}hlschlegel}}, \bibinfo {author} {\bibfnamefont {H.-J.}\ \bibnamefont
  {Eisler}}, \bibinfo {author} {\bibfnamefont {O.~J.~F.}\ \bibnamefont
  {Martin}}, \bibinfo {author} {\bibfnamefont {B.}~\bibnamefont {Hecht}}, \
  and\ \bibinfo {author} {\bibfnamefont {D.~W.}\ \bibnamefont {Pohl}},\ }\href
  {\doibase 10.1126/science.1111886} {\bibfield  {journal} {\bibinfo  {journal}
  {Science}\ }\textbf {\bibinfo {volume} {308}},\ \bibinfo {pages} {1607}
  (\bibinfo {year} {2005})}\BibitemShut {NoStop}%
\bibitem [{\citenamefont {Girard}(2005)}]{girardFieldsNanostructures2005}%
  \BibitemOpen
  \bibfield  {author} {\bibinfo {author} {\bibfnamefont {C.}~\bibnamefont
  {Girard}},\ }\href {\doibase 10.1088/0034-4885/68/8/R05} {\bibfield
  {journal} {\bibinfo  {journal} {Reports on Progress in Physics}\ }\textbf
  {\bibinfo {volume} {68}},\ \bibinfo {pages} {1883} (\bibinfo {year}
  {2005})}\BibitemShut {NoStop}%
\bibitem [{\citenamefont {Novotny}\ and\ \citenamefont
  {Hecht}(2006)}]{novotnyPrinciplesNanooptics2006}%
  \BibitemOpen
  \bibfield  {author} {\bibinfo {author} {\bibfnamefont {L.}~\bibnamefont
  {Novotny}}\ and\ \bibinfo {author} {\bibfnamefont {B.}~\bibnamefont
  {Hecht}},\ }\href@noop {} {\emph {\bibinfo {title} {Principles of
  Nano-Optics}}}\ (\bibinfo  {publisher} {{Cambridge University Press}},\
  \bibinfo {address} {{Cambridge ; New York}},\ \bibinfo {year}
  {2006})\BibitemShut {NoStop}%
\bibitem [{\citenamefont {Kuznetsov}\ \emph {et~al.}(2016)\citenamefont
  {Kuznetsov}, \citenamefont {Miroshnichenko}, \citenamefont {Brongersma},
  \citenamefont {Kivshar},\ and\ \citenamefont
  {Luk'yanchuk}}]{kuznetsovOpticallyResonantDielectric2016}%
  \BibitemOpen
  \bibfield  {author} {\bibinfo {author} {\bibfnamefont {A.~I.}\ \bibnamefont
  {Kuznetsov}}, \bibinfo {author} {\bibfnamefont {A.~E.}\ \bibnamefont
  {Miroshnichenko}}, \bibinfo {author} {\bibfnamefont {M.~L.}\ \bibnamefont
  {Brongersma}}, \bibinfo {author} {\bibfnamefont {Y.~S.}\ \bibnamefont
  {Kivshar}}, \ and\ \bibinfo {author} {\bibfnamefont {B.}~\bibnamefont
  {Luk'yanchuk}},\ }\href {\doibase 10.1126/science.aag2472} {\bibfield
  {journal} {\bibinfo  {journal} {Science}\ }\textbf {\bibinfo {volume} {354}}
  (\bibinfo {year} {2016}),\ 10.1126/science.aag2472}\BibitemShut {NoStop}%
\bibitem [{\citenamefont {Girard}\ and\ \citenamefont
  {Dujardin}(2006)}]{girardNearfieldOpticalProperties2006}%
  \BibitemOpen
  \bibfield  {author} {\bibinfo {author} {\bibfnamefont {C.}~\bibnamefont
  {Girard}}\ and\ \bibinfo {author} {\bibfnamefont {E.}~\bibnamefont
  {Dujardin}},\ }\href {\doibase 10.1088/1464-4258/8/4/S05} {\bibfield
  {journal} {\bibinfo  {journal} {Journal of Optics A: Pure and Applied
  Optics}\ }\textbf {\bibinfo {volume} {8}},\ \bibinfo {pages} {S73} (\bibinfo
  {year} {2006})}\BibitemShut {NoStop}%
\bibitem [{\citenamefont {Pendry}(2000)}]{pendryNegativeRefractionMakes2000}%
  \BibitemOpen
  \bibfield  {author} {\bibinfo {author} {\bibfnamefont {J.~B.}\ \bibnamefont
  {Pendry}},\ }\href {\doibase 10.1103/PhysRevLett.85.3966} {\bibfield
  {journal} {\bibinfo  {journal} {Physical Review Letters}\ }\textbf {\bibinfo
  {volume} {85}},\ \bibinfo {pages} {3966} (\bibinfo {year}
  {2000})}\BibitemShut {NoStop}%
\bibitem [{\citenamefont {Wiecha}\ \emph
  {et~al.}(2017{\natexlab{a}})\citenamefont {Wiecha}, \citenamefont {Cuche},
  \citenamefont {Arbouet}, \citenamefont {Girard}, \citenamefont {{Colas des
  Francs}}, \citenamefont {Lecestre}, \citenamefont {Larrieu}, \citenamefont
  {Fournel}, \citenamefont {Larrey}, \citenamefont {Baron},\ and\ \citenamefont
  {Paillard}}]{wiechaStronglyDirectionalScattering2017}%
  \BibitemOpen
  \bibfield  {author} {\bibinfo {author} {\bibfnamefont {P.~R.}\ \bibnamefont
  {Wiecha}}, \bibinfo {author} {\bibfnamefont {A.}~\bibnamefont {Cuche}},
  \bibinfo {author} {\bibfnamefont {A.}~\bibnamefont {Arbouet}}, \bibinfo
  {author} {\bibfnamefont {C.}~\bibnamefont {Girard}}, \bibinfo {author}
  {\bibfnamefont {G.}~\bibnamefont {{Colas des Francs}}}, \bibinfo {author}
  {\bibfnamefont {A.}~\bibnamefont {Lecestre}}, \bibinfo {author}
  {\bibfnamefont {G.}~\bibnamefont {Larrieu}}, \bibinfo {author} {\bibfnamefont
  {F.}~\bibnamefont {Fournel}}, \bibinfo {author} {\bibfnamefont
  {V.}~\bibnamefont {Larrey}}, \bibinfo {author} {\bibfnamefont
  {T.}~\bibnamefont {Baron}}, \ and\ \bibinfo {author} {\bibfnamefont
  {V.}~\bibnamefont {Paillard}},\ }\href {\doibase
  10.1021/acsphotonics.7b00423} {\bibfield  {journal} {\bibinfo  {journal} {ACS
  Photonics}\ }\textbf {\bibinfo {volume} {4}},\ \bibinfo {pages} {2036}
  (\bibinfo {year} {2017}{\natexlab{a}})}\BibitemShut {NoStop}%
\bibitem [{\citenamefont {Kauranen}\ and\ \citenamefont
  {Zayats}(2012)}]{kauranenNonlinearPlasmonics2012}%
  \BibitemOpen
  \bibfield  {author} {\bibinfo {author} {\bibfnamefont {M.}~\bibnamefont
  {Kauranen}}\ and\ \bibinfo {author} {\bibfnamefont {A.~V.}\ \bibnamefont
  {Zayats}},\ }\href {\doibase 10.1038/nphoton.2012.244} {\bibfield  {journal}
  {\bibinfo  {journal} {Nature Photonics}\ }\textbf {\bibinfo {volume} {6}},\
  \bibinfo {pages} {737} (\bibinfo {year} {2012})}\BibitemShut {NoStop}%
\bibitem [{\citenamefont {Genevet}\ \emph {et~al.}(2017)\citenamefont
  {Genevet}, \citenamefont {Capasso}, \citenamefont {Aieta}, \citenamefont
  {Khorasaninejad},\ and\ \citenamefont
  {Devlin}}]{genevetRecentAdvancesPlanar2017}%
  \BibitemOpen
  \bibfield  {author} {\bibinfo {author} {\bibfnamefont {P.}~\bibnamefont
  {Genevet}}, \bibinfo {author} {\bibfnamefont {F.}~\bibnamefont {Capasso}},
  \bibinfo {author} {\bibfnamefont {F.}~\bibnamefont {Aieta}}, \bibinfo
  {author} {\bibfnamefont {M.}~\bibnamefont {Khorasaninejad}}, \ and\ \bibinfo
  {author} {\bibfnamefont {R.}~\bibnamefont {Devlin}},\ }\href {\doibase
  10.1364/OPTICA.4.000139} {\bibfield  {journal} {\bibinfo  {journal} {Optica}\
  }\textbf {\bibinfo {volume} {4}},\ \bibinfo {pages} {139} (\bibinfo {year}
  {2017})}\BibitemShut {NoStop}%
\bibitem [{\citenamefont {{Colas des Francs}}\ \emph
  {et~al.}(2016)\citenamefont {{Colas des Francs}}, \citenamefont {Barthes},
  \citenamefont {Bouhelier}, \citenamefont {Weeber}, \citenamefont {Dereux},
  \citenamefont {Cuche},\ and\ \citenamefont
  {Girard}}]{colasdesfrancsPlasmonicPurcellFactor2016}%
  \BibitemOpen
  \bibfield  {author} {\bibinfo {author} {\bibfnamefont {G.}~\bibnamefont
  {{Colas des Francs}}}, \bibinfo {author} {\bibfnamefont {J.}~\bibnamefont
  {Barthes}}, \bibinfo {author} {\bibfnamefont {A.}~\bibnamefont {Bouhelier}},
  \bibinfo {author} {\bibfnamefont {J.~C.}\ \bibnamefont {Weeber}}, \bibinfo
  {author} {\bibfnamefont {A.}~\bibnamefont {Dereux}}, \bibinfo {author}
  {\bibfnamefont {A.}~\bibnamefont {Cuche}}, \ and\ \bibinfo {author}
  {\bibfnamefont {C.}~\bibnamefont {Girard}},\ }\href {\doibase
  10.1088/2040-8978/18/9/094005} {\bibfield  {journal} {\bibinfo  {journal}
  {Journal of Optics}\ }\textbf {\bibinfo {volume} {18}},\ \bibinfo {pages}
  {094005} (\bibinfo {year} {2016})}\BibitemShut {NoStop}%
\bibitem [{\citenamefont {Wang}\ \emph
  {et~al.}(2020{\natexlab{a}})\citenamefont {Wang}, \citenamefont {Sciarrino},
  \citenamefont {Laing},\ and\ \citenamefont
  {Thompson}}]{wangIntegratedPhotonicQuantum2020}%
  \BibitemOpen
  \bibfield  {author} {\bibinfo {author} {\bibfnamefont {J.}~\bibnamefont
  {Wang}}, \bibinfo {author} {\bibfnamefont {F.}~\bibnamefont {Sciarrino}},
  \bibinfo {author} {\bibfnamefont {A.}~\bibnamefont {Laing}}, \ and\ \bibinfo
  {author} {\bibfnamefont {M.~G.}\ \bibnamefont {Thompson}},\ }\href {\doibase
  10.1038/s41566-019-0532-1} {\bibfield  {journal} {\bibinfo  {journal} {Nature
  Photonics}\ }\textbf {\bibinfo {volume} {14}},\ \bibinfo {pages} {273}
  (\bibinfo {year} {2020}{\natexlab{a}})}\BibitemShut {NoStop}%
\bibitem [{\citenamefont {Wiecha}\ \emph
  {et~al.}(2019{\natexlab{a}})\citenamefont {Wiecha}, \citenamefont {Majorel},
  \citenamefont {Girard}, \citenamefont {Arbouet}, \citenamefont {Masenelli},
  \citenamefont {Boisron}, \citenamefont {Lecestre}, \citenamefont {Larrieu},
  \citenamefont {Paillard},\ and\ \citenamefont
  {Cuche}}]{wiechaEnhancementElectricMagnetic2019}%
  \BibitemOpen
  \bibfield  {author} {\bibinfo {author} {\bibfnamefont {P.~R.}\ \bibnamefont
  {Wiecha}}, \bibinfo {author} {\bibfnamefont {C.}~\bibnamefont {Majorel}},
  \bibinfo {author} {\bibfnamefont {C.}~\bibnamefont {Girard}}, \bibinfo
  {author} {\bibfnamefont {A.}~\bibnamefont {Arbouet}}, \bibinfo {author}
  {\bibfnamefont {B.}~\bibnamefont {Masenelli}}, \bibinfo {author}
  {\bibfnamefont {O.}~\bibnamefont {Boisron}}, \bibinfo {author} {\bibfnamefont
  {A.}~\bibnamefont {Lecestre}}, \bibinfo {author} {\bibfnamefont
  {G.}~\bibnamefont {Larrieu}}, \bibinfo {author} {\bibfnamefont
  {V.}~\bibnamefont {Paillard}}, \ and\ \bibinfo {author} {\bibfnamefont
  {A.}~\bibnamefont {Cuche}},\ }\href {\doibase 10.1364/AO.58.001682}
  {\bibfield  {journal} {\bibinfo  {journal} {Applied Optics}\ }\textbf
  {\bibinfo {volume} {58}},\ \bibinfo {pages} {1682} (\bibinfo {year}
  {2019}{\natexlab{a}})}\BibitemShut {NoStop}%
\bibitem [{\citenamefont {Hadamard}(1902)}]{hadamardProblemesAuxDerives1902}%
  \BibitemOpen
  \bibfield  {author} {\bibinfo {author} {\bibfnamefont {J.}~\bibnamefont
  {Hadamard}},\ }\href@noop {} {\bibfield  {journal} {\bibinfo  {journal}
  {Princeton University Bulletin}\ }\textbf {\bibinfo {volume} {13}},\ \bibinfo
  {pages} {49} (\bibinfo {year} {1902})}\BibitemShut {NoStop}%
\bibitem [{\citenamefont {Jensen}\ and\ \citenamefont
  {Sigmund}(2011)}]{jensenTopologyOptimizationNanophotonics2011}%
  \BibitemOpen
  \bibfield  {author} {\bibinfo {author} {\bibfnamefont {J.~S.}\ \bibnamefont
  {Jensen}}\ and\ \bibinfo {author} {\bibfnamefont {O.}~\bibnamefont
  {Sigmund}},\ }\href {\doibase 10.1002/lpor.201000014} {\bibfield  {journal}
  {\bibinfo  {journal} {Laser \& Photonics Reviews}\ }\textbf {\bibinfo
  {volume} {5}},\ \bibinfo {pages} {308} (\bibinfo {year} {2011})}\BibitemShut
  {NoStop}%
\bibitem [{\citenamefont {Elsawy}\ \emph {et~al.}(2020)\citenamefont {Elsawy},
  \citenamefont {Lanteri}, \citenamefont {Duvigneau}, \citenamefont {Fan},\
  and\ \citenamefont {Genevet}}]{elsawyNumericalOptimizationMethods2020}%
  \BibitemOpen
  \bibfield  {author} {\bibinfo {author} {\bibfnamefont {M.~M.~R.}\
  \bibnamefont {Elsawy}}, \bibinfo {author} {\bibfnamefont {S.}~\bibnamefont
  {Lanteri}}, \bibinfo {author} {\bibfnamefont {R.}~\bibnamefont {Duvigneau}},
  \bibinfo {author} {\bibfnamefont {J.~A.}\ \bibnamefont {Fan}}, \ and\
  \bibinfo {author} {\bibfnamefont {P.}~\bibnamefont {Genevet}},\ }\href
  {\doibase 10.1002/lpor.201900445} {\bibfield  {journal} {\bibinfo  {journal}
  {Laser \& Photonics Reviews}\ }\textbf {\bibinfo {volume} {14}},\ \bibinfo
  {pages} {1900445} (\bibinfo {year} {2020})}\BibitemShut {NoStop}%
\bibitem [{\citenamefont {Malkiel}\ \emph {et~al.}(2018)\citenamefont
  {Malkiel}, \citenamefont {Mrejen}, \citenamefont {Nagler}, \citenamefont
  {Arieli}, \citenamefont {Wolf},\ and\ \citenamefont
  {Suchowski}}]{malkielPlasmonicNanostructureDesign2018}%
  \BibitemOpen
  \bibfield  {author} {\bibinfo {author} {\bibfnamefont {I.}~\bibnamefont
  {Malkiel}}, \bibinfo {author} {\bibfnamefont {M.}~\bibnamefont {Mrejen}},
  \bibinfo {author} {\bibfnamefont {A.}~\bibnamefont {Nagler}}, \bibinfo
  {author} {\bibfnamefont {U.}~\bibnamefont {Arieli}}, \bibinfo {author}
  {\bibfnamefont {L.}~\bibnamefont {Wolf}}, \ and\ \bibinfo {author}
  {\bibfnamefont {H.}~\bibnamefont {Suchowski}},\ }\href {\doibase
  10.1038/s41377-018-0060-7} {\bibfield  {journal} {\bibinfo  {journal} {Light:
  Science \& Applications}\ }\textbf {\bibinfo {volume} {7}},\ \bibinfo {pages}
  {60} (\bibinfo {year} {2018})}\BibitemShut {NoStop}%
\bibitem [{\citenamefont {Wiecha}\ and\ \citenamefont
  {Muskens}(2020)}]{wiechaDeepLearningMeets2020}%
  \BibitemOpen
  \bibfield  {author} {\bibinfo {author} {\bibfnamefont {P.~R.}\ \bibnamefont
  {Wiecha}}\ and\ \bibinfo {author} {\bibfnamefont {O.~L.}\ \bibnamefont
  {Muskens}},\ }\href {\doibase 10.1021/acs.nanolett.9b03971} {\bibfield
  {journal} {\bibinfo  {journal} {Nano Letters}\ }\textbf {\bibinfo {volume}
  {20}},\ \bibinfo {pages} {329} (\bibinfo {year} {2020})},\ \Eprint
  {http://arxiv.org/abs/1909.12056} {arxiv:1909.12056} \BibitemShut {NoStop}%
\bibitem [{\citenamefont {{Blanchard-Dionne}}\ and\ \citenamefont
  {Martin}(2020)}]{blanchard-dionneTeachingOpticsMachine2020}%
  \BibitemOpen
  \bibfield  {author} {\bibinfo {author} {\bibfnamefont {A.-P.}\ \bibnamefont
  {{Blanchard-Dionne}}}\ and\ \bibinfo {author} {\bibfnamefont {O.~J.~F.}\
  \bibnamefont {Martin}},\ }\href {\doibase 10.1364/OL.390600} {\bibfield
  {journal} {\bibinfo  {journal} {Optics Letters}\ }\textbf {\bibinfo {volume}
  {45}},\ \bibinfo {pages} {2922} (\bibinfo {year} {2020})}\BibitemShut
  {NoStop}%
\bibitem [{\citenamefont {Chen}\ \emph {et~al.}(2022)\citenamefont {Chen},
  \citenamefont {Lupoiu}, \citenamefont {Mao}, \citenamefont {Huang},
  \citenamefont {Jiang}, \citenamefont {Lalanne},\ and\ \citenamefont
  {Fan}}]{chenHighSpeedSimulation2022}%
  \BibitemOpen
  \bibfield  {author} {\bibinfo {author} {\bibfnamefont {M.}~\bibnamefont
  {Chen}}, \bibinfo {author} {\bibfnamefont {R.}~\bibnamefont {Lupoiu}},
  \bibinfo {author} {\bibfnamefont {C.}~\bibnamefont {Mao}}, \bibinfo {author}
  {\bibfnamefont {D.-H.}\ \bibnamefont {Huang}}, \bibinfo {author}
  {\bibfnamefont {J.}~\bibnamefont {Jiang}}, \bibinfo {author} {\bibfnamefont
  {P.}~\bibnamefont {Lalanne}}, \ and\ \bibinfo {author} {\bibfnamefont
  {J.~A.}\ \bibnamefont {Fan}},\ }\href {\doibase 10.1021/acsphotonics.2c00876}
  {\bibfield  {journal} {\bibinfo  {journal} {ACS Photonics}\ }\textbf
  {\bibinfo {volume} {9}},\ \bibinfo {pages} {3110} (\bibinfo {year}
  {2022})}\BibitemShut {NoStop}%
\bibitem [{\citenamefont {Ma}\ \emph {et~al.}(2023)\citenamefont {Ma},
  \citenamefont {Wang},\ and\ \citenamefont
  {Guo}}]{maOptoGPTFoundationModel2023}%
  \BibitemOpen
  \bibfield  {author} {\bibinfo {author} {\bibfnamefont {T.}~\bibnamefont
  {Ma}}, \bibinfo {author} {\bibfnamefont {H.}~\bibnamefont {Wang}}, \ and\
  \bibinfo {author} {\bibfnamefont {L.~J.}\ \bibnamefont {Guo}},\ }\href
  {\doibase 10.48550/arXiv.2304.10294} {\enquote {\bibinfo {title}
  {{{OptoGPT}}: {{A Foundation Model}} for {{Inverse Design}} in {{Optical
  Multilayer Thin Film Structures}}},}\ } (\bibinfo {year} {2023}),\ \Eprint
  {http://arxiv.org/abs/2304.10294} {arxiv:2304.10294 [physics]} \BibitemShut
  {NoStop}%
\bibitem [{\citenamefont {Krizhevsky}\ \emph {et~al.}(2012)\citenamefont
  {Krizhevsky}, \citenamefont {Sutskever},\ and\ \citenamefont
  {Hinton}}]{krizhevskyImageNetClassificationDeep2012}%
  \BibitemOpen
  \bibfield  {author} {\bibinfo {author} {\bibfnamefont {A.}~\bibnamefont
  {Krizhevsky}}, \bibinfo {author} {\bibfnamefont {I.}~\bibnamefont
  {Sutskever}}, \ and\ \bibinfo {author} {\bibfnamefont {G.~E.}\ \bibnamefont
  {Hinton}},\ }\href@noop {} {\bibfield  {journal} {\bibinfo  {journal}
  {Advances in Neural Information Processing Systems 25}\ ,\ \bibinfo {pages}
  {1097}} (\bibinfo {year} {2012})}\BibitemShut {NoStop}%
\bibitem [{\citenamefont {He}\ \emph {et~al.}(2015)\citenamefont {He},
  \citenamefont {Zhang}, \citenamefont {Ren},\ and\ \citenamefont
  {Sun}}]{heDeepResidualLearning2015}%
  \BibitemOpen
  \bibfield  {author} {\bibinfo {author} {\bibfnamefont {K.}~\bibnamefont
  {He}}, \bibinfo {author} {\bibfnamefont {X.}~\bibnamefont {Zhang}}, \bibinfo
  {author} {\bibfnamefont {S.}~\bibnamefont {Ren}}, \ and\ \bibinfo {author}
  {\bibfnamefont {J.}~\bibnamefont {Sun}},\ }\href@noop {} {\bibfield
  {journal} {\bibinfo  {journal} {arXiv:1512.03385 [cs]}\ } (\bibinfo {year}
  {2015})},\ \Eprint {http://arxiv.org/abs/1512.03385} {arxiv:1512.03385 [cs]}
  \BibitemShut {NoStop}%
\bibitem [{\citenamefont {Guo}\ \emph {et~al.}(2016)\citenamefont {Guo},
  \citenamefont {Liu}, \citenamefont {Oerlemans}, \citenamefont {Lao},
  \citenamefont {Wu},\ and\ \citenamefont {Lew}}]{guoDeepLearningVisual2016}%
  \BibitemOpen
  \bibfield  {author} {\bibinfo {author} {\bibfnamefont {Y.}~\bibnamefont
  {Guo}}, \bibinfo {author} {\bibfnamefont {Y.}~\bibnamefont {Liu}}, \bibinfo
  {author} {\bibfnamefont {A.}~\bibnamefont {Oerlemans}}, \bibinfo {author}
  {\bibfnamefont {S.}~\bibnamefont {Lao}}, \bibinfo {author} {\bibfnamefont
  {S.}~\bibnamefont {Wu}}, \ and\ \bibinfo {author} {\bibfnamefont {M.~S.}\
  \bibnamefont {Lew}},\ }\href {\doibase 10.1016/j.neucom.2015.09.116}
  {\bibfield  {journal} {\bibinfo  {journal} {Neurocomputing}\ }\bibinfo
  {series} {Recent {{Developments}} on {{Deep Big Vision}}},\ \textbf {\bibinfo
  {volume} {187}},\ \bibinfo {pages} {27} (\bibinfo {year} {2016})}\BibitemShut
  {NoStop}%
\bibitem [{\citenamefont {Kirillov}\ \emph {et~al.}(2023)\citenamefont
  {Kirillov}, \citenamefont {Mintun}, \citenamefont {Ravi}, \citenamefont
  {Mao}, \citenamefont {Rolland}, \citenamefont {Gustafson}, \citenamefont
  {Xiao}, \citenamefont {Whitehead}, \citenamefont {Berg}, \citenamefont {Lo},
  \citenamefont {Doll{\'a}r},\ and\ \citenamefont
  {Girshick}}]{kirillovSegmentAnything2023}%
  \BibitemOpen
  \bibfield  {author} {\bibinfo {author} {\bibfnamefont {A.}~\bibnamefont
  {Kirillov}}, \bibinfo {author} {\bibfnamefont {E.}~\bibnamefont {Mintun}},
  \bibinfo {author} {\bibfnamefont {N.}~\bibnamefont {Ravi}}, \bibinfo {author}
  {\bibfnamefont {H.}~\bibnamefont {Mao}}, \bibinfo {author} {\bibfnamefont
  {C.}~\bibnamefont {Rolland}}, \bibinfo {author} {\bibfnamefont
  {L.}~\bibnamefont {Gustafson}}, \bibinfo {author} {\bibfnamefont
  {T.}~\bibnamefont {Xiao}}, \bibinfo {author} {\bibfnamefont {S.}~\bibnamefont
  {Whitehead}}, \bibinfo {author} {\bibfnamefont {A.~C.}\ \bibnamefont {Berg}},
  \bibinfo {author} {\bibfnamefont {W.-Y.}\ \bibnamefont {Lo}}, \bibinfo
  {author} {\bibfnamefont {P.}~\bibnamefont {Doll{\'a}r}}, \ and\ \bibinfo
  {author} {\bibfnamefont {R.}~\bibnamefont {Girshick}},\ }\href {\doibase
  10.48550/arXiv.2304.02643} {\enquote {\bibinfo {title} {Segment
  {{Anything}}},}\ } (\bibinfo {year} {2023}),\ \Eprint
  {http://arxiv.org/abs/2304.02643} {arxiv:2304.02643 [cs]} \BibitemShut
  {NoStop}%
\bibitem [{\citenamefont {Sundermeyer}\ \emph {et~al.}(2012)\citenamefont
  {Sundermeyer}, \citenamefont {Schl{\"u}ter},\ and\ \citenamefont
  {Ney}}]{sundermeyerLSTMNeuralNetworks2012}%
  \BibitemOpen
  \bibfield  {author} {\bibinfo {author} {\bibfnamefont {M.}~\bibnamefont
  {Sundermeyer}}, \bibinfo {author} {\bibfnamefont {R.}~\bibnamefont
  {Schl{\"u}ter}}, \ and\ \bibinfo {author} {\bibfnamefont {H.}~\bibnamefont
  {Ney}},\ }in\ \href@noop {} {\emph {\bibinfo {booktitle} {Thirteenth Annual
  Conference of the International Speech Communication Association}}}\
  (\bibinfo {year} {2012})\BibitemShut {NoStop}%
\bibitem [{\citenamefont {Brown}\ \emph {et~al.}(2020)\citenamefont {Brown},
  \citenamefont {Mann}, \citenamefont {Ryder}, \citenamefont {Subbiah},
  \citenamefont {Kaplan}, \citenamefont {Dhariwal}, \citenamefont
  {Neelakantan}, \citenamefont {Shyam}, \citenamefont {Sastry}, \citenamefont
  {Askell}, \citenamefont {Agarwal}, \citenamefont {{Herbert-Voss}},
  \citenamefont {Krueger}, \citenamefont {Henighan}, \citenamefont {Child},
  \citenamefont {Ramesh}, \citenamefont {Ziegler}, \citenamefont {Wu},
  \citenamefont {Winter}, \citenamefont {Hesse}, \citenamefont {Chen},
  \citenamefont {Sigler}, \citenamefont {Litwin}, \citenamefont {Gray},
  \citenamefont {Chess}, \citenamefont {Clark}, \citenamefont {Berner},
  \citenamefont {McCandlish}, \citenamefont {Radford}, \citenamefont
  {Sutskever},\ and\ \citenamefont {Amodei}}]{brownLanguageModelsAre2020}%
  \BibitemOpen
  \bibfield  {author} {\bibinfo {author} {\bibfnamefont {T.~B.}\ \bibnamefont
  {Brown}}, \bibinfo {author} {\bibfnamefont {B.}~\bibnamefont {Mann}},
  \bibinfo {author} {\bibfnamefont {N.}~\bibnamefont {Ryder}}, \bibinfo
  {author} {\bibfnamefont {M.}~\bibnamefont {Subbiah}}, \bibinfo {author}
  {\bibfnamefont {J.}~\bibnamefont {Kaplan}}, \bibinfo {author} {\bibfnamefont
  {P.}~\bibnamefont {Dhariwal}}, \bibinfo {author} {\bibfnamefont
  {A.}~\bibnamefont {Neelakantan}}, \bibinfo {author} {\bibfnamefont
  {P.}~\bibnamefont {Shyam}}, \bibinfo {author} {\bibfnamefont
  {G.}~\bibnamefont {Sastry}}, \bibinfo {author} {\bibfnamefont
  {A.}~\bibnamefont {Askell}}, \bibinfo {author} {\bibfnamefont
  {S.}~\bibnamefont {Agarwal}}, \bibinfo {author} {\bibfnamefont
  {A.}~\bibnamefont {{Herbert-Voss}}}, \bibinfo {author} {\bibfnamefont
  {G.}~\bibnamefont {Krueger}}, \bibinfo {author} {\bibfnamefont
  {T.}~\bibnamefont {Henighan}}, \bibinfo {author} {\bibfnamefont
  {R.}~\bibnamefont {Child}}, \bibinfo {author} {\bibfnamefont
  {A.}~\bibnamefont {Ramesh}}, \bibinfo {author} {\bibfnamefont {D.~M.}\
  \bibnamefont {Ziegler}}, \bibinfo {author} {\bibfnamefont {J.}~\bibnamefont
  {Wu}}, \bibinfo {author} {\bibfnamefont {C.}~\bibnamefont {Winter}}, \bibinfo
  {author} {\bibfnamefont {C.}~\bibnamefont {Hesse}}, \bibinfo {author}
  {\bibfnamefont {M.}~\bibnamefont {Chen}}, \bibinfo {author} {\bibfnamefont
  {E.}~\bibnamefont {Sigler}}, \bibinfo {author} {\bibfnamefont
  {M.}~\bibnamefont {Litwin}}, \bibinfo {author} {\bibfnamefont
  {S.}~\bibnamefont {Gray}}, \bibinfo {author} {\bibfnamefont {B.}~\bibnamefont
  {Chess}}, \bibinfo {author} {\bibfnamefont {J.}~\bibnamefont {Clark}},
  \bibinfo {author} {\bibfnamefont {C.}~\bibnamefont {Berner}}, \bibinfo
  {author} {\bibfnamefont {S.}~\bibnamefont {McCandlish}}, \bibinfo {author}
  {\bibfnamefont {A.}~\bibnamefont {Radford}}, \bibinfo {author} {\bibfnamefont
  {I.}~\bibnamefont {Sutskever}}, \ and\ \bibinfo {author} {\bibfnamefont
  {D.}~\bibnamefont {Amodei}},\ }in\ \href@noop {} {\emph {\bibinfo {booktitle}
  {Advances in {{Neural Information Processing Systems}}}}},\ Vol.\ \bibinfo
  {volume} {300}\ (\bibinfo {year} {2020})\ pp.\ \bibinfo {pages}
  {1877--1901},\ \Eprint {http://arxiv.org/abs/2005.14165} {arxiv:2005.14165}
  \BibitemShut {NoStop}%
\bibitem [{\citenamefont {Otter}\ \emph {et~al.}(2021)\citenamefont {Otter},
  \citenamefont {Medina},\ and\ \citenamefont
  {Kalita}}]{otterSurveyUsagesDeep2021}%
  \BibitemOpen
  \bibfield  {author} {\bibinfo {author} {\bibfnamefont {D.~W.}\ \bibnamefont
  {Otter}}, \bibinfo {author} {\bibfnamefont {J.~R.}\ \bibnamefont {Medina}}, \
  and\ \bibinfo {author} {\bibfnamefont {J.~K.}\ \bibnamefont {Kalita}},\
  }\href {\doibase 10.1109/TNNLS.2020.2979670} {\bibfield  {journal} {\bibinfo
  {journal} {IEEE Transactions on Neural Networks and Learning Systems}\
  }\textbf {\bibinfo {volume} {32}},\ \bibinfo {pages} {604} (\bibinfo {year}
  {2021})}\BibitemShut {NoStop}%
\bibitem [{\citenamefont {Hornik}\ \emph {et~al.}(1989)\citenamefont {Hornik},
  \citenamefont {Stinchcombe},\ and\ \citenamefont
  {White}}]{hornikMultilayerFeedforwardNetworks1989}%
  \BibitemOpen
  \bibfield  {author} {\bibinfo {author} {\bibfnamefont {K.}~\bibnamefont
  {Hornik}}, \bibinfo {author} {\bibfnamefont {M.}~\bibnamefont {Stinchcombe}},
  \ and\ \bibinfo {author} {\bibfnamefont {H.}~\bibnamefont {White}},\ }\href
  {\doibase 10.1016/0893-6080(89)90020-8} {\bibfield  {journal} {\bibinfo
  {journal} {Neural Networks}\ }\textbf {\bibinfo {volume} {2}},\ \bibinfo
  {pages} {359} (\bibinfo {year} {1989})}\BibitemShut {NoStop}%
\bibitem [{\citenamefont {Peurifoy}\ \emph {et~al.}(2018)\citenamefont
  {Peurifoy}, \citenamefont {Shen}, \citenamefont {Jing}, \citenamefont {Yang},
  \citenamefont {{Cano-Renteria}}, \citenamefont {DeLacy}, \citenamefont
  {Joannopoulos}, \citenamefont {Tegmark},\ and\ \citenamefont {Solja{\v
  c}i{\'c}}}]{peurifoyNanophotonicParticleSimulation2018}%
  \BibitemOpen
  \bibfield  {author} {\bibinfo {author} {\bibfnamefont {J.}~\bibnamefont
  {Peurifoy}}, \bibinfo {author} {\bibfnamefont {Y.}~\bibnamefont {Shen}},
  \bibinfo {author} {\bibfnamefont {L.}~\bibnamefont {Jing}}, \bibinfo {author}
  {\bibfnamefont {Y.}~\bibnamefont {Yang}}, \bibinfo {author} {\bibfnamefont
  {F.}~\bibnamefont {{Cano-Renteria}}}, \bibinfo {author} {\bibfnamefont
  {B.~G.}\ \bibnamefont {DeLacy}}, \bibinfo {author} {\bibfnamefont {J.~D.}\
  \bibnamefont {Joannopoulos}}, \bibinfo {author} {\bibfnamefont
  {M.}~\bibnamefont {Tegmark}}, \ and\ \bibinfo {author} {\bibfnamefont
  {M.}~\bibnamefont {Solja{\v c}i{\'c}}},\ }\href {\doibase
  10.1126/sciadv.aar4206} {\bibfield  {journal} {\bibinfo  {journal} {Science
  Advances}\ }\textbf {\bibinfo {volume} {4}},\ \bibinfo {pages} {eaar4206}
  (\bibinfo {year} {2018})}\BibitemShut {NoStop}%
\bibitem [{\citenamefont {{Estrada-Real}}\ \emph {et~al.}(2022)\citenamefont
  {{Estrada-Real}}, \citenamefont {{Khaireh-Walieh}}, \citenamefont
  {Urbaszek},\ and\ \citenamefont
  {Wiecha}}]{estrada-realInverseDesignFlexible2022}%
  \BibitemOpen
  \bibfield  {author} {\bibinfo {author} {\bibfnamefont {A.}~\bibnamefont
  {{Estrada-Real}}}, \bibinfo {author} {\bibfnamefont {A.}~\bibnamefont
  {{Khaireh-Walieh}}}, \bibinfo {author} {\bibfnamefont {B.}~\bibnamefont
  {Urbaszek}}, \ and\ \bibinfo {author} {\bibfnamefont {P.~R.}\ \bibnamefont
  {Wiecha}},\ }\href {\doibase 10.1016/j.photonics.2022.101066} {\bibfield
  {journal} {\bibinfo  {journal} {Photonics and Nanostructures - Fundamentals
  and Applications}\ }\textbf {\bibinfo {volume} {52}},\ \bibinfo {pages}
  {101066} (\bibinfo {year} {2022})},\ \Eprint
  {http://arxiv.org/abs/2207.03431} {arxiv:2207.03431 [physics]} \BibitemShut
  {NoStop}%
\bibitem [{\citenamefont {Jiang}\ and\ \citenamefont
  {Fan}(2019{\natexlab{a}})}]{jiangGlobalOptimizationDielectric2019}%
  \BibitemOpen
  \bibfield  {author} {\bibinfo {author} {\bibfnamefont {J.}~\bibnamefont
  {Jiang}}\ and\ \bibinfo {author} {\bibfnamefont {J.~A.}\ \bibnamefont
  {Fan}},\ }\href {\doibase 10.1021/acs.nanolett.9b01857} {\bibfield  {journal}
  {\bibinfo  {journal} {Nano Letters}\ }\textbf {\bibinfo {volume} {19}},\
  \bibinfo {pages} {5366} (\bibinfo {year} {2019}{\natexlab{a}})}\BibitemShut
  {NoStop}%
\bibitem [{\citenamefont {Jiang}\ and\ \citenamefont
  {Fan}(2019{\natexlab{b}})}]{jiangSimulatorbasedTrainingGenerative2019}%
  \BibitemOpen
  \bibfield  {author} {\bibinfo {author} {\bibfnamefont {J.}~\bibnamefont
  {Jiang}}\ and\ \bibinfo {author} {\bibfnamefont {J.~A.}\ \bibnamefont
  {Fan}},\ }\href {\doibase 10.1515/nanoph-2019-0330} {\bibfield  {journal}
  {\bibinfo  {journal} {Nanophotonics}\ }\textbf {\bibinfo {volume} {9}},\
  \bibinfo {pages} {1059} (\bibinfo {year} {2019}{\natexlab{b}})}\BibitemShut
  {NoStop}%
\bibitem [{\citenamefont {Liu}\ \emph {et~al.}(2018)\citenamefont {Liu},
  \citenamefont {Tan}, \citenamefont {Khoram},\ and\ \citenamefont
  {Yu}}]{liuTrainingDeepNeural2018}%
  \BibitemOpen
  \bibfield  {author} {\bibinfo {author} {\bibfnamefont {D.}~\bibnamefont
  {Liu}}, \bibinfo {author} {\bibfnamefont {Y.}~\bibnamefont {Tan}}, \bibinfo
  {author} {\bibfnamefont {E.}~\bibnamefont {Khoram}}, \ and\ \bibinfo {author}
  {\bibfnamefont {Z.}~\bibnamefont {Yu}},\ }\href {\doibase
  10.1021/acsphotonics.7b01377} {\bibfield  {journal} {\bibinfo  {journal} {ACS
  Photonics}\ }\textbf {\bibinfo {volume} {5}},\ \bibinfo {pages} {1365}
  (\bibinfo {year} {2018})}\BibitemShut {NoStop}%
\bibitem [{\citenamefont {Unni}\ \emph {et~al.}(2020)\citenamefont {Unni},
  \citenamefont {Yao},\ and\ \citenamefont
  {Zheng}}]{unniDeepConvolutionalMixture2020}%
  \BibitemOpen
  \bibfield  {author} {\bibinfo {author} {\bibfnamefont {R.}~\bibnamefont
  {Unni}}, \bibinfo {author} {\bibfnamefont {K.}~\bibnamefont {Yao}}, \ and\
  \bibinfo {author} {\bibfnamefont {Y.}~\bibnamefont {Zheng}},\ }\href
  {\doibase 10.1021/acsphotonics.0c00630} {\bibfield  {journal} {\bibinfo
  {journal} {ACS Photonics}\ }\textbf {\bibinfo {volume} {7}} (\bibinfo {year}
  {2020}),\ 10.1021/acsphotonics.0c00630}\BibitemShut {NoStop}%
\bibitem [{\citenamefont {Dai}\ \emph {et~al.}(2022)\citenamefont {Dai},
  \citenamefont {Sun}, \citenamefont {Yan}, \citenamefont {Muskens},
  \citenamefont {de~Groot}, \citenamefont {Zhu}, \citenamefont {Hu},
  \citenamefont {Duan},\ and\ \citenamefont
  {Huang}}]{daiInverseDesignStructural2022}%
  \BibitemOpen
  \bibfield  {author} {\bibinfo {author} {\bibfnamefont {P.}~\bibnamefont
  {Dai}}, \bibinfo {author} {\bibfnamefont {K.}~\bibnamefont {Sun}}, \bibinfo
  {author} {\bibfnamefont {X.}~\bibnamefont {Yan}}, \bibinfo {author}
  {\bibfnamefont {O.~L.}\ \bibnamefont {Muskens}}, \bibinfo {author}
  {\bibfnamefont {C.~H.~K.}\ \bibnamefont {de~Groot}}, \bibinfo {author}
  {\bibfnamefont {X.}~\bibnamefont {Zhu}}, \bibinfo {author} {\bibfnamefont
  {Y.}~\bibnamefont {Hu}}, \bibinfo {author} {\bibfnamefont {H.}~\bibnamefont
  {Duan}}, \ and\ \bibinfo {author} {\bibfnamefont {R.}~\bibnamefont {Huang}},\
  }\href {\doibase 10.1515/nanoph-2022-0095} {\bibfield  {journal} {\bibinfo
  {journal} {Nanophotonics}\ }\textbf {\bibinfo {volume} {11}},\ \bibinfo
  {pages} {3057} (\bibinfo {year} {2022})}\BibitemShut {NoStop}%
\bibitem [{\citenamefont {Asano}\ and\ \citenamefont
  {Noda}(2019)}]{asanoIterativeOptimizationPhotonic2019}%
  \BibitemOpen
  \bibfield  {author} {\bibinfo {author} {\bibfnamefont {T.}~\bibnamefont
  {Asano}}\ and\ \bibinfo {author} {\bibfnamefont {S.}~\bibnamefont {Noda}},\
  }\href {\doibase 10.1515/nanoph-2019-0308} {\bibfield  {journal} {\bibinfo
  {journal} {Nanophotonics}\ }\textbf {\bibinfo {volume} {8}},\ \bibinfo
  {pages} {2243} (\bibinfo {year} {2019})}\BibitemShut {NoStop}%
\bibitem [{\citenamefont {Zhang}\ \emph {et~al.}(2019)\citenamefont {Zhang},
  \citenamefont {Wang}, \citenamefont {Liu}, \citenamefont {Zhou},
  \citenamefont {Dai}, \citenamefont {Han}, \citenamefont {Zhou},\ and\
  \citenamefont {Xu}}]{zhangEfficientSpectrumPrediction2019}%
  \BibitemOpen
  \bibfield  {author} {\bibinfo {author} {\bibfnamefont {T.}~\bibnamefont
  {Zhang}}, \bibinfo {author} {\bibfnamefont {J.}~\bibnamefont {Wang}},
  \bibinfo {author} {\bibfnamefont {Q.}~\bibnamefont {Liu}}, \bibinfo {author}
  {\bibfnamefont {J.}~\bibnamefont {Zhou}}, \bibinfo {author} {\bibfnamefont
  {J.}~\bibnamefont {Dai}}, \bibinfo {author} {\bibfnamefont {X.}~\bibnamefont
  {Han}}, \bibinfo {author} {\bibfnamefont {Y.}~\bibnamefont {Zhou}}, \ and\
  \bibinfo {author} {\bibfnamefont {K.}~\bibnamefont {Xu}},\ }\href {\doibase
  10.1364/PRJ.7.000368} {\bibfield  {journal} {\bibinfo  {journal} {Photonics
  Research}\ }\textbf {\bibinfo {volume} {7}},\ \bibinfo {pages} {368}
  (\bibinfo {year} {2019})},\ \Eprint {http://arxiv.org/abs/1805.06410}
  {arxiv:1805.06410} \BibitemShut {NoStop}%
\bibitem [{\citenamefont {Tahersima}\ \emph {et~al.}(2019)\citenamefont
  {Tahersima}, \citenamefont {Kojima}, \citenamefont {{Koike-Akino}},
  \citenamefont {Jha}, \citenamefont {Wang}, \citenamefont {Lin},\ and\
  \citenamefont {Parsons}}]{tahersimaDeepNeuralNetwork2019}%
  \BibitemOpen
  \bibfield  {author} {\bibinfo {author} {\bibfnamefont {M.~H.}\ \bibnamefont
  {Tahersima}}, \bibinfo {author} {\bibfnamefont {K.}~\bibnamefont {Kojima}},
  \bibinfo {author} {\bibfnamefont {T.}~\bibnamefont {{Koike-Akino}}}, \bibinfo
  {author} {\bibfnamefont {D.}~\bibnamefont {Jha}}, \bibinfo {author}
  {\bibfnamefont {B.}~\bibnamefont {Wang}}, \bibinfo {author} {\bibfnamefont
  {C.}~\bibnamefont {Lin}}, \ and\ \bibinfo {author} {\bibfnamefont
  {K.}~\bibnamefont {Parsons}},\ }\href {\doibase 10.1038/s41598-018-37952-2}
  {\bibfield  {journal} {\bibinfo  {journal} {Scientific Reports}\ }\textbf
  {\bibinfo {volume} {9}},\ \bibinfo {pages} {1368} (\bibinfo {year}
  {2019})}\BibitemShut {NoStop}%
\bibitem [{\citenamefont {Banerji}\ \emph {et~al.}(2020)\citenamefont
  {Banerji}, \citenamefont {Majumder}, \citenamefont {Hamrick}, \citenamefont
  {Menon},\ and\ \citenamefont
  {{Sensale-Rodriguez}}}]{banerjiMachineLearningEnables2020}%
  \BibitemOpen
  \bibfield  {author} {\bibinfo {author} {\bibfnamefont {S.}~\bibnamefont
  {Banerji}}, \bibinfo {author} {\bibfnamefont {A.}~\bibnamefont {Majumder}},
  \bibinfo {author} {\bibfnamefont {A.}~\bibnamefont {Hamrick}}, \bibinfo
  {author} {\bibfnamefont {R.}~\bibnamefont {Menon}}, \ and\ \bibinfo {author}
  {\bibfnamefont {B.}~\bibnamefont {{Sensale-Rodriguez}}},\ }\href {\doibase
  10.1016/j.nancom.2020.100312} {\bibfield  {journal} {\bibinfo  {journal}
  {Nano Communication Networks}\ }\textbf {\bibinfo {volume} {25}},\ \bibinfo
  {pages} {100312} (\bibinfo {year} {2020})}\BibitemShut {NoStop}%
\bibitem [{\citenamefont {Dinsdale}\ \emph {et~al.}(2021)\citenamefont
  {Dinsdale}, \citenamefont {Wiecha}, \citenamefont {Delaney}, \citenamefont
  {Reynolds}, \citenamefont {Ebert}, \citenamefont {Zeimpekis}, \citenamefont
  {Thomson}, \citenamefont {Reed}, \citenamefont {Lalanne}, \citenamefont
  {Vynck},\ and\ \citenamefont {Muskens}}]{dinsdaleDeepLearningEnabled2021}%
  \BibitemOpen
  \bibfield  {author} {\bibinfo {author} {\bibfnamefont {N.~J.}\ \bibnamefont
  {Dinsdale}}, \bibinfo {author} {\bibfnamefont {P.~R.}\ \bibnamefont
  {Wiecha}}, \bibinfo {author} {\bibfnamefont {M.}~\bibnamefont {Delaney}},
  \bibinfo {author} {\bibfnamefont {J.}~\bibnamefont {Reynolds}}, \bibinfo
  {author} {\bibfnamefont {M.}~\bibnamefont {Ebert}}, \bibinfo {author}
  {\bibfnamefont {I.}~\bibnamefont {Zeimpekis}}, \bibinfo {author}
  {\bibfnamefont {D.~J.}\ \bibnamefont {Thomson}}, \bibinfo {author}
  {\bibfnamefont {G.~T.}\ \bibnamefont {Reed}}, \bibinfo {author}
  {\bibfnamefont {P.}~\bibnamefont {Lalanne}}, \bibinfo {author} {\bibfnamefont
  {K.}~\bibnamefont {Vynck}}, \ and\ \bibinfo {author} {\bibfnamefont {O.~L.}\
  \bibnamefont {Muskens}},\ }\href {\doibase 10.1021/acsphotonics.0c01481}
  {\bibfield  {journal} {\bibinfo  {journal} {ACS Photonics}\ }\textbf
  {\bibinfo {volume} {8}},\ \bibinfo {pages} {283} (\bibinfo {year} {2021})},\
  \Eprint {http://arxiv.org/abs/2009.11810} {arxiv:2009.11810} \BibitemShut
  {NoStop}%
\bibitem [{\citenamefont {Zhou}\ \emph {et~al.}(2019)\citenamefont {Zhou},
  \citenamefont {Huang}, \citenamefont {Yan},\ and\ \citenamefont
  {B{\"u}nzli}}]{zhouEmergingRoleMachine2019}%
  \BibitemOpen
  \bibfield  {author} {\bibinfo {author} {\bibfnamefont {J.}~\bibnamefont
  {Zhou}}, \bibinfo {author} {\bibfnamefont {B.}~\bibnamefont {Huang}},
  \bibinfo {author} {\bibfnamefont {Z.}~\bibnamefont {Yan}}, \ and\ \bibinfo
  {author} {\bibfnamefont {J.-C.~G.}\ \bibnamefont {B{\"u}nzli}},\ }\href
  {\doibase 10.1038/s41377-019-0192-4} {\bibfield  {journal} {\bibinfo
  {journal} {Light: Science \& Applications}\ }\textbf {\bibinfo {volume}
  {8}},\ \bibinfo {pages} {1} (\bibinfo {year} {2019})}\BibitemShut {NoStop}%
\bibitem [{\citenamefont {So}\ \emph {et~al.}(2020)\citenamefont {So},
  \citenamefont {Badloe}, \citenamefont {Noh}, \citenamefont {{Bravo-Abad}},\
  and\ \citenamefont {Rho}}]{soDeepLearningEnabled2020}%
  \BibitemOpen
  \bibfield  {author} {\bibinfo {author} {\bibfnamefont {S.}~\bibnamefont
  {So}}, \bibinfo {author} {\bibfnamefont {T.}~\bibnamefont {Badloe}}, \bibinfo
  {author} {\bibfnamefont {J.}~\bibnamefont {Noh}}, \bibinfo {author}
  {\bibfnamefont {J.}~\bibnamefont {{Bravo-Abad}}}, \ and\ \bibinfo {author}
  {\bibfnamefont {J.}~\bibnamefont {Rho}},\ }\href {\doibase
  10.1515/nanoph-2019-0474} {\bibfield  {journal} {\bibinfo  {journal}
  {Nanophotonics}\ }\textbf {\bibinfo {volume} {9}},\ \bibinfo {pages} {1041}
  (\bibinfo {year} {2020})}\BibitemShut {NoStop}%
\bibitem [{\citenamefont {Jiang}\ \emph {et~al.}(2021)\citenamefont {Jiang},
  \citenamefont {Chen},\ and\ \citenamefont
  {Fan}}]{jiangDeepNeuralNetworks2021}%
  \BibitemOpen
  \bibfield  {author} {\bibinfo {author} {\bibfnamefont {J.}~\bibnamefont
  {Jiang}}, \bibinfo {author} {\bibfnamefont {M.}~\bibnamefont {Chen}}, \ and\
  \bibinfo {author} {\bibfnamefont {J.~A.}\ \bibnamefont {Fan}},\ }\href
  {\doibase 10.1038/s41578-020-00260-1} {\bibfield  {journal} {\bibinfo
  {journal} {Nature Reviews Materials}\ }\textbf {\bibinfo {volume} {6}},\
  \bibinfo {pages} {679} (\bibinfo {year} {2021})},\ \Eprint
  {http://arxiv.org/abs/2007.00084} {arxiv:2007.00084} \BibitemShut {NoStop}%
\bibitem [{\citenamefont {Liu}\ \emph {et~al.}(2021{\natexlab{a}})\citenamefont
  {Liu}, \citenamefont {Zhu}, \citenamefont {Raju},\ and\ \citenamefont
  {Cai}}]{liuTacklingPhotonicInverse2021}%
  \BibitemOpen
  \bibfield  {author} {\bibinfo {author} {\bibfnamefont {Z.}~\bibnamefont
  {Liu}}, \bibinfo {author} {\bibfnamefont {D.}~\bibnamefont {Zhu}}, \bibinfo
  {author} {\bibfnamefont {L.}~\bibnamefont {Raju}}, \ and\ \bibinfo {author}
  {\bibfnamefont {W.}~\bibnamefont {Cai}},\ }\href {\doibase
  10.1002/advs.202002923} {\bibfield  {journal} {\bibinfo  {journal} {Advanced
  Science}\ }\textbf {\bibinfo {volume} {8}},\ \bibinfo {pages} {2002923}
  (\bibinfo {year} {2021}{\natexlab{a}})}\BibitemShut {NoStop}%
\bibitem [{\citenamefont {Wiecha}\ \emph {et~al.}(2021)\citenamefont {Wiecha},
  \citenamefont {Arbouet}, \citenamefont {Girard},\ and\ \citenamefont
  {Muskens}}]{wiechaDeepLearningNanophotonics2021}%
  \BibitemOpen
  \bibfield  {author} {\bibinfo {author} {\bibfnamefont {P.~R.}\ \bibnamefont
  {Wiecha}}, \bibinfo {author} {\bibfnamefont {A.}~\bibnamefont {Arbouet}},
  \bibinfo {author} {\bibfnamefont {C.}~\bibnamefont {Girard}}, \ and\ \bibinfo
  {author} {\bibfnamefont {O.~L.}\ \bibnamefont {Muskens}},\ }\href {\doibase
  10.1364/PRJ.415960} {\bibfield  {journal} {\bibinfo  {journal} {Photonics
  Research}\ }\textbf {\bibinfo {volume} {9}},\ \bibinfo {pages} {B182}
  (\bibinfo {year} {2021})},\ \Eprint {http://arxiv.org/abs/2011.12603}
  {arxiv:2011.12603} \BibitemShut {NoStop}%
\bibitem [{\citenamefont {Deng}\ \emph {et~al.}(2022)\citenamefont {Deng},
  \citenamefont {Ren}, \citenamefont {Malof},\ and\ \citenamefont
  {Padilla}}]{dengDeepInversePhotonic2022}%
  \BibitemOpen
  \bibfield  {author} {\bibinfo {author} {\bibfnamefont {Y.}~\bibnamefont
  {Deng}}, \bibinfo {author} {\bibfnamefont {S.}~\bibnamefont {Ren}}, \bibinfo
  {author} {\bibfnamefont {J.}~\bibnamefont {Malof}}, \ and\ \bibinfo {author}
  {\bibfnamefont {W.~J.}\ \bibnamefont {Padilla}},\ }\href {\doibase
  10.1016/j.photonics.2022.101070} {\bibfield  {journal} {\bibinfo  {journal}
  {Photonics and Nanostructures - Fundamentals and Applications}\ }\textbf
  {\bibinfo {volume} {52}},\ \bibinfo {pages} {101070} (\bibinfo {year}
  {2022})}\BibitemShut {NoStop}%
\bibitem [{\citenamefont {{Kan Yao}}\ and\ \citenamefont {{Yuebing
  Zheng}}(2023)}]{kanyaoNanophotonicsMachineLearning2023}%
  \BibitemOpen
  \bibfield  {author} {\bibinfo {author} {\bibnamefont {{Kan Yao}}}\ and\
  \bibinfo {author} {\bibnamefont {{Yuebing Zheng}}},\ }\href@noop {} {\emph
  {\bibinfo {title} {Nanophotonics and {{Machine Learning}} - {{Concepts}},
  {{Fundamentals}}, and {{Applications}}}}},\ Springer {{Series}} in {{Optical
  Sciences}}\ (\bibinfo  {publisher} {{Springer}},\ \bibinfo {year}
  {2023})\BibitemShut {NoStop}%
\bibitem [{\citenamefont {Ji}\ \emph {et~al.}(2023)\citenamefont {Ji},
  \citenamefont {Chang}, \citenamefont {Xu}, \citenamefont {Gao}, \citenamefont
  {Gr{\"o}blacher}, \citenamefont {Urbach},\ and\ \citenamefont
  {Adam}}]{jiRecentAdvancesMetasurface2023a}%
  \BibitemOpen
  \bibfield  {author} {\bibinfo {author} {\bibfnamefont {W.}~\bibnamefont
  {Ji}}, \bibinfo {author} {\bibfnamefont {J.}~\bibnamefont {Chang}}, \bibinfo
  {author} {\bibfnamefont {H.-X.}\ \bibnamefont {Xu}}, \bibinfo {author}
  {\bibfnamefont {J.~R.}\ \bibnamefont {Gao}}, \bibinfo {author} {\bibfnamefont
  {S.}~\bibnamefont {Gr{\"o}blacher}}, \bibinfo {author} {\bibfnamefont
  {H.~P.}\ \bibnamefont {Urbach}}, \ and\ \bibinfo {author} {\bibfnamefont
  {A.~J.~L.}\ \bibnamefont {Adam}},\ }\href {\doibase
  10.1038/s41377-023-01218-y} {\bibfield  {journal} {\bibinfo  {journal}
  {Light: Science \& Applications}\ }\textbf {\bibinfo {volume} {12}},\
  \bibinfo {pages} {169} (\bibinfo {year} {2023})}\BibitemShut {NoStop}%
\bibitem [{\citenamefont {Schneider}\ \emph {et~al.}(2019)\citenamefont
  {Schneider}, \citenamefont {Garcia~Santiago}, \citenamefont {Soltwisch},
  \citenamefont {Hammerschmidt}, \citenamefont {Burger},\ and\ \citenamefont
  {Rockstuhl}}]{schneiderBenchmarkingFiveGlobal2019}%
  \BibitemOpen
  \bibfield  {author} {\bibinfo {author} {\bibfnamefont {P.-I.}\ \bibnamefont
  {Schneider}}, \bibinfo {author} {\bibfnamefont {X.}~\bibnamefont
  {Garcia~Santiago}}, \bibinfo {author} {\bibfnamefont {V.}~\bibnamefont
  {Soltwisch}}, \bibinfo {author} {\bibfnamefont {M.}~\bibnamefont
  {Hammerschmidt}}, \bibinfo {author} {\bibfnamefont {S.}~\bibnamefont
  {Burger}}, \ and\ \bibinfo {author} {\bibfnamefont {C.}~\bibnamefont
  {Rockstuhl}},\ }\href {\doibase 10.1021/acsphotonics.9b00706} {\bibfield
  {journal} {\bibinfo  {journal} {ACS Photonics}\ }\textbf {\bibinfo {volume}
  {6}},\ \bibinfo {pages} {2726} (\bibinfo {year} {2019})}\BibitemShut
  {NoStop}%
\bibitem [{\citenamefont
  {Hegde}(2020{\natexlab{a}})}]{hegdeDeepLearningNew2020}%
  \BibitemOpen
  \bibfield  {author} {\bibinfo {author} {\bibfnamefont {R.~S.}\ \bibnamefont
  {Hegde}},\ }\href {\doibase 10.1039/C9NA00656G} {\bibfield  {journal}
  {\bibinfo  {journal} {Nanoscale Advances}\ }\textbf {\bibinfo {volume} {2}},\
  \bibinfo {pages} {1007} (\bibinfo {year} {2020}{\natexlab{a}})}\BibitemShut
  {NoStop}%
\bibitem [{\citenamefont {Ren}\ \emph {et~al.}(2022)\citenamefont {Ren},
  \citenamefont {Mahendra}, \citenamefont {Khatib}, \citenamefont {Deng},
  \citenamefont {J.~Padilla},\ and\ \citenamefont
  {M.~Malof}}]{renInverseDeepLearning2022}%
  \BibitemOpen
  \bibfield  {author} {\bibinfo {author} {\bibfnamefont {S.}~\bibnamefont
  {Ren}}, \bibinfo {author} {\bibfnamefont {A.}~\bibnamefont {Mahendra}},
  \bibinfo {author} {\bibfnamefont {O.}~\bibnamefont {Khatib}}, \bibinfo
  {author} {\bibfnamefont {Y.}~\bibnamefont {Deng}}, \bibinfo {author}
  {\bibfnamefont {W.}~\bibnamefont {J.~Padilla}}, \ and\ \bibinfo {author}
  {\bibfnamefont {J.}~\bibnamefont {M.~Malof}},\ }\href {\doibase
  10.1039/D1NR08346E} {\bibfield  {journal} {\bibinfo  {journal} {Nanoscale}\
  }\textbf {\bibinfo {volume} {14}},\ \bibinfo {pages} {3958} (\bibinfo {year}
  {2022})},\ \Eprint {http://arxiv.org/abs/2009.12919} {arxiv:2009.12919}
  \BibitemShut {NoStop}%
\bibitem [{\citenamefont
  {Wiecha}(2023{\natexlab{a}})}]{wiechaNewcomerGuideDeep2023}%
  \BibitemOpen
  \bibfield  {author} {\bibinfo {author} {\bibfnamefont {P.~R.}\ \bibnamefont
  {Wiecha}},\ }\href@noop {} {\enquote {\bibinfo {title} {A newcomer's guide to
  deep learning for inverse design in nano-photonics},}\ }\bibinfo
  {howpublished}
  {https://gitlab.com/wiechapeter/newcomer\_guide\_dl\_inversedesign} (\bibinfo
  {year} {2023}{\natexlab{a}})\BibitemShut {NoStop}%
\bibitem [{\citenamefont {Robbins}\ and\ \citenamefont
  {Monro}(1951)}]{robbinsStochasticApproximationMethod1951}%
  \BibitemOpen
  \bibfield  {author} {\bibinfo {author} {\bibfnamefont {H.}~\bibnamefont
  {Robbins}}\ and\ \bibinfo {author} {\bibfnamefont {S.}~\bibnamefont
  {Monro}},\ }\href {\doibase 10.1214/aoms/1177729586} {\bibfield  {journal}
  {\bibinfo  {journal} {The Annals of Mathematical Statistics}\ }\textbf
  {\bibinfo {volume} {22}},\ \bibinfo {pages} {400} (\bibinfo {year}
  {1951})}\BibitemShut {NoStop}%
\bibitem [{\citenamefont {{Lukas
  Heinrich}}(2020)}]{lukasheinrichPyHEP2020Autodiff2020}%
  \BibitemOpen
  \bibfield  {author} {\bibinfo {author} {\bibnamefont {{Lukas Heinrich}}},\
  }\href@noop {} {\enquote {\bibinfo {title} {{{pyHEP}} 2020 autodiff
  tutorial},}\ } (\bibinfo {year} {2020})\BibitemShut {NoStop}%
\bibitem [{\citenamefont {Goodfellow}\ \emph {et~al.}(2016)\citenamefont
  {Goodfellow}, \citenamefont {Bengio},\ and\ \citenamefont
  {Courville}}]{goodfellowDeepLearning2016}%
  \BibitemOpen
  \bibfield  {author} {\bibinfo {author} {\bibfnamefont {I.}~\bibnamefont
  {Goodfellow}}, \bibinfo {author} {\bibfnamefont {Y.}~\bibnamefont {Bengio}},
  \ and\ \bibinfo {author} {\bibfnamefont {A.}~\bibnamefont {Courville}},\
  }\href@noop {} {\emph {\bibinfo {title} {Deep {{Learning}}}}}\ (\bibinfo
  {publisher} {{MIT Press}},\ \bibinfo {year} {2016})\BibitemShut {NoStop}%
\bibitem [{\citenamefont {Kingma}\ and\ \citenamefont
  {Ba}(2014)}]{kingmaAdamMethodStochastic2014}%
  \BibitemOpen
  \bibfield  {author} {\bibinfo {author} {\bibfnamefont {D.~P.}\ \bibnamefont
  {Kingma}}\ and\ \bibinfo {author} {\bibfnamefont {J.}~\bibnamefont {Ba}},\
  }\href@noop {} {\bibfield  {journal} {\bibinfo  {journal} {arXiv:1412.6980
  [cs]}\ } (\bibinfo {year} {2014})},\ \Eprint {http://arxiv.org/abs/1412.6980}
  {arxiv:1412.6980 [cs]} \BibitemShut {NoStop}%
\bibitem [{\citenamefont {Loshchilov}\ and\ \citenamefont
  {Hutter}(2019)}]{loshchilovDecoupledWeightDecay2019}%
  \BibitemOpen
  \bibfield  {author} {\bibinfo {author} {\bibfnamefont {I.}~\bibnamefont
  {Loshchilov}}\ and\ \bibinfo {author} {\bibfnamefont {F.}~\bibnamefont
  {Hutter}},\ }\href {\doibase 10.48550/arXiv.1711.05101} {\enquote {\bibinfo
  {title} {Decoupled {{Weight Decay Regularization}}},}\ } (\bibinfo {year}
  {2019}),\ \Eprint {http://arxiv.org/abs/1711.05101} {arxiv:1711.05101 [cs,
  math]} \BibitemShut {NoStop}%
\bibitem [{\citenamefont {Paszke}\ \emph {et~al.}(2019)\citenamefont {Paszke},
  \citenamefont {Gross}, \citenamefont {Massa}, \citenamefont {Lerer},
  \citenamefont {Bradbury}, \citenamefont {Chanan}, \citenamefont {Killeen},
  \citenamefont {Lin}, \citenamefont {Gimelshein}, \citenamefont {Antiga},
  \citenamefont {Desmaison}, \citenamefont {K{\"o}pf}, \citenamefont {Yang},
  \citenamefont {DeVito}, \citenamefont {Raison}, \citenamefont {Tejani},
  \citenamefont {Chilamkurthy}, \citenamefont {Steiner}, \citenamefont {Fang},
  \citenamefont {Bai},\ and\ \citenamefont
  {Chintala}}]{paszkePyTorchImperativeStyle2019}%
  \BibitemOpen
  \bibfield  {author} {\bibinfo {author} {\bibfnamefont {A.}~\bibnamefont
  {Paszke}}, \bibinfo {author} {\bibfnamefont {S.}~\bibnamefont {Gross}},
  \bibinfo {author} {\bibfnamefont {F.}~\bibnamefont {Massa}}, \bibinfo
  {author} {\bibfnamefont {A.}~\bibnamefont {Lerer}}, \bibinfo {author}
  {\bibfnamefont {J.}~\bibnamefont {Bradbury}}, \bibinfo {author}
  {\bibfnamefont {G.}~\bibnamefont {Chanan}}, \bibinfo {author} {\bibfnamefont
  {T.}~\bibnamefont {Killeen}}, \bibinfo {author} {\bibfnamefont
  {Z.}~\bibnamefont {Lin}}, \bibinfo {author} {\bibfnamefont {N.}~\bibnamefont
  {Gimelshein}}, \bibinfo {author} {\bibfnamefont {L.}~\bibnamefont {Antiga}},
  \bibinfo {author} {\bibfnamefont {A.}~\bibnamefont {Desmaison}}, \bibinfo
  {author} {\bibfnamefont {A.}~\bibnamefont {K{\"o}pf}}, \bibinfo {author}
  {\bibfnamefont {E.}~\bibnamefont {Yang}}, \bibinfo {author} {\bibfnamefont
  {Z.}~\bibnamefont {DeVito}}, \bibinfo {author} {\bibfnamefont
  {M.}~\bibnamefont {Raison}}, \bibinfo {author} {\bibfnamefont
  {A.}~\bibnamefont {Tejani}}, \bibinfo {author} {\bibfnamefont
  {S.}~\bibnamefont {Chilamkurthy}}, \bibinfo {author} {\bibfnamefont
  {B.}~\bibnamefont {Steiner}}, \bibinfo {author} {\bibfnamefont
  {L.}~\bibnamefont {Fang}}, \bibinfo {author} {\bibfnamefont {J.}~\bibnamefont
  {Bai}}, \ and\ \bibinfo {author} {\bibfnamefont {S.}~\bibnamefont
  {Chintala}},\ }\href {\doibase 10.48550/arXiv.1912.01703} {\enquote {\bibinfo
  {title} {{{PyTorch}}: {{An Imperative Style}}, {{High-Performance Deep
  Learning Library}}},}\ } (\bibinfo {year} {2019}),\ \Eprint
  {http://arxiv.org/abs/1912.01703} {arxiv:1912.01703 [cs, stat]} \BibitemShut
  {NoStop}%
\bibitem [{\citenamefont {Abadi}\ \emph {et~al.}(2015)\citenamefont {Abadi},
  \citenamefont {Agarwal}, \citenamefont {Barham}, \citenamefont {Brevdo},
  \citenamefont {Chen}, \citenamefont {Citro}, \citenamefont {Corrado},
  \citenamefont {Davis}, \citenamefont {Dean}, \citenamefont {Devin},
  \citenamefont {Ghemawat}, \citenamefont {Goodfellow}, \citenamefont {Harp},
  \citenamefont {Irving}, \citenamefont {Isard}, \citenamefont {Jia},
  \citenamefont {Jozefowicz}, \citenamefont {Kaiser}, \citenamefont {Kudlur},
  \citenamefont {Levenberg}, \citenamefont {Man{\'e}}, \citenamefont {Monga},
  \citenamefont {Moore}, \citenamefont {Murray}, \citenamefont {Olah},
  \citenamefont {Schuster}, \citenamefont {Shlens}, \citenamefont {Steiner},
  \citenamefont {Sutskever}, \citenamefont {Talwar}, \citenamefont {Tucker},
  \citenamefont {Vanhoucke}, \citenamefont {Vasudevan}, \citenamefont
  {Vi{\'e}gas}, \citenamefont {Vinyals}, \citenamefont {Warden}, \citenamefont
  {Wattenberg}, \citenamefont {Wicke}, \citenamefont {Yu},\ and\ \citenamefont
  {Zheng}}]{abadiTensorFlowLargeScaleMachine2015}%
  \BibitemOpen
  \bibfield  {author} {\bibinfo {author} {\bibfnamefont {M.}~\bibnamefont
  {Abadi}}, \bibinfo {author} {\bibfnamefont {A.}~\bibnamefont {Agarwal}},
  \bibinfo {author} {\bibfnamefont {P.}~\bibnamefont {Barham}}, \bibinfo
  {author} {\bibfnamefont {E.}~\bibnamefont {Brevdo}}, \bibinfo {author}
  {\bibfnamefont {Z.}~\bibnamefont {Chen}}, \bibinfo {author} {\bibfnamefont
  {C.}~\bibnamefont {Citro}}, \bibinfo {author} {\bibfnamefont {G.~S.}\
  \bibnamefont {Corrado}}, \bibinfo {author} {\bibfnamefont {A.}~\bibnamefont
  {Davis}}, \bibinfo {author} {\bibfnamefont {J.}~\bibnamefont {Dean}},
  \bibinfo {author} {\bibfnamefont {M.}~\bibnamefont {Devin}}, \bibinfo
  {author} {\bibfnamefont {S.}~\bibnamefont {Ghemawat}}, \bibinfo {author}
  {\bibfnamefont {I.}~\bibnamefont {Goodfellow}}, \bibinfo {author}
  {\bibfnamefont {A.}~\bibnamefont {Harp}}, \bibinfo {author} {\bibfnamefont
  {G.}~\bibnamefont {Irving}}, \bibinfo {author} {\bibfnamefont
  {M.}~\bibnamefont {Isard}}, \bibinfo {author} {\bibfnamefont
  {Y.}~\bibnamefont {Jia}}, \bibinfo {author} {\bibfnamefont {R.}~\bibnamefont
  {Jozefowicz}}, \bibinfo {author} {\bibfnamefont {L.}~\bibnamefont {Kaiser}},
  \bibinfo {author} {\bibfnamefont {M.}~\bibnamefont {Kudlur}}, \bibinfo
  {author} {\bibfnamefont {J.}~\bibnamefont {Levenberg}}, \bibinfo {author}
  {\bibfnamefont {D.}~\bibnamefont {Man{\'e}}}, \bibinfo {author}
  {\bibfnamefont {R.}~\bibnamefont {Monga}}, \bibinfo {author} {\bibfnamefont
  {S.}~\bibnamefont {Moore}}, \bibinfo {author} {\bibfnamefont
  {D.}~\bibnamefont {Murray}}, \bibinfo {author} {\bibfnamefont
  {C.}~\bibnamefont {Olah}}, \bibinfo {author} {\bibfnamefont {M.}~\bibnamefont
  {Schuster}}, \bibinfo {author} {\bibfnamefont {J.}~\bibnamefont {Shlens}},
  \bibinfo {author} {\bibfnamefont {B.}~\bibnamefont {Steiner}}, \bibinfo
  {author} {\bibfnamefont {I.}~\bibnamefont {Sutskever}}, \bibinfo {author}
  {\bibfnamefont {K.}~\bibnamefont {Talwar}}, \bibinfo {author} {\bibfnamefont
  {P.}~\bibnamefont {Tucker}}, \bibinfo {author} {\bibfnamefont
  {V.}~\bibnamefont {Vanhoucke}}, \bibinfo {author} {\bibfnamefont
  {V.}~\bibnamefont {Vasudevan}}, \bibinfo {author} {\bibfnamefont
  {F.}~\bibnamefont {Vi{\'e}gas}}, \bibinfo {author} {\bibfnamefont
  {O.}~\bibnamefont {Vinyals}}, \bibinfo {author} {\bibfnamefont
  {P.}~\bibnamefont {Warden}}, \bibinfo {author} {\bibfnamefont
  {M.}~\bibnamefont {Wattenberg}}, \bibinfo {author} {\bibfnamefont
  {M.}~\bibnamefont {Wicke}}, \bibinfo {author} {\bibfnamefont
  {Y.}~\bibnamefont {Yu}}, \ and\ \bibinfo {author} {\bibfnamefont
  {X.}~\bibnamefont {Zheng}},\ }\href@noop {} {\bibfield  {journal} {\bibinfo
  {journal} {https://www.tensorflow.org/}\ } (\bibinfo {year}
  {2015})}\BibitemShut {NoStop}%
\bibitem [{\citenamefont {Chollet}(2017)}]{cholletDeepLearningPython2017}%
  \BibitemOpen
  \bibfield  {author} {\bibinfo {author} {\bibfnamefont {F.}~\bibnamefont
  {Chollet}},\ }\href@noop {} {\emph {\bibinfo {title} {Deep {{Learning}} with
  {{Python}}}}}\ (\bibinfo  {publisher} {{Manning Publications Company}},\
  \bibinfo {year} {2017})\BibitemShut {NoStop}%
\bibitem [{\citenamefont {Heek}\ \emph {et~al.}(2023)\citenamefont {Heek},
  \citenamefont {Levskaya}, \citenamefont {Oliver}, \citenamefont {Ritter},
  \citenamefont {Rondepierre}, \citenamefont {Steiner},\ and\ \citenamefont
  {{van Zee}}}]{heekFlaxNeuralNetwork2023}%
  \BibitemOpen
  \bibfield  {author} {\bibinfo {author} {\bibfnamefont {J.}~\bibnamefont
  {Heek}}, \bibinfo {author} {\bibfnamefont {A.}~\bibnamefont {Levskaya}},
  \bibinfo {author} {\bibfnamefont {A.}~\bibnamefont {Oliver}}, \bibinfo
  {author} {\bibfnamefont {M.}~\bibnamefont {Ritter}}, \bibinfo {author}
  {\bibfnamefont {B.}~\bibnamefont {Rondepierre}}, \bibinfo {author}
  {\bibfnamefont {A.}~\bibnamefont {Steiner}}, \ and\ \bibinfo {author}
  {\bibfnamefont {M.}~\bibnamefont {{van Zee}}},\ }\href@noop {} {\enquote
  {\bibinfo {title} {Flax: {{A}} neural network library and ecosystem for
  {{JAX}}},}\ } (\bibinfo {year} {2023})\BibitemShut {NoStop}%
\bibitem [{\citenamefont {Chen}\ \emph {et~al.}(2015)\citenamefont {Chen},
  \citenamefont {Li}, \citenamefont {Li}, \citenamefont {Lin}, \citenamefont
  {Wang}, \citenamefont {Wang}, \citenamefont {Xiao}, \citenamefont {Xu},
  \citenamefont {Zhang},\ and\ \citenamefont
  {Zhang}}]{chenMXNetFlexibleEfficient2015}%
  \BibitemOpen
  \bibfield  {author} {\bibinfo {author} {\bibfnamefont {T.}~\bibnamefont
  {Chen}}, \bibinfo {author} {\bibfnamefont {M.}~\bibnamefont {Li}}, \bibinfo
  {author} {\bibfnamefont {Y.}~\bibnamefont {Li}}, \bibinfo {author}
  {\bibfnamefont {M.}~\bibnamefont {Lin}}, \bibinfo {author} {\bibfnamefont
  {N.}~\bibnamefont {Wang}}, \bibinfo {author} {\bibfnamefont {M.}~\bibnamefont
  {Wang}}, \bibinfo {author} {\bibfnamefont {T.}~\bibnamefont {Xiao}}, \bibinfo
  {author} {\bibfnamefont {B.}~\bibnamefont {Xu}}, \bibinfo {author}
  {\bibfnamefont {C.}~\bibnamefont {Zhang}}, \ and\ \bibinfo {author}
  {\bibfnamefont {Z.}~\bibnamefont {Zhang}},\ }\href {\doibase
  10.48550/arXiv.1512.01274} {\enquote {\bibinfo {title} {{{MXNet}}: {{A
  Flexible}} and {{Efficient Machine Learning Library}} for {{Heterogeneous
  Distributed Systems}}},}\ } (\bibinfo {year} {2015}),\ \Eprint
  {http://arxiv.org/abs/1512.01274} {arxiv:1512.01274 [cs]} \BibitemShut
  {NoStop}%
\bibitem [{\citenamefont {Kaplan}\ \emph {et~al.}(2020)\citenamefont {Kaplan},
  \citenamefont {McCandlish}, \citenamefont {Henighan}, \citenamefont {Brown},
  \citenamefont {Chess}, \citenamefont {Child}, \citenamefont {Gray},
  \citenamefont {Radford}, \citenamefont {Wu},\ and\ \citenamefont
  {Amodei}}]{kaplanScalingLawsNeural2020}%
  \BibitemOpen
  \bibfield  {author} {\bibinfo {author} {\bibfnamefont {J.}~\bibnamefont
  {Kaplan}}, \bibinfo {author} {\bibfnamefont {S.}~\bibnamefont {McCandlish}},
  \bibinfo {author} {\bibfnamefont {T.}~\bibnamefont {Henighan}}, \bibinfo
  {author} {\bibfnamefont {T.~B.}\ \bibnamefont {Brown}}, \bibinfo {author}
  {\bibfnamefont {B.}~\bibnamefont {Chess}}, \bibinfo {author} {\bibfnamefont
  {R.}~\bibnamefont {Child}}, \bibinfo {author} {\bibfnamefont
  {S.}~\bibnamefont {Gray}}, \bibinfo {author} {\bibfnamefont {A.}~\bibnamefont
  {Radford}}, \bibinfo {author} {\bibfnamefont {J.}~\bibnamefont {Wu}}, \ and\
  \bibinfo {author} {\bibfnamefont {D.}~\bibnamefont {Amodei}},\ }\href
  {\doibase 10.48550/arXiv.2001.08361} {\enquote {\bibinfo {title} {Scaling
  {{Laws}} for {{Neural Language Models}}},}\ } (\bibinfo {year} {2020}),\
  \Eprint {http://arxiv.org/abs/2001.08361} {arxiv:2001.08361 [cs, stat]}
  \BibitemShut {NoStop}%
\bibitem [{\citenamefont {Yu}\ \emph {et~al.}(2022)\citenamefont {Yu},
  \citenamefont {Xu}, \citenamefont {Koh}, \citenamefont {Luong}, \citenamefont
  {Baid}, \citenamefont {Wang}, \citenamefont {Vasudevan}, \citenamefont {Ku},
  \citenamefont {Yang}, \citenamefont {Ayan}, \citenamefont {Hutchinson},
  \citenamefont {Han}, \citenamefont {Parekh}, \citenamefont {Li},
  \citenamefont {Zhang}, \citenamefont {Baldridge},\ and\ \citenamefont
  {Wu}}]{yuScalingAutoregressiveModels2022}%
  \BibitemOpen
  \bibfield  {author} {\bibinfo {author} {\bibfnamefont {J.}~\bibnamefont
  {Yu}}, \bibinfo {author} {\bibfnamefont {Y.}~\bibnamefont {Xu}}, \bibinfo
  {author} {\bibfnamefont {J.~Y.}\ \bibnamefont {Koh}}, \bibinfo {author}
  {\bibfnamefont {T.}~\bibnamefont {Luong}}, \bibinfo {author} {\bibfnamefont
  {G.}~\bibnamefont {Baid}}, \bibinfo {author} {\bibfnamefont {Z.}~\bibnamefont
  {Wang}}, \bibinfo {author} {\bibfnamefont {V.}~\bibnamefont {Vasudevan}},
  \bibinfo {author} {\bibfnamefont {A.}~\bibnamefont {Ku}}, \bibinfo {author}
  {\bibfnamefont {Y.}~\bibnamefont {Yang}}, \bibinfo {author} {\bibfnamefont
  {B.~K.}\ \bibnamefont {Ayan}}, \bibinfo {author} {\bibfnamefont
  {B.}~\bibnamefont {Hutchinson}}, \bibinfo {author} {\bibfnamefont
  {W.}~\bibnamefont {Han}}, \bibinfo {author} {\bibfnamefont {Z.}~\bibnamefont
  {Parekh}}, \bibinfo {author} {\bibfnamefont {X.}~\bibnamefont {Li}}, \bibinfo
  {author} {\bibfnamefont {H.}~\bibnamefont {Zhang}}, \bibinfo {author}
  {\bibfnamefont {J.}~\bibnamefont {Baldridge}}, \ and\ \bibinfo {author}
  {\bibfnamefont {Y.}~\bibnamefont {Wu}},\ }\href {\doibase
  10.48550/arXiv.2206.10789} {\enquote {\bibinfo {title} {Scaling
  {{Autoregressive Models}} for {{Content-Rich Text-to-Image Generation}}},}\ }
  (\bibinfo {year} {2022}),\ \Eprint {http://arxiv.org/abs/2206.10789}
  {arxiv:2206.10789 [cs]} \BibitemShut {NoStop}%
\bibitem [{\citenamefont {Caruana}(1997)}]{caruanaMultitaskLearning1997}%
  \BibitemOpen
  \bibfield  {author} {\bibinfo {author} {\bibfnamefont {R.}~\bibnamefont
  {Caruana}},\ }\href {\doibase 10.1023/A:1007379606734} {\bibfield  {journal}
  {\bibinfo  {journal} {Machine Learning}\ }\textbf {\bibinfo {volume} {28}},\
  \bibinfo {pages} {41} (\bibinfo {year} {1997})}\BibitemShut {NoStop}%
\bibitem [{\citenamefont {Wu}(2020)}]{wuAcceleratingSelfPlayLearning2020}%
  \BibitemOpen
  \bibfield  {author} {\bibinfo {author} {\bibfnamefont {D.~J.}\ \bibnamefont
  {Wu}},\ }\href {\doibase 10.48550/arXiv.1902.10565} {\enquote {\bibinfo
  {title} {Accelerating {{Self-Play Learning}} in {{Go}}},}\ } (\bibinfo {year}
  {2020}),\ \Eprint {http://arxiv.org/abs/1902.10565} {arxiv:1902.10565 [cs,
  stat]} \BibitemShut {NoStop}%
\bibitem [{\citenamefont {Karras}\ \emph {et~al.}(2020)\citenamefont {Karras},
  \citenamefont {Laine}, \citenamefont {Aittala}, \citenamefont {Hellsten},
  \citenamefont {Lehtinen},\ and\ \citenamefont
  {Aila}}]{karrasAnalyzingImprovingImage2020}%
  \BibitemOpen
  \bibfield  {author} {\bibinfo {author} {\bibfnamefont {T.}~\bibnamefont
  {Karras}}, \bibinfo {author} {\bibfnamefont {S.}~\bibnamefont {Laine}},
  \bibinfo {author} {\bibfnamefont {M.}~\bibnamefont {Aittala}}, \bibinfo
  {author} {\bibfnamefont {J.}~\bibnamefont {Hellsten}}, \bibinfo {author}
  {\bibfnamefont {J.}~\bibnamefont {Lehtinen}}, \ and\ \bibinfo {author}
  {\bibfnamefont {T.}~\bibnamefont {Aila}},\ }\href {\doibase
  10.48550/arXiv.1912.04958} {\enquote {\bibinfo {title} {Analyzing and
  {{Improving}} the {{Image Quality}} of {{StyleGAN}}},}\ } (\bibinfo {year}
  {2020}),\ \Eprint {http://arxiv.org/abs/1912.04958} {arxiv:1912.04958 [cs,
  eess, stat]} \BibitemShut {NoStop}%
\bibitem [{\citenamefont {Rombach}\ \emph {et~al.}(2022)\citenamefont
  {Rombach}, \citenamefont {Blattmann}, \citenamefont {Lorenz}, \citenamefont
  {Esser},\ and\ \citenamefont
  {Ommer}}]{rombachHighResolutionImageSynthesis2022}%
  \BibitemOpen
  \bibfield  {author} {\bibinfo {author} {\bibfnamefont {R.}~\bibnamefont
  {Rombach}}, \bibinfo {author} {\bibfnamefont {A.}~\bibnamefont {Blattmann}},
  \bibinfo {author} {\bibfnamefont {D.}~\bibnamefont {Lorenz}}, \bibinfo
  {author} {\bibfnamefont {P.}~\bibnamefont {Esser}}, \ and\ \bibinfo {author}
  {\bibfnamefont {B.}~\bibnamefont {Ommer}},\ }\href {\doibase
  10.48550/arXiv.2112.10752} {\enquote {\bibinfo {title} {High-{{Resolution
  Image Synthesis}} with {{Latent Diffusion Models}}},}\ } (\bibinfo {year}
  {2022}),\ \Eprint {http://arxiv.org/abs/2112.10752} {arxiv:2112.10752 [cs]}
  \BibitemShut {NoStop}%
\bibitem [{\citenamefont {Kingma}\ and\ \citenamefont
  {Welling}(2019)}]{kingmaIntroductionVariationalAutoencoders2019}%
  \BibitemOpen
  \bibfield  {author} {\bibinfo {author} {\bibfnamefont {D.~P.}\ \bibnamefont
  {Kingma}}\ and\ \bibinfo {author} {\bibfnamefont {M.}~\bibnamefont
  {Welling}},\ }\href {\doibase 10.1561/2200000056} {\bibfield  {journal}
  {\bibinfo  {journal} {Foundations and Trends in Machine Learning}\ }\textbf
  {\bibinfo {volume} {12}},\ \bibinfo {pages} {307} (\bibinfo {year} {2019})},\
  \Eprint {http://arxiv.org/abs/1906.02691} {arxiv:1906.02691} \BibitemShut
  {NoStop}%
\bibitem [{\citenamefont {{Khaireh-Walieh}}\ \emph {et~al.}(2023)\citenamefont
  {{Khaireh-Walieh}}, \citenamefont {Arnoult}, \citenamefont {Plissard},\ and\
  \citenamefont {Wiecha}}]{khaireh-waliehMonitoringMBESubstrate2023}%
  \BibitemOpen
  \bibfield  {author} {\bibinfo {author} {\bibfnamefont {A.}~\bibnamefont
  {{Khaireh-Walieh}}}, \bibinfo {author} {\bibfnamefont {A.}~\bibnamefont
  {Arnoult}}, \bibinfo {author} {\bibfnamefont {S.}~\bibnamefont {Plissard}}, \
  and\ \bibinfo {author} {\bibfnamefont {P.~R.}\ \bibnamefont {Wiecha}},\
  }\href {\doibase 10.1021/acs.cgd.2c01132} {\bibfield  {journal} {\bibinfo
  {journal} {Crystal Growth \& Design}\ }\textbf {\bibinfo {volume} {23}},\
  \bibinfo {pages} {892} (\bibinfo {year} {2023})}\BibitemShut {NoStop}%
\bibitem [{\citenamefont {Melati}\ \emph {et~al.}(2019)\citenamefont {Melati},
  \citenamefont {Grinberg}, \citenamefont {Kamandar~Dezfouli}, \citenamefont
  {Janz}, \citenamefont {Cheben}, \citenamefont {Schmid}, \citenamefont
  {{S{\'a}nchez-Postigo}},\ and\ \citenamefont
  {Xu}}]{melatiMappingGlobalDesign2019}%
  \BibitemOpen
  \bibfield  {author} {\bibinfo {author} {\bibfnamefont {D.}~\bibnamefont
  {Melati}}, \bibinfo {author} {\bibfnamefont {Y.}~\bibnamefont {Grinberg}},
  \bibinfo {author} {\bibfnamefont {M.}~\bibnamefont {Kamandar~Dezfouli}},
  \bibinfo {author} {\bibfnamefont {S.}~\bibnamefont {Janz}}, \bibinfo {author}
  {\bibfnamefont {P.}~\bibnamefont {Cheben}}, \bibinfo {author} {\bibfnamefont
  {J.~H.}\ \bibnamefont {Schmid}}, \bibinfo {author} {\bibfnamefont
  {A.}~\bibnamefont {{S{\'a}nchez-Postigo}}}, \ and\ \bibinfo {author}
  {\bibfnamefont {D.-X.}\ \bibnamefont {Xu}},\ }\href {\doibase
  10.1038/s41467-019-12698-1} {\bibfield  {journal} {\bibinfo  {journal}
  {Nature Communications}\ }\textbf {\bibinfo {volume} {10}},\ \bibinfo {pages}
  {4775} (\bibinfo {year} {2019})}\BibitemShut {NoStop}%
\bibitem [{\citenamefont {Kiarashinejad}\ \emph {et~al.}(2020)\citenamefont
  {Kiarashinejad}, \citenamefont {Zandehshahvar}, \citenamefont
  {Abdollahramezani}, \citenamefont {Hemmatyar}, \citenamefont
  {Pourabolghasem},\ and\ \citenamefont
  {Adibi}}]{kiarashinejadKnowledgeDiscoveryNanophotonics2020}%
  \BibitemOpen
  \bibfield  {author} {\bibinfo {author} {\bibfnamefont {Y.}~\bibnamefont
  {Kiarashinejad}}, \bibinfo {author} {\bibfnamefont {M.}~\bibnamefont
  {Zandehshahvar}}, \bibinfo {author} {\bibfnamefont {S.}~\bibnamefont
  {Abdollahramezani}}, \bibinfo {author} {\bibfnamefont {O.}~\bibnamefont
  {Hemmatyar}}, \bibinfo {author} {\bibfnamefont {R.}~\bibnamefont
  {Pourabolghasem}}, \ and\ \bibinfo {author} {\bibfnamefont {A.}~\bibnamefont
  {Adibi}},\ }\href {\doibase 10.1002/aisy.201900132} {\bibfield  {journal}
  {\bibinfo  {journal} {Advanced Intelligent Systems}\ }\textbf {\bibinfo
  {volume} {2}},\ \bibinfo {pages} {1900132} (\bibinfo {year} {2020})},\
  \Eprint {http://arxiv.org/abs/1909.07330} {arxiv:1909.07330} \BibitemShut
  {NoStop}%
\bibitem [{\citenamefont {Zandehshahvar}\ \emph {et~al.}(2022)\citenamefont
  {Zandehshahvar}, \citenamefont {Kiarashinejad}, \citenamefont {Zhu},
  \citenamefont {Maleki}, \citenamefont {Brown},\ and\ \citenamefont
  {Adibi}}]{zandehshahvarManifoldLearningKnowledge2022}%
  \BibitemOpen
  \bibfield  {author} {\bibinfo {author} {\bibfnamefont {M.}~\bibnamefont
  {Zandehshahvar}}, \bibinfo {author} {\bibfnamefont {Y.}~\bibnamefont
  {Kiarashinejad}}, \bibinfo {author} {\bibfnamefont {M.}~\bibnamefont {Zhu}},
  \bibinfo {author} {\bibfnamefont {H.}~\bibnamefont {Maleki}}, \bibinfo
  {author} {\bibfnamefont {T.}~\bibnamefont {Brown}}, \ and\ \bibinfo {author}
  {\bibfnamefont {A.}~\bibnamefont {Adibi}},\ }\href {\doibase
  10.1021/acsphotonics.1c01888} {\bibfield  {journal} {\bibinfo  {journal} {ACS
  Photonics}\ }\textbf {\bibinfo {volume} {9}},\ \bibinfo {pages} {714}
  (\bibinfo {year} {2022})},\ \Eprint {http://arxiv.org/abs/2102.04454}
  {arxiv:2102.04454} \BibitemShut {NoStop}%
\bibitem [{\citenamefont {Bachmann}\ \emph {et~al.}(2022)\citenamefont
  {Bachmann}, \citenamefont {Mizrahi}, \citenamefont {Atanov},\ and\
  \citenamefont {Zamir}}]{bachmannMultiMAEMultimodalMultitask2022}%
  \BibitemOpen
  \bibfield  {author} {\bibinfo {author} {\bibfnamefont {R.}~\bibnamefont
  {Bachmann}}, \bibinfo {author} {\bibfnamefont {D.}~\bibnamefont {Mizrahi}},
  \bibinfo {author} {\bibfnamefont {A.}~\bibnamefont {Atanov}}, \ and\ \bibinfo
  {author} {\bibfnamefont {A.}~\bibnamefont {Zamir}},\ }in\ \href {\doibase
  10.1007/978-3-031-19836-6_20} {\emph {\bibinfo {booktitle} {Computer
  {{Vision}} \textendash{} {{ECCV}} 2022}}},\ \bibinfo {series and number}
  {Lecture {{Notes}} in {{Computer Science}}},\ \bibinfo {editor} {edited by\
  \bibinfo {editor} {\bibfnamefont {S.}~\bibnamefont {Avidan}}, \bibinfo
  {editor} {\bibfnamefont {G.}~\bibnamefont {Brostow}}, \bibinfo {editor}
  {\bibfnamefont {M.}~\bibnamefont {Ciss{\'e}}}, \bibinfo {editor}
  {\bibfnamefont {G.~M.}\ \bibnamefont {Farinella}}, \ and\ \bibinfo {editor}
  {\bibfnamefont {T.}~\bibnamefont {Hassner}}}\ (\bibinfo  {publisher}
  {{Springer Nature Switzerland}},\ \bibinfo {address} {{Cham}},\ \bibinfo
  {year} {2022})\ pp.\ \bibinfo {pages} {348--367}\BibitemShut {NoStop}%
\bibitem [{\citenamefont {Liu}\ \emph {et~al.}(2023)\citenamefont {Liu},
  \citenamefont {Sun}, \citenamefont {Xue}, \citenamefont {Zhang},
  \citenamefont {Yen},\ and\ \citenamefont
  {Tan}}]{liuSurveyEvolutionaryNeural2023}%
  \BibitemOpen
  \bibfield  {author} {\bibinfo {author} {\bibfnamefont {Y.}~\bibnamefont
  {Liu}}, \bibinfo {author} {\bibfnamefont {Y.}~\bibnamefont {Sun}}, \bibinfo
  {author} {\bibfnamefont {B.}~\bibnamefont {Xue}}, \bibinfo {author}
  {\bibfnamefont {M.}~\bibnamefont {Zhang}}, \bibinfo {author} {\bibfnamefont
  {G.~G.}\ \bibnamefont {Yen}}, \ and\ \bibinfo {author} {\bibfnamefont
  {K.~C.}\ \bibnamefont {Tan}},\ }\href {\doibase 10.1109/TNNLS.2021.3100554}
  {\bibfield  {journal} {\bibinfo  {journal} {IEEE Transactions on Neural
  Networks and Learning Systems}\ }\textbf {\bibinfo {volume} {34}},\ \bibinfo
  {pages} {550} (\bibinfo {year} {2023})}\BibitemShut {NoStop}%
\bibitem [{\citenamefont {Li}\ and\ \citenamefont
  {Talwalkar}(2020)}]{liRandomSearchReproducibility2020}%
  \BibitemOpen
  \bibfield  {author} {\bibinfo {author} {\bibfnamefont {L.}~\bibnamefont
  {Li}}\ and\ \bibinfo {author} {\bibfnamefont {A.}~\bibnamefont {Talwalkar}},\
  }in\ \href@noop {} {\emph {\bibinfo {booktitle} {Proceedings of {{The}} 35th
  {{Uncertainty}} in {{Artificial Intelligence Conference}}}}}\ (\bibinfo
  {publisher} {{PMLR}},\ \bibinfo {year} {2020})\ pp.\ \bibinfo {pages}
  {367--377}\BibitemShut {NoStop}%
\bibitem [{\citenamefont {Pham}\ \emph {et~al.}(2018)\citenamefont {Pham},
  \citenamefont {Guan}, \citenamefont {Zoph}, \citenamefont {Le},\ and\
  \citenamefont {Dean}}]{phamEfficientNeuralArchitecture2018}%
  \BibitemOpen
  \bibfield  {author} {\bibinfo {author} {\bibfnamefont {H.}~\bibnamefont
  {Pham}}, \bibinfo {author} {\bibfnamefont {M.}~\bibnamefont {Guan}}, \bibinfo
  {author} {\bibfnamefont {B.}~\bibnamefont {Zoph}}, \bibinfo {author}
  {\bibfnamefont {Q.}~\bibnamefont {Le}}, \ and\ \bibinfo {author}
  {\bibfnamefont {J.}~\bibnamefont {Dean}},\ }in\ \href@noop {} {\emph
  {\bibinfo {booktitle} {Proceedings of the 35th {{International Conference}}
  on {{Machine Learning}}}}}\ (\bibinfo  {publisher} {{PMLR}},\ \bibinfo {year}
  {2018})\ pp.\ \bibinfo {pages} {4095--4104}\BibitemShut {NoStop}%
\bibitem [{\citenamefont {Real}\ \emph {et~al.}(2019)\citenamefont {Real},
  \citenamefont {Aggarwal}, \citenamefont {Huang},\ and\ \citenamefont
  {Le}}]{realRegularizedEvolutionImage2019}%
  \BibitemOpen
  \bibfield  {author} {\bibinfo {author} {\bibfnamefont {E.}~\bibnamefont
  {Real}}, \bibinfo {author} {\bibfnamefont {A.}~\bibnamefont {Aggarwal}},
  \bibinfo {author} {\bibfnamefont {Y.}~\bibnamefont {Huang}}, \ and\ \bibinfo
  {author} {\bibfnamefont {Q.~V.}\ \bibnamefont {Le}},\ }in\ \href {\doibase
  10.1609/aaai.v33i01.33014780} {\emph {\bibinfo {booktitle} {Proceedings of
  the {{Thirty-Third AAAI Conference}} on {{Artificial Intelligence}} and
  {{Thirty-First Innovative Applications}} of {{Artificial Intelligence
  Conference}} and {{Ninth AAAI Symposium}} on {{Educational Advances}} in
  {{Artificial Intelligence}}}}},\ \bibinfo {series and number}
  {{{AAAI}}'19/{{IAAI}}'19/{{EAAI}}'19}\ (\bibinfo  {publisher} {{AAAI
  Press}},\ \bibinfo {address} {{Honolulu, Hawaii, USA}},\ \bibinfo {year}
  {2019})\ pp.\ \bibinfo {pages} {4780--4789}\BibitemShut {NoStop}%
\bibitem [{\citenamefont {Hammerschmidt}\ \emph {et~al.}(2018)\citenamefont
  {Hammerschmidt}, \citenamefont {Schneider}, \citenamefont {Santiago},
  \citenamefont {Zschiedrich}, \citenamefont {Weiser},\ and\ \citenamefont
  {Burger}}]{hammerschmidtSolvingInverseProblems2018}%
  \BibitemOpen
  \bibfield  {author} {\bibinfo {author} {\bibfnamefont {M.}~\bibnamefont
  {Hammerschmidt}}, \bibinfo {author} {\bibfnamefont {P.-I.}\ \bibnamefont
  {Schneider}}, \bibinfo {author} {\bibfnamefont {X.~G.}\ \bibnamefont
  {Santiago}}, \bibinfo {author} {\bibfnamefont {L.}~\bibnamefont
  {Zschiedrich}}, \bibinfo {author} {\bibfnamefont {M.}~\bibnamefont {Weiser}},
  \ and\ \bibinfo {author} {\bibfnamefont {S.}~\bibnamefont {Burger}},\ }in\
  \href {\doibase 10.1117/12.2315468} {\emph {\bibinfo {booktitle}
  {Computational {{Optics II}}}}},\ Vol.\ \bibinfo {volume} {10694}\ (\bibinfo
  {publisher} {{SPIE}},\ \bibinfo {year} {2018})\ pp.\ \bibinfo {pages}
  {38--45}\BibitemShut {NoStop}%
\bibitem [{\citenamefont {{Garcia-Santiago}}\ \emph {et~al.}(2021)\citenamefont
  {{Garcia-Santiago}}, \citenamefont {Burger}, \citenamefont {Rockstuhl},\ and\
  \citenamefont {Schneider}}]{garcia-santiagoBayesianOptimizationImproved2021}%
  \BibitemOpen
  \bibfield  {author} {\bibinfo {author} {\bibfnamefont {X.}~\bibnamefont
  {{Garcia-Santiago}}}, \bibinfo {author} {\bibfnamefont {S.}~\bibnamefont
  {Burger}}, \bibinfo {author} {\bibfnamefont {C.}~\bibnamefont {Rockstuhl}}, \
  and\ \bibinfo {author} {\bibfnamefont {P.-I.}\ \bibnamefont {Schneider}},\
  }\href@noop {} {\bibfield  {journal} {\bibinfo  {journal} {Journal of
  Lightwave Technology}\ }\textbf {\bibinfo {volume} {39}},\ \bibinfo {pages}
  {167} (\bibinfo {year} {2021})}\BibitemShut {NoStop}%
\bibitem [{\citenamefont {Wu}\ \emph {et~al.}(2021)\citenamefont {Wu},
  \citenamefont {Arrivault}, \citenamefont {Durufl{\'e}}, \citenamefont {Gras},
  \citenamefont {Binkowski}, \citenamefont {Burger}, \citenamefont {Yan},\ and\
  \citenamefont {Lalanne}}]{wuEfficientHybridMethod2021}%
  \BibitemOpen
  \bibfield  {author} {\bibinfo {author} {\bibfnamefont {T.}~\bibnamefont
  {Wu}}, \bibinfo {author} {\bibfnamefont {D.}~\bibnamefont {Arrivault}},
  \bibinfo {author} {\bibfnamefont {M.}~\bibnamefont {Durufl{\'e}}}, \bibinfo
  {author} {\bibfnamefont {A.}~\bibnamefont {Gras}}, \bibinfo {author}
  {\bibfnamefont {F.}~\bibnamefont {Binkowski}}, \bibinfo {author}
  {\bibfnamefont {S.}~\bibnamefont {Burger}}, \bibinfo {author} {\bibfnamefont
  {W.}~\bibnamefont {Yan}}, \ and\ \bibinfo {author} {\bibfnamefont
  {P.}~\bibnamefont {Lalanne}},\ }\href {\doibase 10.1364/JOSAA.428224}
  {\bibfield  {journal} {\bibinfo  {journal} {JOSA A}\ }\textbf {\bibinfo
  {volume} {38}},\ \bibinfo {pages} {1224} (\bibinfo {year}
  {2021})}\BibitemShut {NoStop}%
\bibitem [{\citenamefont {Elsawy}\ \emph {et~al.}(2021)\citenamefont {Elsawy},
  \citenamefont {Gourdin}, \citenamefont {Binois}, \citenamefont {Duvigneau},
  \citenamefont {Felbacq}, \citenamefont {Khadir}, \citenamefont {Genevet},\
  and\ \citenamefont {Lanteri}}]{elsawyMultiobjectiveStatisticalLearning2021}%
  \BibitemOpen
  \bibfield  {author} {\bibinfo {author} {\bibfnamefont {M.~M.~R.}\
  \bibnamefont {Elsawy}}, \bibinfo {author} {\bibfnamefont {A.}~\bibnamefont
  {Gourdin}}, \bibinfo {author} {\bibfnamefont {M.}~\bibnamefont {Binois}},
  \bibinfo {author} {\bibfnamefont {R.}~\bibnamefont {Duvigneau}}, \bibinfo
  {author} {\bibfnamefont {D.}~\bibnamefont {Felbacq}}, \bibinfo {author}
  {\bibfnamefont {S.}~\bibnamefont {Khadir}}, \bibinfo {author} {\bibfnamefont
  {P.}~\bibnamefont {Genevet}}, \ and\ \bibinfo {author} {\bibfnamefont
  {S.}~\bibnamefont {Lanteri}},\ }\href {\doibase 10.1021/acsphotonics.1c00753}
  {\bibfield  {journal} {\bibinfo  {journal} {ACS Photonics}\ }\textbf
  {\bibinfo {volume} {8}},\ \bibinfo {pages} {2498} (\bibinfo {year}
  {2021})}\BibitemShut {NoStop}%
\bibitem [{\citenamefont {Goodfellow}\ \emph {et~al.}(2015)\citenamefont
  {Goodfellow}, \citenamefont {Shlens},\ and\ \citenamefont
  {Szegedy}}]{goodfellowExplainingHarnessingAdversarial2015}%
  \BibitemOpen
  \bibfield  {author} {\bibinfo {author} {\bibfnamefont {I.~J.}\ \bibnamefont
  {Goodfellow}}, \bibinfo {author} {\bibfnamefont {J.}~\bibnamefont {Shlens}},
  \ and\ \bibinfo {author} {\bibfnamefont {C.}~\bibnamefont {Szegedy}},\ }\href
  {\doibase 10.48550/arXiv.1412.6572} {\enquote {\bibinfo {title} {Explaining
  and {{Harnessing Adversarial Examples}}},}\ } (\bibinfo {year} {2015}),\
  \Eprint {http://arxiv.org/abs/1412.6572} {arxiv:1412.6572 [cs, stat]}
  \BibitemShut {NoStop}%
\bibitem [{\citenamefont {Simonyan}\ and\ \citenamefont
  {Zisserman}(2015)}]{simonyanVeryDeepConvolutional2015}%
  \BibitemOpen
  \bibfield  {author} {\bibinfo {author} {\bibfnamefont {K.}~\bibnamefont
  {Simonyan}}\ and\ \bibinfo {author} {\bibfnamefont {A.}~\bibnamefont
  {Zisserman}},\ }\href {\doibase 10.48550/arXiv.1409.1556} {\enquote {\bibinfo
  {title} {Very {{Deep Convolutional Networks}} for {{Large-Scale Image
  Recognition}}},}\ } (\bibinfo {year} {2015}),\ \Eprint
  {http://arxiv.org/abs/1409.1556} {arxiv:1409.1556 [cs]} \BibitemShut
  {NoStop}%
\bibitem [{\citenamefont {Vaswani}\ \emph {et~al.}(2017)\citenamefont
  {Vaswani}, \citenamefont {Shazeer}, \citenamefont {Parmar}, \citenamefont
  {Uszkoreit}, \citenamefont {Jones}, \citenamefont {Gomez}, \citenamefont
  {Kaiser},\ and\ \citenamefont {Polosukhin}}]{vaswaniAttentionAllYou2017}%
  \BibitemOpen
  \bibfield  {author} {\bibinfo {author} {\bibfnamefont {A.}~\bibnamefont
  {Vaswani}}, \bibinfo {author} {\bibfnamefont {N.}~\bibnamefont {Shazeer}},
  \bibinfo {author} {\bibfnamefont {N.}~\bibnamefont {Parmar}}, \bibinfo
  {author} {\bibfnamefont {J.}~\bibnamefont {Uszkoreit}}, \bibinfo {author}
  {\bibfnamefont {L.}~\bibnamefont {Jones}}, \bibinfo {author} {\bibfnamefont
  {A.~N.}\ \bibnamefont {Gomez}}, \bibinfo {author} {\bibfnamefont
  {L.}~\bibnamefont {Kaiser}}, \ and\ \bibinfo {author} {\bibfnamefont
  {I.}~\bibnamefont {Polosukhin}},\ }\href {\doibase 10.48550/arXiv.1706.03762}
  {\enquote {\bibinfo {title} {Attention {{Is All You Need}}},}\ } (\bibinfo
  {year} {2017}),\ \Eprint {http://arxiv.org/abs/1706.03762} {arxiv:1706.03762
  [cs]} \BibitemShut {NoStop}%
\bibitem [{\citenamefont
  {Fukushima}(1980)}]{fukushimaNeocognitronSelforganizingNeural1980}%
  \BibitemOpen
  \bibfield  {author} {\bibinfo {author} {\bibfnamefont {K.}~\bibnamefont
  {Fukushima}},\ }\href {\doibase 10.1007/BF00344251} {\bibfield  {journal}
  {\bibinfo  {journal} {Biological Cybernetics}\ }\textbf {\bibinfo {volume}
  {36}},\ \bibinfo {pages} {193} (\bibinfo {year} {1980})}\BibitemShut
  {NoStop}%
\bibitem [{\citenamefont {Atlas}\ \emph {et~al.}(1987)\citenamefont {Atlas},
  \citenamefont {Homma},\ and\ \citenamefont
  {Marks}}]{atlasArtificialNeuralNetwork1987}%
  \BibitemOpen
  \bibfield  {author} {\bibinfo {author} {\bibfnamefont {L.}~\bibnamefont
  {Atlas}}, \bibinfo {author} {\bibfnamefont {T.}~\bibnamefont {Homma}}, \ and\
  \bibinfo {author} {\bibfnamefont {R.}~\bibnamefont {Marks}},\ }in\ \href@noop
  {} {\emph {\bibinfo {booktitle} {Neural {{Information Processing
  Systems}}}}},\ Vol.~\bibinfo {volume} {0}\ (\bibinfo  {publisher} {{American
  Institute of Physics}},\ \bibinfo {year} {1987})\BibitemShut {NoStop}%
\bibitem [{\citenamefont {LeCun}\ \emph {et~al.}(1989)\citenamefont {LeCun},
  \citenamefont {Boser}, \citenamefont {Denker}, \citenamefont {Henderson},
  \citenamefont {Howard}, \citenamefont {Hubbard},\ and\ \citenamefont
  {Jackel}}]{lecunBackpropagationAppliedHandwritten1989}%
  \BibitemOpen
  \bibfield  {author} {\bibinfo {author} {\bibfnamefont {Y.}~\bibnamefont
  {LeCun}}, \bibinfo {author} {\bibfnamefont {B.}~\bibnamefont {Boser}},
  \bibinfo {author} {\bibfnamefont {J.~S.}\ \bibnamefont {Denker}}, \bibinfo
  {author} {\bibfnamefont {D.}~\bibnamefont {Henderson}}, \bibinfo {author}
  {\bibfnamefont {R.~E.}\ \bibnamefont {Howard}}, \bibinfo {author}
  {\bibfnamefont {W.}~\bibnamefont {Hubbard}}, \ and\ \bibinfo {author}
  {\bibfnamefont {L.~D.}\ \bibnamefont {Jackel}},\ }\href {\doibase
  10.1162/neco.1989.1.4.541} {\bibfield  {journal} {\bibinfo  {journal} {Neural
  Computation}\ }\textbf {\bibinfo {volume} {1}},\ \bibinfo {pages} {541}
  (\bibinfo {year} {1989})}\BibitemShut {NoStop}%
\bibitem [{\citenamefont {Zhang}\ \emph {et~al.}(1990)\citenamefont {Zhang},
  \citenamefont {Itoh}, \citenamefont {Tanida},\ and\ \citenamefont
  {Ichioka}}]{zhangParallelDistributedProcessing1990}%
  \BibitemOpen
  \bibfield  {author} {\bibinfo {author} {\bibfnamefont {W.}~\bibnamefont
  {Zhang}}, \bibinfo {author} {\bibfnamefont {K.}~\bibnamefont {Itoh}},
  \bibinfo {author} {\bibfnamefont {J.}~\bibnamefont {Tanida}}, \ and\ \bibinfo
  {author} {\bibfnamefont {Y.}~\bibnamefont {Ichioka}},\ }\href {\doibase
  10.1364/AO.29.004790} {\bibfield  {journal} {\bibinfo  {journal} {Applied
  Optics}\ }\textbf {\bibinfo {volume} {29}},\ \bibinfo {pages} {4790}
  (\bibinfo {year} {1990})}\BibitemShut {NoStop}%
\bibitem [{\citenamefont {Hubel}\ and\ \citenamefont
  {Wiesel}(1959)}]{hubelReceptiveFieldsSingle1959}%
  \BibitemOpen
  \bibfield  {author} {\bibinfo {author} {\bibfnamefont {D.~H.}\ \bibnamefont
  {Hubel}}\ and\ \bibinfo {author} {\bibfnamefont {T.~N.}\ \bibnamefont
  {Wiesel}},\ }\href {\doibase 10.1113/jphysiol.1959.sp006308} {\bibfield
  {journal} {\bibinfo  {journal} {The Journal of Physiology}\ }\textbf
  {\bibinfo {volume} {148}},\ \bibinfo {pages} {574} (\bibinfo {year}
  {1959})}\BibitemShut {NoStop}%
\bibitem [{\citenamefont {Lowe}(1999)}]{loweObjectRecognitionLocal1999}%
  \BibitemOpen
  \bibfield  {author} {\bibinfo {author} {\bibfnamefont {D.}~\bibnamefont
  {Lowe}},\ }in\ \href {\doibase 10.1109/ICCV.1999.790410} {\emph {\bibinfo
  {booktitle} {Proceedings of the {{Seventh IEEE International Conference}} on
  {{Computer Vision}}}}},\ Vol.~\bibinfo {volume} {2}\ (\bibinfo {year}
  {1999})\ pp.\ \bibinfo {pages} {1150--1157 vol.2}\BibitemShut {NoStop}%
\bibitem [{\citenamefont {Li}\ \emph {et~al.}(2022)\citenamefont {Li},
  \citenamefont {Liu}, \citenamefont {Yang}, \citenamefont {Peng},\ and\
  \citenamefont {Zhou}}]{liSurveyConvolutionalNeural2022}%
  \BibitemOpen
  \bibfield  {author} {\bibinfo {author} {\bibfnamefont {Z.}~\bibnamefont
  {Li}}, \bibinfo {author} {\bibfnamefont {F.}~\bibnamefont {Liu}}, \bibinfo
  {author} {\bibfnamefont {W.}~\bibnamefont {Yang}}, \bibinfo {author}
  {\bibfnamefont {S.}~\bibnamefont {Peng}}, \ and\ \bibinfo {author}
  {\bibfnamefont {J.}~\bibnamefont {Zhou}},\ }\href {\doibase
  10.1109/TNNLS.2021.3084827} {\bibfield  {journal} {\bibinfo  {journal} {IEEE
  Transactions on Neural Networks and Learning Systems}\ }\textbf {\bibinfo
  {volume} {33}},\ \bibinfo {pages} {6999} (\bibinfo {year}
  {2022})}\BibitemShut {NoStop}%
\bibitem [{\citenamefont {He}\ \emph {et~al.}(2016)\citenamefont {He},
  \citenamefont {Zhang}, \citenamefont {Ren},\ and\ \citenamefont
  {Sun}}]{heIdentityMappingsDeep2016}%
  \BibitemOpen
  \bibfield  {author} {\bibinfo {author} {\bibfnamefont {K.}~\bibnamefont
  {He}}, \bibinfo {author} {\bibfnamefont {X.}~\bibnamefont {Zhang}}, \bibinfo
  {author} {\bibfnamefont {S.}~\bibnamefont {Ren}}, \ and\ \bibinfo {author}
  {\bibfnamefont {J.}~\bibnamefont {Sun}},\ }in\ \href {\doibase
  10.48550/arXiv.1603.05027} {\emph {\bibinfo {booktitle} {{{ECCV}} 2016}}}\
  (\bibinfo  {publisher} {{arXiv}},\ \bibinfo {year} {2016})\ \Eprint
  {http://arxiv.org/abs/1603.05027} {arxiv:1603.05027 [cs]} \BibitemShut
  {NoStop}%
\bibitem [{\citenamefont {Huang}\ \emph {et~al.}(2016)\citenamefont {Huang},
  \citenamefont {Sun}, \citenamefont {Liu}, \citenamefont {Sedra},\ and\
  \citenamefont {Weinberger}}]{huangDeepNetworksStochastic2016}%
  \BibitemOpen
  \bibfield  {author} {\bibinfo {author} {\bibfnamefont {G.}~\bibnamefont
  {Huang}}, \bibinfo {author} {\bibfnamefont {Y.}~\bibnamefont {Sun}}, \bibinfo
  {author} {\bibfnamefont {Z.}~\bibnamefont {Liu}}, \bibinfo {author}
  {\bibfnamefont {D.}~\bibnamefont {Sedra}}, \ and\ \bibinfo {author}
  {\bibfnamefont {K.}~\bibnamefont {Weinberger}},\ }\href {\doibase
  10.48550/arXiv.1603.09382} {\enquote {\bibinfo {title} {Deep {{Networks}}
  with {{Stochastic Depth}}},}\ } (\bibinfo {year} {2016}),\ \Eprint
  {http://arxiv.org/abs/1603.09382} {arxiv:1603.09382 [cs]} \BibitemShut
  {NoStop}%
\bibitem [{\citenamefont {Lin}\ \emph {et~al.}(2014)\citenamefont {Lin},
  \citenamefont {Chen},\ and\ \citenamefont {Yan}}]{linNetworkNetwork2014}%
  \BibitemOpen
  \bibfield  {author} {\bibinfo {author} {\bibfnamefont {M.}~\bibnamefont
  {Lin}}, \bibinfo {author} {\bibfnamefont {Q.}~\bibnamefont {Chen}}, \ and\
  \bibinfo {author} {\bibfnamefont {S.}~\bibnamefont {Yan}},\ }\href {\doibase
  10.48550/arXiv.1312.4400} {\enquote {\bibinfo {title} {Network {{In
  Network}}},}\ } (\bibinfo {year} {2014}),\ \Eprint
  {http://arxiv.org/abs/1312.4400} {arxiv:1312.4400 [cs]} \BibitemShut
  {NoStop}%
\bibitem [{\citenamefont {Hochreiter}\ and\ \citenamefont
  {Schmidhuber}(1997)}]{hochreiterLongShortTermMemory1997}%
  \BibitemOpen
  \bibfield  {author} {\bibinfo {author} {\bibfnamefont {S.}~\bibnamefont
  {Hochreiter}}\ and\ \bibinfo {author} {\bibfnamefont {J.}~\bibnamefont
  {Schmidhuber}},\ }\href {\doibase 10.1162/neco.1997.9.8.1735} {\bibfield
  {journal} {\bibinfo  {journal} {Neural Computation}\ }\textbf {\bibinfo
  {volume} {9}},\ \bibinfo {pages} {1735} (\bibinfo {year} {1997})}\BibitemShut
  {NoStop}%
\bibitem [{\citenamefont {Hochreiter}\ \emph {et~al.}(2001)\citenamefont
  {Hochreiter}, \citenamefont {Bengio}, \citenamefont {Frasconi},\ and\
  \citenamefont {Schmidhuber}}]{hochreiterGradientFlowRecurrent2001}%
  \BibitemOpen
  \bibfield  {author} {\bibinfo {author} {\bibfnamefont {S.}~\bibnamefont
  {Hochreiter}}, \bibinfo {author} {\bibfnamefont {Y.}~\bibnamefont {Bengio}},
  \bibinfo {author} {\bibfnamefont {P.}~\bibnamefont {Frasconi}}, \ and\
  \bibinfo {author} {\bibfnamefont {J.}~\bibnamefont {Schmidhuber}},\ }in\
  \href@noop {} {\emph {\bibinfo {booktitle} {A Field Guide to Dynamical
  Recurrent Neural Networks}}},\ \bibinfo {editor} {edited by\ \bibinfo
  {editor} {\bibfnamefont {S.~C.}\ \bibnamefont {Kremer}}\ and\ \bibinfo
  {editor} {\bibfnamefont {J.~F.}\ \bibnamefont {Kolen}}}\ (\bibinfo
  {publisher} {{IEEE Press}},\ \bibinfo {year} {2001})\BibitemShut {NoStop}%
\bibitem [{\citenamefont {Sutskever}\ \emph {et~al.}(2014)\citenamefont
  {Sutskever}, \citenamefont {Vinyals},\ and\ \citenamefont
  {Le}}]{sutskeverSequenceSequenceLearning2014}%
  \BibitemOpen
  \bibfield  {author} {\bibinfo {author} {\bibfnamefont {I.}~\bibnamefont
  {Sutskever}}, \bibinfo {author} {\bibfnamefont {O.}~\bibnamefont {Vinyals}},
  \ and\ \bibinfo {author} {\bibfnamefont {Q.~V.}\ \bibnamefont {Le}},\ }in\
  \href@noop {} {\emph {\bibinfo {booktitle} {Advances in Neural Information
  Processing Systems}}},\ Vol.~\bibinfo {volume} {27},\ \bibinfo {editor}
  {edited by\ \bibinfo {editor} {\bibfnamefont {Z.}~\bibnamefont {Ghahramani}},
  \bibinfo {editor} {\bibfnamefont {M.}~\bibnamefont {Welling}}, \bibinfo
  {editor} {\bibfnamefont {C.}~\bibnamefont {Cortes}}, \bibinfo {editor}
  {\bibfnamefont {N.}~\bibnamefont {Lawrence}}, \ and\ \bibinfo {editor}
  {\bibfnamefont {K.}~\bibnamefont {Weinberger}}}\ (\bibinfo  {publisher}
  {{Curran Associates, Inc.}},\ \bibinfo {year} {2014})\ \Eprint
  {http://arxiv.org/abs/1409.3215} {arxiv:1409.3215 [cs]} \BibitemShut
  {NoStop}%
\bibitem [{\citenamefont {Lakew}\ \emph {et~al.}(2018)\citenamefont {Lakew},
  \citenamefont {Cettolo},\ and\ \citenamefont
  {Federico}}]{lakewComparisonTransformerRecurrent2018}%
  \BibitemOpen
  \bibfield  {author} {\bibinfo {author} {\bibfnamefont {S.~M.}\ \bibnamefont
  {Lakew}}, \bibinfo {author} {\bibfnamefont {M.}~\bibnamefont {Cettolo}}, \
  and\ \bibinfo {author} {\bibfnamefont {M.}~\bibnamefont {Federico}},\ }\href
  {\doibase 10.48550/arXiv.1806.06957} {\enquote {\bibinfo {title} {A
  {{Comparison}} of {{Transformer}} and {{Recurrent Neural Networks}} on
  {{Multilingual Neural Machine Translation}}},}\ } (\bibinfo {year} {2018}),\
  \Eprint {http://arxiv.org/abs/1806.06957} {arxiv:1806.06957 [cs]}
  \BibitemShut {NoStop}%
\bibitem [{\citenamefont {Wolf}\ \emph {et~al.}(2020)\citenamefont {Wolf},
  \citenamefont {Debut}, \citenamefont {Sanh}, \citenamefont {Chaumond},
  \citenamefont {Delangue}, \citenamefont {Moi}, \citenamefont {Cistac},
  \citenamefont {Rault}, \citenamefont {Louf}, \citenamefont {Funtowicz},
  \citenamefont {Davison}, \citenamefont {Shleifer}, \citenamefont {{von
  Platen}}, \citenamefont {Ma}, \citenamefont {Jernite}, \citenamefont {Plu},
  \citenamefont {Xu}, \citenamefont {Le~Scao}, \citenamefont {Gugger},
  \citenamefont {Drame}, \citenamefont {Lhoest},\ and\ \citenamefont
  {Rush}}]{wolfTransformersStateoftheArtNatural2020}%
  \BibitemOpen
  \bibfield  {author} {\bibinfo {author} {\bibfnamefont {T.}~\bibnamefont
  {Wolf}}, \bibinfo {author} {\bibfnamefont {L.}~\bibnamefont {Debut}},
  \bibinfo {author} {\bibfnamefont {V.}~\bibnamefont {Sanh}}, \bibinfo {author}
  {\bibfnamefont {J.}~\bibnamefont {Chaumond}}, \bibinfo {author}
  {\bibfnamefont {C.}~\bibnamefont {Delangue}}, \bibinfo {author}
  {\bibfnamefont {A.}~\bibnamefont {Moi}}, \bibinfo {author} {\bibfnamefont
  {P.}~\bibnamefont {Cistac}}, \bibinfo {author} {\bibfnamefont
  {T.}~\bibnamefont {Rault}}, \bibinfo {author} {\bibfnamefont
  {R.}~\bibnamefont {Louf}}, \bibinfo {author} {\bibfnamefont {M.}~\bibnamefont
  {Funtowicz}}, \bibinfo {author} {\bibfnamefont {J.}~\bibnamefont {Davison}},
  \bibinfo {author} {\bibfnamefont {S.}~\bibnamefont {Shleifer}}, \bibinfo
  {author} {\bibfnamefont {P.}~\bibnamefont {{von Platen}}}, \bibinfo {author}
  {\bibfnamefont {C.}~\bibnamefont {Ma}}, \bibinfo {author} {\bibfnamefont
  {Y.}~\bibnamefont {Jernite}}, \bibinfo {author} {\bibfnamefont
  {J.}~\bibnamefont {Plu}}, \bibinfo {author} {\bibfnamefont {C.}~\bibnamefont
  {Xu}}, \bibinfo {author} {\bibfnamefont {T.}~\bibnamefont {Le~Scao}},
  \bibinfo {author} {\bibfnamefont {S.}~\bibnamefont {Gugger}}, \bibinfo
  {author} {\bibfnamefont {M.}~\bibnamefont {Drame}}, \bibinfo {author}
  {\bibfnamefont {Q.}~\bibnamefont {Lhoest}}, \ and\ \bibinfo {author}
  {\bibfnamefont {A.}~\bibnamefont {Rush}},\ }in\ \href {\doibase
  10.18653/v1/2020.emnlp-demos.6} {\emph {\bibinfo {booktitle} {Proceedings of
  the 2020 {{Conference}} on {{Empirical Methods}} in {{Natural Language
  Processing}}: {{System Demonstrations}}}}}\ (\bibinfo  {publisher}
  {{Association for Computational Linguistics}},\ \bibinfo {address}
  {{Online}},\ \bibinfo {year} {2020})\ pp.\ \bibinfo {pages}
  {38--45}\BibitemShut {NoStop}%
\bibitem [{\citenamefont {Scarselli}\ \emph {et~al.}(2009)\citenamefont
  {Scarselli}, \citenamefont {Gori}, \citenamefont {Tsoi}, \citenamefont
  {Hagenbuchner},\ and\ \citenamefont
  {Monfardini}}]{scarselliGraphNeuralNetwork2009}%
  \BibitemOpen
  \bibfield  {author} {\bibinfo {author} {\bibfnamefont {F.}~\bibnamefont
  {Scarselli}}, \bibinfo {author} {\bibfnamefont {M.}~\bibnamefont {Gori}},
  \bibinfo {author} {\bibfnamefont {A.~C.}\ \bibnamefont {Tsoi}}, \bibinfo
  {author} {\bibfnamefont {M.}~\bibnamefont {Hagenbuchner}}, \ and\ \bibinfo
  {author} {\bibfnamefont {G.}~\bibnamefont {Monfardini}},\ }\href {\doibase
  10.1109/TNN.2008.2005605} {\bibfield  {journal} {\bibinfo  {journal} {IEEE
  Transactions on Neural Networks}\ }\textbf {\bibinfo {volume} {20}},\
  \bibinfo {pages} {61} (\bibinfo {year} {2009})}\BibitemShut {NoStop}%
\bibitem [{\citenamefont {Bronstein}\ \emph {et~al.}(2021)\citenamefont
  {Bronstein}, \citenamefont {Bruna}, \citenamefont {Cohen},\ and\
  \citenamefont {Veli{\v c}kovi{\'c}}}]{bronsteinGeometricDeepLearning2021}%
  \BibitemOpen
  \bibfield  {author} {\bibinfo {author} {\bibfnamefont {M.~M.}\ \bibnamefont
  {Bronstein}}, \bibinfo {author} {\bibfnamefont {J.}~\bibnamefont {Bruna}},
  \bibinfo {author} {\bibfnamefont {T.}~\bibnamefont {Cohen}}, \ and\ \bibinfo
  {author} {\bibfnamefont {P.}~\bibnamefont {Veli{\v c}kovi{\'c}}},\ }\href
  {\doibase 10.48550/arXiv.2104.13478} {\enquote {\bibinfo {title} {Geometric
  {{Deep Learning}}: {{Grids}}, {{Groups}}, {{Graphs}}, {{Geodesics}}, and
  {{Gauges}}},}\ } (\bibinfo {year} {2021}),\ \Eprint
  {http://arxiv.org/abs/2104.13478} {arxiv:2104.13478 [cs, stat]} \BibitemShut
  {NoStop}%
\bibitem [{\citenamefont {Kipf}\ and\ \citenamefont
  {Welling}(2017)}]{kipfSemiSupervisedClassificationGraph2017}%
  \BibitemOpen
  \bibfield  {author} {\bibinfo {author} {\bibfnamefont {T.~N.}\ \bibnamefont
  {Kipf}}\ and\ \bibinfo {author} {\bibfnamefont {M.}~\bibnamefont {Welling}},\
  }\href {\doibase 10.48550/arXiv.1609.02907} {\enquote {\bibinfo {title}
  {Semi-{{Supervised Classification}} with {{Graph Convolutional Networks}}},}\
  } (\bibinfo {year} {2017}),\ \Eprint {http://arxiv.org/abs/1609.02907}
  {arxiv:1609.02907 [cs, stat]} \BibitemShut {NoStop}%
\bibitem [{\citenamefont {Veli{\v c}kovi{\'c}}\ \emph
  {et~al.}(2018)\citenamefont {Veli{\v c}kovi{\'c}}, \citenamefont {Cucurull},
  \citenamefont {Casanova}, \citenamefont {Romero}, \citenamefont {Li{\`o}},\
  and\ \citenamefont {Bengio}}]{velickovicGraphAttentionNetworks2018}%
  \BibitemOpen
  \bibfield  {author} {\bibinfo {author} {\bibfnamefont {P.}~\bibnamefont
  {Veli{\v c}kovi{\'c}}}, \bibinfo {author} {\bibfnamefont {G.}~\bibnamefont
  {Cucurull}}, \bibinfo {author} {\bibfnamefont {A.}~\bibnamefont {Casanova}},
  \bibinfo {author} {\bibfnamefont {A.}~\bibnamefont {Romero}}, \bibinfo
  {author} {\bibfnamefont {P.}~\bibnamefont {Li{\`o}}}, \ and\ \bibinfo
  {author} {\bibfnamefont {Y.}~\bibnamefont {Bengio}},\ }\href {\doibase
  10.48550/arXiv.1710.10903} {\enquote {\bibinfo {title} {Graph {{Attention
  Networks}}},}\ } (\bibinfo {year} {2018}),\ \Eprint
  {http://arxiv.org/abs/1710.10903} {arxiv:1710.10903 [cs, stat]} \BibitemShut
  {NoStop}%
\bibitem [{\citenamefont {Li}\ \emph {et~al.}(2017)\citenamefont {Li},
  \citenamefont {Tarlow}, \citenamefont {Brockschmidt},\ and\ \citenamefont
  {Zemel}}]{liGatedGraphSequence2017}%
  \BibitemOpen
  \bibfield  {author} {\bibinfo {author} {\bibfnamefont {Y.}~\bibnamefont
  {Li}}, \bibinfo {author} {\bibfnamefont {D.}~\bibnamefont {Tarlow}}, \bibinfo
  {author} {\bibfnamefont {M.}~\bibnamefont {Brockschmidt}}, \ and\ \bibinfo
  {author} {\bibfnamefont {R.}~\bibnamefont {Zemel}},\ }\href {\doibase
  10.48550/arXiv.1511.05493} {\enquote {\bibinfo {title} {Gated {{Graph
  Sequence Neural Networks}}},}\ } (\bibinfo {year} {2017}),\ \Eprint
  {http://arxiv.org/abs/1511.05493} {arxiv:1511.05493 [cs, stat]} \BibitemShut
  {NoStop}%
\bibitem [{\citenamefont {Deshpande}\ \emph {et~al.}(2023)\citenamefont
  {Deshpande}, \citenamefont {Bordas},\ and\ \citenamefont
  {Lengiewicz}}]{deshpandeMAgNETGraphUNet2023}%
  \BibitemOpen
  \bibfield  {author} {\bibinfo {author} {\bibfnamefont {S.}~\bibnamefont
  {Deshpande}}, \bibinfo {author} {\bibfnamefont {S.~P.~A.}\ \bibnamefont
  {Bordas}}, \ and\ \bibinfo {author} {\bibfnamefont {J.}~\bibnamefont
  {Lengiewicz}},\ }\href {\doibase 10.48550/arXiv.2211.00713} {\enquote
  {\bibinfo {title} {{{MAgNET}}: {{A Graph U-Net Architecture}} for
  {{Mesh-Based Simulations}}},}\ } (\bibinfo {year} {2023}),\ \Eprint
  {http://arxiv.org/abs/2211.00713} {arxiv:2211.00713 [cs]} \BibitemShut
  {NoStop}%
\bibitem [{\citenamefont {Khoram}\ \emph {et~al.}(2023)\citenamefont {Khoram},
  \citenamefont {Wu}, \citenamefont {Qu}, \citenamefont {Zhou},\ and\
  \citenamefont {Yu}}]{khoramGraphNeuralNetworks2023}%
  \BibitemOpen
  \bibfield  {author} {\bibinfo {author} {\bibfnamefont {E.}~\bibnamefont
  {Khoram}}, \bibinfo {author} {\bibfnamefont {Z.}~\bibnamefont {Wu}}, \bibinfo
  {author} {\bibfnamefont {Y.}~\bibnamefont {Qu}}, \bibinfo {author}
  {\bibfnamefont {M.}~\bibnamefont {Zhou}}, \ and\ \bibinfo {author}
  {\bibfnamefont {Z.}~\bibnamefont {Yu}},\ }\href {\doibase
  10.1021/acsphotonics.2c01019} {\bibfield  {journal} {\bibinfo  {journal} {ACS
  Photonics}\ }\textbf {\bibinfo {volume} {10}},\ \bibinfo {pages} {892}
  (\bibinfo {year} {2023})}\BibitemShut {NoStop}%
\bibitem [{\citenamefont {Kuhn}\ \emph {et~al.}(2023)\citenamefont {Kuhn},
  \citenamefont {Rep{\"a}n},\ and\ \citenamefont
  {Rockstuhl}}]{kuhnExploitingGraphNeural2023}%
  \BibitemOpen
  \bibfield  {author} {\bibinfo {author} {\bibfnamefont {L.}~\bibnamefont
  {Kuhn}}, \bibinfo {author} {\bibfnamefont {T.}~\bibnamefont {Rep{\"a}n}}, \
  and\ \bibinfo {author} {\bibfnamefont {C.}~\bibnamefont {Rockstuhl}},\ }\href
  {\doibase 10.1063/5.0139004} {\bibfield  {journal} {\bibinfo  {journal} {APL
  Photonics}\ }\textbf {\bibinfo {volume} {8}},\ \bibinfo {pages} {036109}
  (\bibinfo {year} {2023})}\BibitemShut {NoStop}%
\bibitem [{\citenamefont {Bahdanau}\ \emph {et~al.}(2016)\citenamefont
  {Bahdanau}, \citenamefont {Cho},\ and\ \citenamefont
  {Bengio}}]{bahdanauNeuralMachineTranslation2016}%
  \BibitemOpen
  \bibfield  {author} {\bibinfo {author} {\bibfnamefont {D.}~\bibnamefont
  {Bahdanau}}, \bibinfo {author} {\bibfnamefont {K.}~\bibnamefont {Cho}}, \
  and\ \bibinfo {author} {\bibfnamefont {Y.}~\bibnamefont {Bengio}},\ }\href
  {\doibase 10.48550/arXiv.1409.0473} {\enquote {\bibinfo {title} {Neural
  {{Machine Translation}} by {{Jointly Learning}} to {{Align}} and
  {{Translate}}},}\ } (\bibinfo {year} {2016}),\ \Eprint
  {http://arxiv.org/abs/1409.0473} {arxiv:1409.0473 [cs, stat]} \BibitemShut
  {NoStop}%
\bibitem [{\citenamefont {Cordonnier}\ \emph {et~al.}(2020)\citenamefont
  {Cordonnier}, \citenamefont {Loukas},\ and\ \citenamefont
  {Jaggi}}]{cordonnierRelationshipSelfAttentionConvolutional2020}%
  \BibitemOpen
  \bibfield  {author} {\bibinfo {author} {\bibfnamefont {J.-B.}\ \bibnamefont
  {Cordonnier}}, \bibinfo {author} {\bibfnamefont {A.}~\bibnamefont {Loukas}},
  \ and\ \bibinfo {author} {\bibfnamefont {M.}~\bibnamefont {Jaggi}},\ }\href
  {\doibase 10.48550/arXiv.1911.03584} {\enquote {\bibinfo {title} {On the
  {{Relationship}} between {{Self-Attention}} and {{Convolutional Layers}}},}\
  } (\bibinfo {year} {2020}),\ \Eprint {http://arxiv.org/abs/1911.03584}
  {arxiv:1911.03584 [cs, stat]} \BibitemShut {NoStop}%
\bibitem [{\citenamefont {Dosovitskiy}\ \emph {et~al.}(2021)\citenamefont
  {Dosovitskiy}, \citenamefont {Beyer}, \citenamefont {Kolesnikov},
  \citenamefont {Weissenborn}, \citenamefont {Zhai}, \citenamefont
  {Unterthiner}, \citenamefont {Dehghani}, \citenamefont {Minderer},
  \citenamefont {Heigold}, \citenamefont {Gelly}, \citenamefont {Uszkoreit},\
  and\ \citenamefont {Houlsby}}]{dosovitskiyImageWorth16x162021}%
  \BibitemOpen
  \bibfield  {author} {\bibinfo {author} {\bibfnamefont {A.}~\bibnamefont
  {Dosovitskiy}}, \bibinfo {author} {\bibfnamefont {L.}~\bibnamefont {Beyer}},
  \bibinfo {author} {\bibfnamefont {A.}~\bibnamefont {Kolesnikov}}, \bibinfo
  {author} {\bibfnamefont {D.}~\bibnamefont {Weissenborn}}, \bibinfo {author}
  {\bibfnamefont {X.}~\bibnamefont {Zhai}}, \bibinfo {author} {\bibfnamefont
  {T.}~\bibnamefont {Unterthiner}}, \bibinfo {author} {\bibfnamefont
  {M.}~\bibnamefont {Dehghani}}, \bibinfo {author} {\bibfnamefont
  {M.}~\bibnamefont {Minderer}}, \bibinfo {author} {\bibfnamefont
  {G.}~\bibnamefont {Heigold}}, \bibinfo {author} {\bibfnamefont
  {S.}~\bibnamefont {Gelly}}, \bibinfo {author} {\bibfnamefont
  {J.}~\bibnamefont {Uszkoreit}}, \ and\ \bibinfo {author} {\bibfnamefont
  {N.}~\bibnamefont {Houlsby}},\ }\href {\doibase 10.48550/arXiv.2010.11929}
  {\enquote {\bibinfo {title} {An {{Image}} is {{Worth}} 16x16 {{Words}}:
  {{Transformers}} for {{Image Recognition}} at {{Scale}}},}\ } (\bibinfo
  {year} {2021}),\ \Eprint {http://arxiv.org/abs/2010.11929} {arxiv:2010.11929
  [cs]} \BibitemShut {NoStop}%
\bibitem [{\citenamefont {Liu}\ \emph {et~al.}(2021{\natexlab{b}})\citenamefont
  {Liu}, \citenamefont {Lin}, \citenamefont {Cao}, \citenamefont {Hu},
  \citenamefont {Wei}, \citenamefont {Zhang}, \citenamefont {Lin},\ and\
  \citenamefont {Guo}}]{liuSwinTransformerHierarchical2021}%
  \BibitemOpen
  \bibfield  {author} {\bibinfo {author} {\bibfnamefont {Z.}~\bibnamefont
  {Liu}}, \bibinfo {author} {\bibfnamefont {Y.}~\bibnamefont {Lin}}, \bibinfo
  {author} {\bibfnamefont {Y.}~\bibnamefont {Cao}}, \bibinfo {author}
  {\bibfnamefont {H.}~\bibnamefont {Hu}}, \bibinfo {author} {\bibfnamefont
  {Y.}~\bibnamefont {Wei}}, \bibinfo {author} {\bibfnamefont {Z.}~\bibnamefont
  {Zhang}}, \bibinfo {author} {\bibfnamefont {S.}~\bibnamefont {Lin}}, \ and\
  \bibinfo {author} {\bibfnamefont {B.}~\bibnamefont {Guo}},\ }\href {\doibase
  10.48550/arXiv.2103.14030} {\enquote {\bibinfo {title} {Swin {{Transformer}}:
  {{Hierarchical Vision Transformer}} using {{Shifted Windows}}},}\ } (\bibinfo
  {year} {2021}{\natexlab{b}}),\ \Eprint {http://arxiv.org/abs/2103.14030}
  {arxiv:2103.14030 [cs]} \BibitemShut {NoStop}%
\bibitem [{\citenamefont {Naseer}\ \emph {et~al.}(2021)\citenamefont {Naseer},
  \citenamefont {Ranasinghe}, \citenamefont {Khan}, \citenamefont {Hayat},
  \citenamefont {Khan},\ and\ \citenamefont
  {Yang}}]{naseerIntriguingPropertiesVision2021}%
  \BibitemOpen
  \bibfield  {author} {\bibinfo {author} {\bibfnamefont {M.}~\bibnamefont
  {Naseer}}, \bibinfo {author} {\bibfnamefont {K.}~\bibnamefont {Ranasinghe}},
  \bibinfo {author} {\bibfnamefont {S.}~\bibnamefont {Khan}}, \bibinfo {author}
  {\bibfnamefont {M.}~\bibnamefont {Hayat}}, \bibinfo {author} {\bibfnamefont
  {F.~S.}\ \bibnamefont {Khan}}, \ and\ \bibinfo {author} {\bibfnamefont
  {M.-H.}\ \bibnamefont {Yang}},\ }\href {\doibase 10.48550/arXiv.2105.10497}
  {\enquote {\bibinfo {title} {Intriguing {{Properties}} of {{Vision
  Transformers}}},}\ } (\bibinfo {year} {2021}),\ \Eprint
  {http://arxiv.org/abs/2105.10497} {arxiv:2105.10497 [cs]} \BibitemShut
  {NoStop}%
\bibitem [{\citenamefont {Xiao}\ \emph {et~al.}(2021)\citenamefont {Xiao},
  \citenamefont {Singh}, \citenamefont {Mintun}, \citenamefont {Darrell},
  \citenamefont {Doll{\'a}r},\ and\ \citenamefont
  {Girshick}}]{xiaoEarlyConvolutionsHelp2021}%
  \BibitemOpen
  \bibfield  {author} {\bibinfo {author} {\bibfnamefont {T.}~\bibnamefont
  {Xiao}}, \bibinfo {author} {\bibfnamefont {M.}~\bibnamefont {Singh}},
  \bibinfo {author} {\bibfnamefont {E.}~\bibnamefont {Mintun}}, \bibinfo
  {author} {\bibfnamefont {T.}~\bibnamefont {Darrell}}, \bibinfo {author}
  {\bibfnamefont {P.}~\bibnamefont {Doll{\'a}r}}, \ and\ \bibinfo {author}
  {\bibfnamefont {R.}~\bibnamefont {Girshick}},\ }\href {\doibase
  10.48550/arXiv.2106.14881} {\enquote {\bibinfo {title} {Early {{Convolutions
  Help Transformers See Better}}},}\ } (\bibinfo {year} {2021}),\ \Eprint
  {http://arxiv.org/abs/2106.14881} {arxiv:2106.14881 [cs]} \BibitemShut
  {NoStop}%
\bibitem [{\citenamefont {Dai}\ \emph {et~al.}(2021)\citenamefont {Dai},
  \citenamefont {Liu}, \citenamefont {Le},\ and\ \citenamefont
  {Tan}}]{daiCoAtNetMarryingConvolution2021}%
  \BibitemOpen
  \bibfield  {author} {\bibinfo {author} {\bibfnamefont {Z.}~\bibnamefont
  {Dai}}, \bibinfo {author} {\bibfnamefont {H.}~\bibnamefont {Liu}}, \bibinfo
  {author} {\bibfnamefont {Q.~V.}\ \bibnamefont {Le}}, \ and\ \bibinfo {author}
  {\bibfnamefont {M.}~\bibnamefont {Tan}},\ }\href {\doibase
  10.48550/arXiv.2106.04803} {\enquote {\bibinfo {title} {{{CoAtNet}}:
  {{Marrying Convolution}} and {{Attention}} for {{All Data Sizes}}},}\ }
  (\bibinfo {year} {2021}),\ \Eprint {http://arxiv.org/abs/2106.04803}
  {arxiv:2106.04803 [cs]} \BibitemShut {NoStop}%
\bibitem [{\citenamefont {Liu}\ \emph {et~al.}(2022)\citenamefont {Liu},
  \citenamefont {Mao}, \citenamefont {Wu}, \citenamefont {Feichtenhofer},
  \citenamefont {Darrell},\ and\ \citenamefont {Xie}}]{liuConvNet2020s2022}%
  \BibitemOpen
  \bibfield  {author} {\bibinfo {author} {\bibfnamefont {Z.}~\bibnamefont
  {Liu}}, \bibinfo {author} {\bibfnamefont {H.}~\bibnamefont {Mao}}, \bibinfo
  {author} {\bibfnamefont {C.-Y.}\ \bibnamefont {Wu}}, \bibinfo {author}
  {\bibfnamefont {C.}~\bibnamefont {Feichtenhofer}}, \bibinfo {author}
  {\bibfnamefont {T.}~\bibnamefont {Darrell}}, \ and\ \bibinfo {author}
  {\bibfnamefont {S.}~\bibnamefont {Xie}},\ }\href {\doibase
  10.48550/arXiv.2201.03545} {\enquote {\bibinfo {title} {A {{ConvNet}} for the
  2020s},}\ } (\bibinfo {year} {2022}),\ \Eprint
  {http://arxiv.org/abs/2201.03545} {arxiv:2201.03545 [cs]} \BibitemShut
  {NoStop}%
\bibitem [{\citenamefont {Lee}\ \emph {et~al.}(2021)\citenamefont {Lee},
  \citenamefont {Lee},\ and\ \citenamefont
  {Song}}]{leeVisionTransformerSmallSize2021}%
  \BibitemOpen
  \bibfield  {author} {\bibinfo {author} {\bibfnamefont {S.~H.}\ \bibnamefont
  {Lee}}, \bibinfo {author} {\bibfnamefont {S.}~\bibnamefont {Lee}}, \ and\
  \bibinfo {author} {\bibfnamefont {B.~C.}\ \bibnamefont {Song}},\ }\href@noop
  {} {\enquote {\bibinfo {title} {Vision {{Transformer}} for {{Small-Size
  Datasets}}},}\ } (\bibinfo {year} {2021}),\ \Eprint
  {http://arxiv.org/abs/2112.13492} {arxiv:2112.13492} \BibitemShut {NoStop}%
\bibitem [{\citenamefont {Xie}\ \emph {et~al.}(2017)\citenamefont {Xie},
  \citenamefont {Girshick}, \citenamefont {Doll{\'a}r}, \citenamefont {Tu},\
  and\ \citenamefont {He}}]{xieAggregatedResidualTransformations2017}%
  \BibitemOpen
  \bibfield  {author} {\bibinfo {author} {\bibfnamefont {S.}~\bibnamefont
  {Xie}}, \bibinfo {author} {\bibfnamefont {R.}~\bibnamefont {Girshick}},
  \bibinfo {author} {\bibfnamefont {P.}~\bibnamefont {Doll{\'a}r}}, \bibinfo
  {author} {\bibfnamefont {Z.}~\bibnamefont {Tu}}, \ and\ \bibinfo {author}
  {\bibfnamefont {K.}~\bibnamefont {He}},\ }\href {\doibase
  10.48550/arXiv.1611.05431} {\enquote {\bibinfo {title} {Aggregated {{Residual
  Transformations}} for {{Deep Neural Networks}}},}\ } (\bibinfo {year}
  {2017}),\ \Eprint {http://arxiv.org/abs/1611.05431} {arxiv:1611.05431 [cs]}
  \BibitemShut {NoStop}%
\bibitem [{\citenamefont
  {Hegde}(2020{\natexlab{b}})}]{hegdePhotonicsInverseDesign2020}%
  \BibitemOpen
  \bibfield  {author} {\bibinfo {author} {\bibfnamefont {R.~S.}\ \bibnamefont
  {Hegde}},\ }\href {\doibase 10.1109/JSTQE.2019.2933796} {\bibfield  {journal}
  {\bibinfo  {journal} {IEEE Journal of Selected Topics in Quantum
  Electronics}\ }\textbf {\bibinfo {volume} {26}},\ \bibinfo {pages} {1}
  (\bibinfo {year} {2020}{\natexlab{b}})}\BibitemShut {NoStop}%
\bibitem [{\citenamefont {Huang}\ \emph {et~al.}(2017)\citenamefont {Huang},
  \citenamefont {Papernot}, \citenamefont {Goodfellow}, \citenamefont {Duan},\
  and\ \citenamefont {Abbeel}}]{huangAdversarialAttacksNeural2017}%
  \BibitemOpen
  \bibfield  {author} {\bibinfo {author} {\bibfnamefont {S.}~\bibnamefont
  {Huang}}, \bibinfo {author} {\bibfnamefont {N.}~\bibnamefont {Papernot}},
  \bibinfo {author} {\bibfnamefont {I.}~\bibnamefont {Goodfellow}}, \bibinfo
  {author} {\bibfnamefont {Y.}~\bibnamefont {Duan}}, \ and\ \bibinfo {author}
  {\bibfnamefont {P.}~\bibnamefont {Abbeel}},\ }\href {\doibase
  10.48550/arXiv.1702.02284} {\enquote {\bibinfo {title} {Adversarial
  {{Attacks}} on {{Neural Network Policies}}},}\ } (\bibinfo {year} {2017}),\
  \Eprint {http://arxiv.org/abs/1702.02284} {arxiv:1702.02284 [cs, stat]}
  \BibitemShut {NoStop}%
\bibitem [{\citenamefont {Deng}\ \emph {et~al.}(2021)\citenamefont {Deng},
  \citenamefont {Ren}, \citenamefont {Fan}, \citenamefont {Malof},\ and\
  \citenamefont {Padilla}}]{dengNeuraladjointMethodInverse2021}%
  \BibitemOpen
  \bibfield  {author} {\bibinfo {author} {\bibfnamefont {Y.}~\bibnamefont
  {Deng}}, \bibinfo {author} {\bibfnamefont {S.}~\bibnamefont {Ren}}, \bibinfo
  {author} {\bibfnamefont {K.}~\bibnamefont {Fan}}, \bibinfo {author}
  {\bibfnamefont {J.~M.}\ \bibnamefont {Malof}}, \ and\ \bibinfo {author}
  {\bibfnamefont {W.~J.}\ \bibnamefont {Padilla}},\ }\href {\doibase
  10.1364/OE.419138} {\bibfield  {journal} {\bibinfo  {journal} {Optics
  Express}\ }\textbf {\bibinfo {volume} {29}},\ \bibinfo {pages} {7526}
  (\bibinfo {year} {2021})}\BibitemShut {NoStop}%
\bibitem [{\citenamefont {Luk{\v s}i{\v c}}\ \emph {et~al.}(2019)\citenamefont
  {Luk{\v s}i{\v c}}, \citenamefont {Tanevski}, \citenamefont {D{\v z}eroski},\
  and\ \citenamefont
  {Todorovski}}]{luksicMetaModelFrameworkSurrogateBased2019}%
  \BibitemOpen
  \bibfield  {author} {\bibinfo {author} {\bibfnamefont {{\v Z}.}~\bibnamefont
  {Luk{\v s}i{\v c}}}, \bibinfo {author} {\bibfnamefont {J.}~\bibnamefont
  {Tanevski}}, \bibinfo {author} {\bibfnamefont {S.}~\bibnamefont {D{\v
  z}eroski}}, \ and\ \bibinfo {author} {\bibfnamefont {L.}~\bibnamefont
  {Todorovski}},\ }\href {\doibase 10.1109/ACCESS.2019.2959846} {\bibfield
  {journal} {\bibinfo  {journal} {IEEE Access}\ }\textbf {\bibinfo {volume}
  {7}},\ \bibinfo {pages} {181829} (\bibinfo {year} {2019})}\BibitemShut
  {NoStop}%
\bibitem [{\citenamefont {Khowaja}\ \emph {et~al.}(2021)\citenamefont
  {Khowaja}, \citenamefont {Shcherbatyy},\ and\ \citenamefont
  {H{\"a}rdle}}]{khowajaSurrogateModelsOptimization2021}%
  \BibitemOpen
  \bibfield  {author} {\bibinfo {author} {\bibfnamefont {K.}~\bibnamefont
  {Khowaja}}, \bibinfo {author} {\bibfnamefont {M.}~\bibnamefont
  {Shcherbatyy}}, \ and\ \bibinfo {author} {\bibfnamefont {W.~K.}\ \bibnamefont
  {H{\"a}rdle}},\ }\href {\doibase 10.48550/arXiv.2101.10189} {\enquote
  {\bibinfo {title} {Surrogate {{Models}} for {{Optimization}} of {{Dynamical
  Systems}}},}\ } (\bibinfo {year} {2021}),\ \Eprint
  {http://arxiv.org/abs/2101.10189} {arxiv:2101.10189 [math, stat]}
  \BibitemShut {NoStop}%
\bibitem [{\citenamefont {Hu}\ \emph {et~al.}(2020)\citenamefont {Hu},
  \citenamefont {Chen}, \citenamefont {Nair},\ and\ \citenamefont
  {Sudjianto}}]{huSurrogateLocallyInterpretableModels2020}%
  \BibitemOpen
  \bibfield  {author} {\bibinfo {author} {\bibfnamefont {L.}~\bibnamefont
  {Hu}}, \bibinfo {author} {\bibfnamefont {J.}~\bibnamefont {Chen}}, \bibinfo
  {author} {\bibfnamefont {V.~N.}\ \bibnamefont {Nair}}, \ and\ \bibinfo
  {author} {\bibfnamefont {A.}~\bibnamefont {Sudjianto}},\ }\href {\doibase
  10.48550/arXiv.2007.14528} {\enquote {\bibinfo {title} {Surrogate
  {{Locally-Interpretable Models}} with {{Supervised Machine Learning
  Algorithms}}},}\ } (\bibinfo {year} {2020}),\ \Eprint
  {http://arxiv.org/abs/2007.14528} {arxiv:2007.14528 [cs, stat]} \BibitemShut
  {NoStop}%
\bibitem [{\citenamefont {Popov}\ and\ \citenamefont
  {Sandu}(2021)}]{popovMultifidelityEnsembleKalman2021}%
  \BibitemOpen
  \bibfield  {author} {\bibinfo {author} {\bibfnamefont {A.~A.}\ \bibnamefont
  {Popov}}\ and\ \bibinfo {author} {\bibfnamefont {A.}~\bibnamefont {Sandu}},\
  }\href {\doibase 10.48550/arXiv.2102.13025} {\enquote {\bibinfo {title}
  {Multifidelity {{Ensemble Kalman Filtering Using Surrogate Models Defined}}
  by {{Physics-Informed Autoencoders}}},}\ } (\bibinfo {year} {2021}),\ \Eprint
  {http://arxiv.org/abs/2102.13025} {arxiv:2102.13025 [cs, math]} \BibitemShut
  {NoStop}%
\bibitem [{\citenamefont {Dave}\ \emph {et~al.}(2020)\citenamefont {Dave},
  \citenamefont {Wilson},\ and\ \citenamefont
  {Sun}}]{daveDeepSurrogateModels2020}%
  \BibitemOpen
  \bibfield  {author} {\bibinfo {author} {\bibfnamefont {A.~J.}\ \bibnamefont
  {Dave}}, \bibinfo {author} {\bibfnamefont {J.}~\bibnamefont {Wilson}}, \ and\
  \bibinfo {author} {\bibfnamefont {K.}~\bibnamefont {Sun}},\ }\href {\doibase
  10.48550/arXiv.2007.05435} {\enquote {\bibinfo {title} {Deep {{Surrogate
  Models}} for {{Multi-dimensional Regression}} of {{Reactor Power}}},}\ }
  (\bibinfo {year} {2020}),\ \Eprint {http://arxiv.org/abs/2007.05435}
  {arxiv:2007.05435 [physics]} \BibitemShut {NoStop}%
\bibitem [{\citenamefont {Wen}\ \emph {et~al.}(2020)\citenamefont {Wen},
  \citenamefont {Jiang},\ and\ \citenamefont
  {Fan}}]{wenRobustFreeformMetasurface2020}%
  \BibitemOpen
  \bibfield  {author} {\bibinfo {author} {\bibfnamefont {F.}~\bibnamefont
  {Wen}}, \bibinfo {author} {\bibfnamefont {J.}~\bibnamefont {Jiang}}, \ and\
  \bibinfo {author} {\bibfnamefont {J.~A.}\ \bibnamefont {Fan}},\ }\href
  {\doibase 10.1021/acsphotonics.0c00539} {\bibfield  {journal} {\bibinfo
  {journal} {ACS Photonics}\ }\textbf {\bibinfo {volume} {7}},\ \bibinfo
  {pages} {2098} (\bibinfo {year} {2020})},\ \Eprint
  {http://arxiv.org/abs/1911.13029} {arxiv:1911.13029} \BibitemShut {NoStop}%
\bibitem [{\citenamefont {Zeiler}\ and\ \citenamefont
  {Fergus}(2014)}]{zeilerVisualizingUnderstandingConvolutional2014}%
  \BibitemOpen
  \bibfield  {author} {\bibinfo {author} {\bibfnamefont {M.~D.}\ \bibnamefont
  {Zeiler}}\ and\ \bibinfo {author} {\bibfnamefont {R.}~\bibnamefont
  {Fergus}},\ }in\ \href {\doibase 10.1007/978-3-319-10590-1_53} {\emph
  {\bibinfo {booktitle} {Computer {{Vision}} \textendash{} {{ECCV}} 2014}}},\
  \bibinfo {series and number} {Lecture {{Notes}} in {{Computer Science}}},\
  \bibinfo {editor} {edited by\ \bibinfo {editor} {\bibfnamefont
  {D.}~\bibnamefont {Fleet}}, \bibinfo {editor} {\bibfnamefont
  {T.}~\bibnamefont {Pajdla}}, \bibinfo {editor} {\bibfnamefont
  {B.}~\bibnamefont {Schiele}}, \ and\ \bibinfo {editor} {\bibfnamefont
  {T.}~\bibnamefont {Tuytelaars}}}\ (\bibinfo  {publisher} {{Springer
  International Publishing}},\ \bibinfo {address} {{Cham}},\ \bibinfo {year}
  {2014})\ pp.\ \bibinfo {pages} {818--833}\BibitemShut {NoStop}%
\bibitem [{\citenamefont {Ronneberger}\ \emph {et~al.}(2015)\citenamefont
  {Ronneberger}, \citenamefont {Fischer},\ and\ \citenamefont
  {Brox}}]{ronnebergerUNetConvolutionalNetworks2015}%
  \BibitemOpen
  \bibfield  {author} {\bibinfo {author} {\bibfnamefont {O.}~\bibnamefont
  {Ronneberger}}, \bibinfo {author} {\bibfnamefont {P.}~\bibnamefont
  {Fischer}}, \ and\ \bibinfo {author} {\bibfnamefont {T.}~\bibnamefont
  {Brox}},\ }\href@noop {} {\bibfield  {journal} {\bibinfo  {journal}
  {arXiv:1505.04597 [cs]}\ } (\bibinfo {year} {2015})},\ \Eprint
  {http://arxiv.org/abs/1505.04597} {arxiv:1505.04597 [cs]} \BibitemShut
  {NoStop}%
\bibitem [{\citenamefont {Provost}\ \emph {et~al.}(1999)\citenamefont
  {Provost}, \citenamefont {Jensen},\ and\ \citenamefont
  {Oates}}]{provostEfficientProgressiveSampling1999}%
  \BibitemOpen
  \bibfield  {author} {\bibinfo {author} {\bibfnamefont {F.}~\bibnamefont
  {Provost}}, \bibinfo {author} {\bibfnamefont {D.}~\bibnamefont {Jensen}}, \
  and\ \bibinfo {author} {\bibfnamefont {T.}~\bibnamefont {Oates}},\ }in\ \href
  {\doibase 10.1145/312129.312188} {\emph {\bibinfo {booktitle} {Proceedings of
  the Fifth {{ACM SIGKDD}} International Conference on {{Knowledge}} Discovery
  and Data Mining}}},\ \bibinfo {series and number} {{{KDD}} '99}\ (\bibinfo
  {publisher} {{Association for Computing Machinery}},\ \bibinfo {address}
  {{New York, NY, USA}},\ \bibinfo {year} {1999})\ pp.\ \bibinfo {pages}
  {23--32}\BibitemShut {NoStop}%
\bibitem [{\citenamefont {Bierkens}\ \emph {et~al.}(2019)\citenamefont
  {Bierkens}, \citenamefont {Fearnhead},\ and\ \citenamefont
  {Roberts}}]{bierkensZigZagProcessSuperefficient2019}%
  \BibitemOpen
  \bibfield  {author} {\bibinfo {author} {\bibfnamefont {J.}~\bibnamefont
  {Bierkens}}, \bibinfo {author} {\bibfnamefont {P.}~\bibnamefont {Fearnhead}},
  \ and\ \bibinfo {author} {\bibfnamefont {G.}~\bibnamefont {Roberts}},\ }\href
  {\doibase 10.1214/18-AOS1715} {\bibfield  {journal} {\bibinfo  {journal} {The
  Annals of Statistics}\ }\textbf {\bibinfo {volume} {47}},\ \bibinfo {pages}
  {1288} (\bibinfo {year} {2019})}\BibitemShut {NoStop}%
\bibitem [{\citenamefont {Renardy}\ \emph {et~al.}(2021)\citenamefont
  {Renardy}, \citenamefont {Joslyn}, \citenamefont {Millar},\ and\
  \citenamefont {Kirschner}}]{renardySobolNotSobol2021}%
  \BibitemOpen
  \bibfield  {author} {\bibinfo {author} {\bibfnamefont {M.}~\bibnamefont
  {Renardy}}, \bibinfo {author} {\bibfnamefont {L.~R.}\ \bibnamefont {Joslyn}},
  \bibinfo {author} {\bibfnamefont {J.~A.}\ \bibnamefont {Millar}}, \ and\
  \bibinfo {author} {\bibfnamefont {D.~E.}\ \bibnamefont {Kirschner}},\ }\href
  {\doibase 10.1016/j.mbs.2021.108593} {\bibfield  {journal} {\bibinfo
  {journal} {Mathematical Biosciences}\ }\textbf {\bibinfo {volume} {337}},\
  \bibinfo {pages} {108593} (\bibinfo {year} {2021})}\BibitemShut {NoStop}%
\bibitem [{\citenamefont {Yeo}\ and\ \citenamefont
  {Johnson}(2000)}]{yeoNewFamilyPower2000}%
  \BibitemOpen
  \bibfield  {author} {\bibinfo {author} {\bibfnamefont {I.-K.}\ \bibnamefont
  {Yeo}}\ and\ \bibinfo {author} {\bibfnamefont {R.~A.}\ \bibnamefont
  {Johnson}},\ }\href@noop {} {\bibfield  {journal} {\bibinfo  {journal}
  {Biometrika}\ }\textbf {\bibinfo {volume} {87}},\ \bibinfo {pages} {954}
  (\bibinfo {year} {2000})},\ \Eprint {http://arxiv.org/abs/2673623} {2673623}
  \BibitemShut {NoStop}%
\bibitem [{\citenamefont
  {Karvanen}(2006)}]{karvanenEstimationQuantileMixtures2006}%
  \BibitemOpen
  \bibfield  {author} {\bibinfo {author} {\bibfnamefont {J.}~\bibnamefont
  {Karvanen}},\ }\href {\doibase 10.1016/j.csda.2005.09.014} {\bibfield
  {journal} {\bibinfo  {journal} {Computational Statistics \& Data Analysis}\
  }\textbf {\bibinfo {volume} {51}},\ \bibinfo {pages} {947} (\bibinfo {year}
  {2006})}\BibitemShut {NoStop}%
\bibitem [{\citenamefont {Khatib}\ \emph {et~al.}(2022)\citenamefont {Khatib},
  \citenamefont {Ren}, \citenamefont {Malof},\ and\ \citenamefont
  {Padilla}}]{khatibLearningPhysicsAllDielectric2022}%
  \BibitemOpen
  \bibfield  {author} {\bibinfo {author} {\bibfnamefont {O.}~\bibnamefont
  {Khatib}}, \bibinfo {author} {\bibfnamefont {S.}~\bibnamefont {Ren}},
  \bibinfo {author} {\bibfnamefont {J.}~\bibnamefont {Malof}}, \ and\ \bibinfo
  {author} {\bibfnamefont {W.~J.}\ \bibnamefont {Padilla}},\ }\href {\doibase
  10.1002/adom.202200097} {\bibfield  {journal} {\bibinfo  {journal} {Advanced
  Optical Materials}\ }\textbf {\bibinfo {volume} {10}},\ \bibinfo {pages}
  {2200097} (\bibinfo {year} {2022})}\BibitemShut {NoStop}%
\bibitem [{\citenamefont {Luo}\ \emph {et~al.}(2017)\citenamefont {Luo},
  \citenamefont {Li}, \citenamefont {Urtasun},\ and\ \citenamefont
  {Zemel}}]{luoUnderstandingEffectiveReceptive2017}%
  \BibitemOpen
  \bibfield  {author} {\bibinfo {author} {\bibfnamefont {W.}~\bibnamefont
  {Luo}}, \bibinfo {author} {\bibfnamefont {Y.}~\bibnamefont {Li}}, \bibinfo
  {author} {\bibfnamefont {R.}~\bibnamefont {Urtasun}}, \ and\ \bibinfo
  {author} {\bibfnamefont {R.}~\bibnamefont {Zemel}},\ }\href {\doibase
  10.48550/arXiv.1701.04128} {\enquote {\bibinfo {title} {Understanding the
  {{Effective Receptive Field}} in {{Deep Convolutional Neural Networks}}},}\ }
  (\bibinfo {year} {2017}),\ \Eprint {http://arxiv.org/abs/1701.04128}
  {arxiv:1701.04128 [cs]} \BibitemShut {NoStop}%
\bibitem [{\citenamefont {Srivastava}\ \emph {et~al.}(2014)\citenamefont
  {Srivastava}, \citenamefont {Hinton}, \citenamefont {Krizhevsky},
  \citenamefont {Sutskever},\ and\ \citenamefont
  {Salakhutdinov}}]{srivastavaDropoutSimpleWay2014}%
  \BibitemOpen
  \bibfield  {author} {\bibinfo {author} {\bibfnamefont {N.}~\bibnamefont
  {Srivastava}}, \bibinfo {author} {\bibfnamefont {G.}~\bibnamefont {Hinton}},
  \bibinfo {author} {\bibfnamefont {A.}~\bibnamefont {Krizhevsky}}, \bibinfo
  {author} {\bibfnamefont {I.}~\bibnamefont {Sutskever}}, \ and\ \bibinfo
  {author} {\bibfnamefont {R.}~\bibnamefont {Salakhutdinov}},\ }\href@noop {}
  {\bibfield  {journal} {\bibinfo  {journal} {Journal of Machine Learning
  Research}\ }\textbf {\bibinfo {volume} {15}},\ \bibinfo {pages} {1929}
  (\bibinfo {year} {2014})}\BibitemShut {NoStop}%
\bibitem [{\citenamefont {Belkin}\ \emph {et~al.}(2019)\citenamefont {Belkin},
  \citenamefont {Hsu}, \citenamefont {Ma},\ and\ \citenamefont
  {Mandal}}]{belkinReconcilingModernMachinelearning2019}%
  \BibitemOpen
  \bibfield  {author} {\bibinfo {author} {\bibfnamefont {M.}~\bibnamefont
  {Belkin}}, \bibinfo {author} {\bibfnamefont {D.}~\bibnamefont {Hsu}},
  \bibinfo {author} {\bibfnamefont {S.}~\bibnamefont {Ma}}, \ and\ \bibinfo
  {author} {\bibfnamefont {S.}~\bibnamefont {Mandal}},\ }\href {\doibase
  10.1073/pnas.1903070116} {\bibfield  {journal} {\bibinfo  {journal}
  {Proceedings of the National Academy of Sciences}\ }\textbf {\bibinfo
  {volume} {116}},\ \bibinfo {pages} {15849} (\bibinfo {year}
  {2019})}\BibitemShut {NoStop}%
\bibitem [{\citenamefont {Loog}\ \emph {et~al.}(2020)\citenamefont {Loog},
  \citenamefont {Viering}, \citenamefont {Mey}, \citenamefont {Krijthe},\ and\
  \citenamefont {Tax}}]{loogBriefPrehistoryDouble2020}%
  \BibitemOpen
  \bibfield  {author} {\bibinfo {author} {\bibfnamefont {M.}~\bibnamefont
  {Loog}}, \bibinfo {author} {\bibfnamefont {T.}~\bibnamefont {Viering}},
  \bibinfo {author} {\bibfnamefont {A.}~\bibnamefont {Mey}}, \bibinfo {author}
  {\bibfnamefont {J.~H.}\ \bibnamefont {Krijthe}}, \ and\ \bibinfo {author}
  {\bibfnamefont {D.~M.~J.}\ \bibnamefont {Tax}},\ }\href {\doibase
  10.1073/pnas.2001875117} {\bibfield  {journal} {\bibinfo  {journal}
  {Proceedings of the National Academy of Sciences}\ }\textbf {\bibinfo
  {volume} {117}},\ \bibinfo {pages} {10625} (\bibinfo {year}
  {2020})}\BibitemShut {NoStop}%
\bibitem [{\citenamefont {Nakkiran}\ \emph {et~al.}(2021)\citenamefont
  {Nakkiran}, \citenamefont {Kaplun}, \citenamefont {Bansal}, \citenamefont
  {Yang}, \citenamefont {Barak},\ and\ \citenamefont
  {Sutskever}}]{nakkiranDeepDoubleDescent2021}%
  \BibitemOpen
  \bibfield  {author} {\bibinfo {author} {\bibfnamefont {P.}~\bibnamefont
  {Nakkiran}}, \bibinfo {author} {\bibfnamefont {G.}~\bibnamefont {Kaplun}},
  \bibinfo {author} {\bibfnamefont {Y.}~\bibnamefont {Bansal}}, \bibinfo
  {author} {\bibfnamefont {T.}~\bibnamefont {Yang}}, \bibinfo {author}
  {\bibfnamefont {B.}~\bibnamefont {Barak}}, \ and\ \bibinfo {author}
  {\bibfnamefont {I.}~\bibnamefont {Sutskever}},\ }\href {\doibase
  10.1088/1742-5468/ac3a74} {\bibfield  {journal} {\bibinfo  {journal} {Journal
  of Statistical Mechanics: Theory and Experiment}\ }\textbf {\bibinfo {volume}
  {2021}},\ \bibinfo {pages} {124003} (\bibinfo {year} {2021})}\BibitemShut
  {NoStop}%
\bibitem [{\citenamefont {Schaeffer}\ \emph {et~al.}(2023)\citenamefont
  {Schaeffer}, \citenamefont {Khona}, \citenamefont {Robertson}, \citenamefont
  {Boopathy}, \citenamefont {Pistunova}, \citenamefont {Rocks}, \citenamefont
  {Fiete},\ and\ \citenamefont
  {Koyejo}}]{schaefferDoubleDescentDemystified2023}%
  \BibitemOpen
  \bibfield  {author} {\bibinfo {author} {\bibfnamefont {R.}~\bibnamefont
  {Schaeffer}}, \bibinfo {author} {\bibfnamefont {M.}~\bibnamefont {Khona}},
  \bibinfo {author} {\bibfnamefont {Z.}~\bibnamefont {Robertson}}, \bibinfo
  {author} {\bibfnamefont {A.}~\bibnamefont {Boopathy}}, \bibinfo {author}
  {\bibfnamefont {K.}~\bibnamefont {Pistunova}}, \bibinfo {author}
  {\bibfnamefont {J.~W.}\ \bibnamefont {Rocks}}, \bibinfo {author}
  {\bibfnamefont {I.~R.}\ \bibnamefont {Fiete}}, \ and\ \bibinfo {author}
  {\bibfnamefont {O.}~\bibnamefont {Koyejo}},\ }\href {\doibase
  10.48550/arXiv.2303.14151} {\enquote {\bibinfo {title} {Double {{Descent
  Demystified}}: {{Identifying}}, {{Interpreting}} \& {{Ablating}} the
  {{Sources}} of a {{Deep Learning Puzzle}}},}\ } (\bibinfo {year} {2023}),\
  \Eprint {http://arxiv.org/abs/2303.14151} {arxiv:2303.14151 [cs, stat]}
  \BibitemShut {NoStop}%
\bibitem [{\citenamefont {Bubeck}\ \emph {et~al.}(2023)\citenamefont {Bubeck},
  \citenamefont {Chandrasekaran}, \citenamefont {Eldan}, \citenamefont
  {Gehrke}, \citenamefont {Horvitz}, \citenamefont {Kamar}, \citenamefont
  {Lee}, \citenamefont {Lee}, \citenamefont {Li}, \citenamefont {Lundberg},
  \citenamefont {Nori}, \citenamefont {Palangi}, \citenamefont {Ribeiro},\ and\
  \citenamefont {Zhang}}]{bubeckSparksArtificialGeneral2023}%
  \BibitemOpen
  \bibfield  {author} {\bibinfo {author} {\bibfnamefont {S.}~\bibnamefont
  {Bubeck}}, \bibinfo {author} {\bibfnamefont {V.}~\bibnamefont
  {Chandrasekaran}}, \bibinfo {author} {\bibfnamefont {R.}~\bibnamefont
  {Eldan}}, \bibinfo {author} {\bibfnamefont {J.}~\bibnamefont {Gehrke}},
  \bibinfo {author} {\bibfnamefont {E.}~\bibnamefont {Horvitz}}, \bibinfo
  {author} {\bibfnamefont {E.}~\bibnamefont {Kamar}}, \bibinfo {author}
  {\bibfnamefont {P.}~\bibnamefont {Lee}}, \bibinfo {author} {\bibfnamefont
  {Y.~T.}\ \bibnamefont {Lee}}, \bibinfo {author} {\bibfnamefont
  {Y.}~\bibnamefont {Li}}, \bibinfo {author} {\bibfnamefont {S.}~\bibnamefont
  {Lundberg}}, \bibinfo {author} {\bibfnamefont {H.}~\bibnamefont {Nori}},
  \bibinfo {author} {\bibfnamefont {H.}~\bibnamefont {Palangi}}, \bibinfo
  {author} {\bibfnamefont {M.~T.}\ \bibnamefont {Ribeiro}}, \ and\ \bibinfo
  {author} {\bibfnamefont {Y.}~\bibnamefont {Zhang}},\ }\href {\doibase
  10.48550/arXiv.2303.12712} {\enquote {\bibinfo {title} {Sparks of
  {{Artificial General Intelligence}}: {{Early}} experiments with {{GPT-4}}},}\
  } (\bibinfo {year} {2023}),\ \Eprint {http://arxiv.org/abs/2303.12712}
  {arxiv:2303.12712 [cs]} \BibitemShut {NoStop}%
\bibitem [{\citenamefont {Bommasani}\ \emph {et~al.}(2022)\citenamefont
  {Bommasani}, \citenamefont {Hudson}, \citenamefont {Adeli}, \citenamefont
  {Altman}, \citenamefont {Arora}, \citenamefont {{von Arx}}, \citenamefont
  {Bernstein}, \citenamefont {Bohg}, \citenamefont {Bosselut}, \citenamefont
  {Brunskill}, \citenamefont {Brynjolfsson}, \citenamefont {Buch},
  \citenamefont {Card}, \citenamefont {Castellon}, \citenamefont {Chatterji},
  \citenamefont {Chen}, \citenamefont {Creel}, \citenamefont {Davis},
  \citenamefont {Demszky}, \citenamefont {Donahue}, \citenamefont {Doumbouya},
  \citenamefont {Durmus}, \citenamefont {Ermon}, \citenamefont {Etchemendy},
  \citenamefont {Ethayarajh}, \citenamefont {{Fei-Fei}}, \citenamefont {Finn},
  \citenamefont {Gale}, \citenamefont {Gillespie}, \citenamefont {Goel},
  \citenamefont {Goodman}, \citenamefont {Grossman}, \citenamefont {Guha},
  \citenamefont {Hashimoto}, \citenamefont {Henderson}, \citenamefont {Hewitt},
  \citenamefont {Ho}, \citenamefont {Hong}, \citenamefont {Hsu}, \citenamefont
  {Huang}, \citenamefont {Icard}, \citenamefont {Jain}, \citenamefont
  {Jurafsky}, \citenamefont {Kalluri}, \citenamefont {Karamcheti},
  \citenamefont {Keeling}, \citenamefont {Khani}, \citenamefont {Khattab},
  \citenamefont {Koh}, \citenamefont {Krass}, \citenamefont {Krishna},
  \citenamefont {Kuditipudi}, \citenamefont {Kumar}, \citenamefont {Ladhak},
  \citenamefont {Lee}, \citenamefont {Lee}, \citenamefont {Leskovec},
  \citenamefont {Levent}, \citenamefont {Li}, \citenamefont {Li}, \citenamefont
  {Ma}, \citenamefont {Malik}, \citenamefont {Manning}, \citenamefont
  {Mirchandani}, \citenamefont {Mitchell}, \citenamefont {Munyikwa},
  \citenamefont {Nair}, \citenamefont {Narayan}, \citenamefont {Narayanan},
  \citenamefont {Newman}, \citenamefont {Nie}, \citenamefont {Niebles},
  \citenamefont {Nilforoshan}, \citenamefont {Nyarko}, \citenamefont {Ogut},
  \citenamefont {Orr}, \citenamefont {Papadimitriou}, \citenamefont {Park},
  \citenamefont {Piech}, \citenamefont {Portelance}, \citenamefont {Potts},
  \citenamefont {Raghunathan}, \citenamefont {Reich}, \citenamefont {Ren},
  \citenamefont {Rong}, \citenamefont {Roohani}, \citenamefont {Ruiz},
  \citenamefont {Ryan}, \citenamefont {R{\'e}}, \citenamefont {Sadigh},
  \citenamefont {Sagawa}, \citenamefont {Santhanam}, \citenamefont {Shih},
  \citenamefont {Srinivasan}, \citenamefont {Tamkin}, \citenamefont {Taori},
  \citenamefont {Thomas}, \citenamefont {Tram{\`e}r}, \citenamefont {Wang},
  \citenamefont {Wang}, \citenamefont {Wu}, \citenamefont {Wu}, \citenamefont
  {Wu}, \citenamefont {Xie}, \citenamefont {Yasunaga}, \citenamefont {You},
  \citenamefont {Zaharia}, \citenamefont {Zhang}, \citenamefont {Zhang},
  \citenamefont {Zhang}, \citenamefont {Zhang}, \citenamefont {Zheng},
  \citenamefont {Zhou},\ and\ \citenamefont
  {Liang}}]{bommasaniOpportunitiesRisksFoundation2022}%
  \BibitemOpen
  \bibfield  {author} {\bibinfo {author} {\bibfnamefont {R.}~\bibnamefont
  {Bommasani}}, \bibinfo {author} {\bibfnamefont {D.~A.}\ \bibnamefont
  {Hudson}}, \bibinfo {author} {\bibfnamefont {E.}~\bibnamefont {Adeli}},
  \bibinfo {author} {\bibfnamefont {R.}~\bibnamefont {Altman}}, \bibinfo
  {author} {\bibfnamefont {S.}~\bibnamefont {Arora}}, \bibinfo {author}
  {\bibfnamefont {S.}~\bibnamefont {{von Arx}}}, \bibinfo {author}
  {\bibfnamefont {M.~S.}\ \bibnamefont {Bernstein}}, \bibinfo {author}
  {\bibfnamefont {J.}~\bibnamefont {Bohg}}, \bibinfo {author} {\bibfnamefont
  {A.}~\bibnamefont {Bosselut}}, \bibinfo {author} {\bibfnamefont
  {E.}~\bibnamefont {Brunskill}}, \bibinfo {author} {\bibfnamefont
  {E.}~\bibnamefont {Brynjolfsson}}, \bibinfo {author} {\bibfnamefont
  {S.}~\bibnamefont {Buch}}, \bibinfo {author} {\bibfnamefont {D.}~\bibnamefont
  {Card}}, \bibinfo {author} {\bibfnamefont {R.}~\bibnamefont {Castellon}},
  \bibinfo {author} {\bibfnamefont {N.}~\bibnamefont {Chatterji}}, \bibinfo
  {author} {\bibfnamefont {A.}~\bibnamefont {Chen}}, \bibinfo {author}
  {\bibfnamefont {K.}~\bibnamefont {Creel}}, \bibinfo {author} {\bibfnamefont
  {J.~Q.}\ \bibnamefont {Davis}}, \bibinfo {author} {\bibfnamefont
  {D.}~\bibnamefont {Demszky}}, \bibinfo {author} {\bibfnamefont
  {C.}~\bibnamefont {Donahue}}, \bibinfo {author} {\bibfnamefont
  {M.}~\bibnamefont {Doumbouya}}, \bibinfo {author} {\bibfnamefont
  {E.}~\bibnamefont {Durmus}}, \bibinfo {author} {\bibfnamefont
  {S.}~\bibnamefont {Ermon}}, \bibinfo {author} {\bibfnamefont
  {J.}~\bibnamefont {Etchemendy}}, \bibinfo {author} {\bibfnamefont
  {K.}~\bibnamefont {Ethayarajh}}, \bibinfo {author} {\bibfnamefont
  {L.}~\bibnamefont {{Fei-Fei}}}, \bibinfo {author} {\bibfnamefont
  {C.}~\bibnamefont {Finn}}, \bibinfo {author} {\bibfnamefont {T.}~\bibnamefont
  {Gale}}, \bibinfo {author} {\bibfnamefont {L.}~\bibnamefont {Gillespie}},
  \bibinfo {author} {\bibfnamefont {K.}~\bibnamefont {Goel}}, \bibinfo {author}
  {\bibfnamefont {N.}~\bibnamefont {Goodman}}, \bibinfo {author} {\bibfnamefont
  {S.}~\bibnamefont {Grossman}}, \bibinfo {author} {\bibfnamefont
  {N.}~\bibnamefont {Guha}}, \bibinfo {author} {\bibfnamefont {T.}~\bibnamefont
  {Hashimoto}}, \bibinfo {author} {\bibfnamefont {P.}~\bibnamefont
  {Henderson}}, \bibinfo {author} {\bibfnamefont {J.}~\bibnamefont {Hewitt}},
  \bibinfo {author} {\bibfnamefont {D.~E.}\ \bibnamefont {Ho}}, \bibinfo
  {author} {\bibfnamefont {J.}~\bibnamefont {Hong}}, \bibinfo {author}
  {\bibfnamefont {K.}~\bibnamefont {Hsu}}, \bibinfo {author} {\bibfnamefont
  {J.}~\bibnamefont {Huang}}, \bibinfo {author} {\bibfnamefont
  {T.}~\bibnamefont {Icard}}, \bibinfo {author} {\bibfnamefont
  {S.}~\bibnamefont {Jain}}, \bibinfo {author} {\bibfnamefont {D.}~\bibnamefont
  {Jurafsky}}, \bibinfo {author} {\bibfnamefont {P.}~\bibnamefont {Kalluri}},
  \bibinfo {author} {\bibfnamefont {S.}~\bibnamefont {Karamcheti}}, \bibinfo
  {author} {\bibfnamefont {G.}~\bibnamefont {Keeling}}, \bibinfo {author}
  {\bibfnamefont {F.}~\bibnamefont {Khani}}, \bibinfo {author} {\bibfnamefont
  {O.}~\bibnamefont {Khattab}}, \bibinfo {author} {\bibfnamefont {P.~W.}\
  \bibnamefont {Koh}}, \bibinfo {author} {\bibfnamefont {M.}~\bibnamefont
  {Krass}}, \bibinfo {author} {\bibfnamefont {R.}~\bibnamefont {Krishna}},
  \bibinfo {author} {\bibfnamefont {R.}~\bibnamefont {Kuditipudi}}, \bibinfo
  {author} {\bibfnamefont {A.}~\bibnamefont {Kumar}}, \bibinfo {author}
  {\bibfnamefont {F.}~\bibnamefont {Ladhak}}, \bibinfo {author} {\bibfnamefont
  {M.}~\bibnamefont {Lee}}, \bibinfo {author} {\bibfnamefont {T.}~\bibnamefont
  {Lee}}, \bibinfo {author} {\bibfnamefont {J.}~\bibnamefont {Leskovec}},
  \bibinfo {author} {\bibfnamefont {I.}~\bibnamefont {Levent}}, \bibinfo
  {author} {\bibfnamefont {X.~L.}\ \bibnamefont {Li}}, \bibinfo {author}
  {\bibfnamefont {X.}~\bibnamefont {Li}}, \bibinfo {author} {\bibfnamefont
  {T.}~\bibnamefont {Ma}}, \bibinfo {author} {\bibfnamefont {A.}~\bibnamefont
  {Malik}}, \bibinfo {author} {\bibfnamefont {C.~D.}\ \bibnamefont {Manning}},
  \bibinfo {author} {\bibfnamefont {S.}~\bibnamefont {Mirchandani}}, \bibinfo
  {author} {\bibfnamefont {E.}~\bibnamefont {Mitchell}}, \bibinfo {author}
  {\bibfnamefont {Z.}~\bibnamefont {Munyikwa}}, \bibinfo {author}
  {\bibfnamefont {S.}~\bibnamefont {Nair}}, \bibinfo {author} {\bibfnamefont
  {A.}~\bibnamefont {Narayan}}, \bibinfo {author} {\bibfnamefont
  {D.}~\bibnamefont {Narayanan}}, \bibinfo {author} {\bibfnamefont
  {B.}~\bibnamefont {Newman}}, \bibinfo {author} {\bibfnamefont
  {A.}~\bibnamefont {Nie}}, \bibinfo {author} {\bibfnamefont {J.~C.}\
  \bibnamefont {Niebles}}, \bibinfo {author} {\bibfnamefont {H.}~\bibnamefont
  {Nilforoshan}}, \bibinfo {author} {\bibfnamefont {J.}~\bibnamefont {Nyarko}},
  \bibinfo {author} {\bibfnamefont {G.}~\bibnamefont {Ogut}}, \bibinfo {author}
  {\bibfnamefont {L.}~\bibnamefont {Orr}}, \bibinfo {author} {\bibfnamefont
  {I.}~\bibnamefont {Papadimitriou}}, \bibinfo {author} {\bibfnamefont {J.~S.}\
  \bibnamefont {Park}}, \bibinfo {author} {\bibfnamefont {C.}~\bibnamefont
  {Piech}}, \bibinfo {author} {\bibfnamefont {E.}~\bibnamefont {Portelance}},
  \bibinfo {author} {\bibfnamefont {C.}~\bibnamefont {Potts}}, \bibinfo
  {author} {\bibfnamefont {A.}~\bibnamefont {Raghunathan}}, \bibinfo {author}
  {\bibfnamefont {R.}~\bibnamefont {Reich}}, \bibinfo {author} {\bibfnamefont
  {H.}~\bibnamefont {Ren}}, \bibinfo {author} {\bibfnamefont {F.}~\bibnamefont
  {Rong}}, \bibinfo {author} {\bibfnamefont {Y.}~\bibnamefont {Roohani}},
  \bibinfo {author} {\bibfnamefont {C.}~\bibnamefont {Ruiz}}, \bibinfo {author}
  {\bibfnamefont {J.}~\bibnamefont {Ryan}}, \bibinfo {author} {\bibfnamefont
  {C.}~\bibnamefont {R{\'e}}}, \bibinfo {author} {\bibfnamefont
  {D.}~\bibnamefont {Sadigh}}, \bibinfo {author} {\bibfnamefont
  {S.}~\bibnamefont {Sagawa}}, \bibinfo {author} {\bibfnamefont
  {K.}~\bibnamefont {Santhanam}}, \bibinfo {author} {\bibfnamefont
  {A.}~\bibnamefont {Shih}}, \bibinfo {author} {\bibfnamefont {K.}~\bibnamefont
  {Srinivasan}}, \bibinfo {author} {\bibfnamefont {A.}~\bibnamefont {Tamkin}},
  \bibinfo {author} {\bibfnamefont {R.}~\bibnamefont {Taori}}, \bibinfo
  {author} {\bibfnamefont {A.~W.}\ \bibnamefont {Thomas}}, \bibinfo {author}
  {\bibfnamefont {F.}~\bibnamefont {Tram{\`e}r}}, \bibinfo {author}
  {\bibfnamefont {R.~E.}\ \bibnamefont {Wang}}, \bibinfo {author}
  {\bibfnamefont {W.}~\bibnamefont {Wang}}, \bibinfo {author} {\bibfnamefont
  {B.}~\bibnamefont {Wu}}, \bibinfo {author} {\bibfnamefont {J.}~\bibnamefont
  {Wu}}, \bibinfo {author} {\bibfnamefont {Y.}~\bibnamefont {Wu}}, \bibinfo
  {author} {\bibfnamefont {S.~M.}\ \bibnamefont {Xie}}, \bibinfo {author}
  {\bibfnamefont {M.}~\bibnamefont {Yasunaga}}, \bibinfo {author}
  {\bibfnamefont {J.}~\bibnamefont {You}}, \bibinfo {author} {\bibfnamefont
  {M.}~\bibnamefont {Zaharia}}, \bibinfo {author} {\bibfnamefont
  {M.}~\bibnamefont {Zhang}}, \bibinfo {author} {\bibfnamefont
  {T.}~\bibnamefont {Zhang}}, \bibinfo {author} {\bibfnamefont
  {X.}~\bibnamefont {Zhang}}, \bibinfo {author} {\bibfnamefont
  {Y.}~\bibnamefont {Zhang}}, \bibinfo {author} {\bibfnamefont
  {L.}~\bibnamefont {Zheng}}, \bibinfo {author} {\bibfnamefont
  {K.}~\bibnamefont {Zhou}}, \ and\ \bibinfo {author} {\bibfnamefont
  {P.}~\bibnamefont {Liang}},\ }\href {\doibase 10.48550/arXiv.2108.07258}
  {\enquote {\bibinfo {title} {On the {{Opportunities}} and {{Risks}} of
  {{Foundation Models}}},}\ } (\bibinfo {year} {2022}),\ \Eprint
  {http://arxiv.org/abs/2108.07258} {arxiv:2108.07258 [cs]} \BibitemShut
  {NoStop}%
\bibitem [{\citenamefont {Ioffe}\ and\ \citenamefont
  {Szegedy}(2015)}]{ioffeBatchNormalizationAccelerating2015}%
  \BibitemOpen
  \bibfield  {author} {\bibinfo {author} {\bibfnamefont {S.}~\bibnamefont
  {Ioffe}}\ and\ \bibinfo {author} {\bibfnamefont {C.}~\bibnamefont
  {Szegedy}},\ }\href@noop {} {\bibfield  {journal} {\bibinfo  {journal}
  {arXiv:1502.03167 [cs]}\ } (\bibinfo {year} {2015})},\ \Eprint
  {http://arxiv.org/abs/1502.03167} {arxiv:1502.03167 [cs]} \BibitemShut
  {NoStop}%
\bibitem [{\citenamefont {Mianjy}\ \emph {et~al.}(2018)\citenamefont {Mianjy},
  \citenamefont {Arora},\ and\ \citenamefont
  {Vidal}}]{mianjyImplicitBiasDropout2018}%
  \BibitemOpen
  \bibfield  {author} {\bibinfo {author} {\bibfnamefont {P.}~\bibnamefont
  {Mianjy}}, \bibinfo {author} {\bibfnamefont {R.}~\bibnamefont {Arora}}, \
  and\ \bibinfo {author} {\bibfnamefont {R.}~\bibnamefont {Vidal}},\ }\href
  {\doibase 10.48550/arXiv.1806.09777} {\enquote {\bibinfo {title} {On the
  {{Implicit Bias}} of {{Dropout}}},}\ } (\bibinfo {year} {2018}),\ \Eprint
  {http://arxiv.org/abs/1806.09777} {arxiv:1806.09777 [cs, stat]} \BibitemShut
  {NoStop}%
\bibitem [{\citenamefont {Li}\ \emph {et~al.}(2018)\citenamefont {Li},
  \citenamefont {Chen}, \citenamefont {Hu},\ and\ \citenamefont
  {Yang}}]{liUnderstandingDisharmonyDropout2018}%
  \BibitemOpen
  \bibfield  {author} {\bibinfo {author} {\bibfnamefont {X.}~\bibnamefont
  {Li}}, \bibinfo {author} {\bibfnamefont {S.}~\bibnamefont {Chen}}, \bibinfo
  {author} {\bibfnamefont {X.}~\bibnamefont {Hu}}, \ and\ \bibinfo {author}
  {\bibfnamefont {J.}~\bibnamefont {Yang}},\ }\href {\doibase
  10.48550/arXiv.1801.05134} {\enquote {\bibinfo {title} {Understanding the
  {{Disharmony}} between {{Dropout}} and {{Batch Normalization}} by {{Variance
  Shift}}},}\ } (\bibinfo {year} {2018}),\ \Eprint
  {http://arxiv.org/abs/1801.05134} {arxiv:1801.05134 [cs, stat]} \BibitemShut
  {NoStop}%
\bibitem [{\citenamefont {Brock}\ \emph {et~al.}(2021)\citenamefont {Brock},
  \citenamefont {De},\ and\ \citenamefont
  {Smith}}]{brockCharacterizingSignalPropagation2021}%
  \BibitemOpen
  \bibfield  {author} {\bibinfo {author} {\bibfnamefont {A.}~\bibnamefont
  {Brock}}, \bibinfo {author} {\bibfnamefont {S.}~\bibnamefont {De}}, \ and\
  \bibinfo {author} {\bibfnamefont {S.~L.}\ \bibnamefont {Smith}},\ }\href
  {\doibase 10.48550/arXiv.2101.08692} {\enquote {\bibinfo {title}
  {Characterizing signal propagation to close the performance gap in
  unnormalized {{ResNets}}},}\ } (\bibinfo {year} {2021}),\ \Eprint
  {http://arxiv.org/abs/2101.08692} {arxiv:2101.08692 [cs, stat]} \BibitemShut
  {NoStop}%
\bibitem [{\citenamefont {Lian}\ and\ \citenamefont
  {Liu}(2019)}]{lianRevisitBatchNormalization2019}%
  \BibitemOpen
  \bibfield  {author} {\bibinfo {author} {\bibfnamefont {X.}~\bibnamefont
  {Lian}}\ and\ \bibinfo {author} {\bibfnamefont {J.}~\bibnamefont {Liu}},\
  }in\ \href@noop {} {\emph {\bibinfo {booktitle} {Proceedings of the
  {{Twenty-Second International Conference}} on {{Artificial Intelligence}} and
  {{Statistics}}}}}\ (\bibinfo  {publisher} {{PMLR}},\ \bibinfo {year} {2019})\
  pp.\ \bibinfo {pages} {3254--3263}\BibitemShut {NoStop}%
\bibitem [{\citenamefont {{\"O}zg{\"u}r}\ and\ \citenamefont
  {Nar}(2020)}]{ozgurEffectDropoutLayer2020}%
  \BibitemOpen
  \bibfield  {author} {\bibinfo {author} {\bibfnamefont {A.}~\bibnamefont
  {{\"O}zg{\"u}r}}\ and\ \bibinfo {author} {\bibfnamefont {F.}~\bibnamefont
  {Nar}},\ }in\ \href {\doibase 10.1109/SIU49456.2020.9302054} {\emph {\bibinfo
  {booktitle} {2020 28th {{Signal Processing}} and {{Communications
  Applications Conference}} ({{SIU}})}}}\ (\bibinfo {year} {2020})\ pp.\
  \bibinfo {pages} {1--4}\BibitemShut {NoStop}%
\bibitem [{\citenamefont {Wu}\ and\ \citenamefont
  {Johnson}(2021)}]{wuRethinkingBatchBatchNorm2021}%
  \BibitemOpen
  \bibfield  {author} {\bibinfo {author} {\bibfnamefont {Y.}~\bibnamefont
  {Wu}}\ and\ \bibinfo {author} {\bibfnamefont {J.}~\bibnamefont {Johnson}},\
  }\href {\doibase 10.48550/arXiv.2105.07576} {\enquote {\bibinfo {title}
  {Rethinking "{{Batch}}" in {{BatchNorm}}},}\ } (\bibinfo {year} {2021}),\
  \Eprint {http://arxiv.org/abs/2105.07576} {arxiv:2105.07576 [cs]}
  \BibitemShut {NoStop}%
\bibitem [{\citenamefont {LeCun}\ \emph {et~al.}(1998)\citenamefont {LeCun},
  \citenamefont {Bottou}, \citenamefont {Orr},\ and\ \citenamefont
  {M{\"u}ller}}]{lecunEfficientBackProp1998}%
  \BibitemOpen
  \bibfield  {author} {\bibinfo {author} {\bibfnamefont {Y.}~\bibnamefont
  {LeCun}}, \bibinfo {author} {\bibfnamefont {L.}~\bibnamefont {Bottou}},
  \bibinfo {author} {\bibfnamefont {G.~B.}\ \bibnamefont {Orr}}, \ and\
  \bibinfo {author} {\bibfnamefont {K.~R.}\ \bibnamefont {M{\"u}ller}},\ }in\
  \href {\doibase 10.1007/3-540-49430-8_2} {\emph {\bibinfo {booktitle} {Neural
  {{Networks}}: {{Tricks}} of the {{Trade}}}}},\ \bibinfo {series and number}
  {Lecture {{Notes}} in {{Computer Science}}},\ \bibinfo {editor} {edited by\
  \bibinfo {editor} {\bibfnamefont {G.~B.}\ \bibnamefont {Orr}}\ and\ \bibinfo
  {editor} {\bibfnamefont {K.-R.}\ \bibnamefont {M{\"u}ller}}}\ (\bibinfo
  {publisher} {{Springer}},\ \bibinfo {address} {{Berlin, Heidelberg}},\
  \bibinfo {year} {1998})\ pp.\ \bibinfo {pages} {9--50}\BibitemShut {NoStop}%
\bibitem [{\citenamefont {Keskar}\ \emph {et~al.}(2017)\citenamefont {Keskar},
  \citenamefont {Mudigere}, \citenamefont {Nocedal}, \citenamefont
  {Smelyanskiy},\ and\ \citenamefont
  {Tang}}]{keskarLargeBatchTrainingDeep2017}%
  \BibitemOpen
  \bibfield  {author} {\bibinfo {author} {\bibfnamefont {N.~S.}\ \bibnamefont
  {Keskar}}, \bibinfo {author} {\bibfnamefont {D.}~\bibnamefont {Mudigere}},
  \bibinfo {author} {\bibfnamefont {J.}~\bibnamefont {Nocedal}}, \bibinfo
  {author} {\bibfnamefont {M.}~\bibnamefont {Smelyanskiy}}, \ and\ \bibinfo
  {author} {\bibfnamefont {P.~T.~P.}\ \bibnamefont {Tang}},\ }\href {\doibase
  10.48550/arXiv.1609.04836} {\enquote {\bibinfo {title} {On {{Large-Batch
  Training}} for {{Deep Learning}}: {{Generalization Gap}} and {{Sharp
  Minima}}},}\ } (\bibinfo {year} {2017}),\ \Eprint
  {http://arxiv.org/abs/1609.04836} {arxiv:1609.04836 [cs, math]} \BibitemShut
  {NoStop}%
\bibitem [{\citenamefont {Masters}\ and\ \citenamefont
  {Luschi}(2018)}]{mastersRevisitingSmallBatch2018}%
  \BibitemOpen
  \bibfield  {author} {\bibinfo {author} {\bibfnamefont {D.}~\bibnamefont
  {Masters}}\ and\ \bibinfo {author} {\bibfnamefont {C.}~\bibnamefont
  {Luschi}},\ }\href {\doibase 10.48550/arXiv.1804.07612} {\enquote {\bibinfo
  {title} {Revisiting {{Small Batch Training}} for {{Deep Neural Networks}}},}\
  } (\bibinfo {year} {2018}),\ \Eprint {http://arxiv.org/abs/1804.07612}
  {arxiv:1804.07612 [cs, stat]} \BibitemShut {NoStop}%
\bibitem [{\citenamefont {Smith}\ \emph {et~al.}(2018)\citenamefont {Smith},
  \citenamefont {Kindermans}, \citenamefont {Ying},\ and\ \citenamefont
  {Le}}]{smithDonDecayLearning2018}%
  \BibitemOpen
  \bibfield  {author} {\bibinfo {author} {\bibfnamefont {S.~L.}\ \bibnamefont
  {Smith}}, \bibinfo {author} {\bibfnamefont {P.-J.}\ \bibnamefont
  {Kindermans}}, \bibinfo {author} {\bibfnamefont {C.}~\bibnamefont {Ying}}, \
  and\ \bibinfo {author} {\bibfnamefont {Q.~V.}\ \bibnamefont {Le}},\
  }\href@noop {} {\bibfield  {journal} {\bibinfo  {journal} {arXiv:1711.00489
  [cs, stat]}\ } (\bibinfo {year} {2018})},\ \Eprint
  {http://arxiv.org/abs/1711.00489} {arxiv:1711.00489 [cs, stat]} \BibitemShut
  {NoStop}%
\bibitem [{\citenamefont {Fournier}\ and\ \citenamefont
  {Aloise}(2019)}]{fournierEmpiricalComparisonAutoencoders2019}%
  \BibitemOpen
  \bibfield  {author} {\bibinfo {author} {\bibfnamefont {Q.}~\bibnamefont
  {Fournier}}\ and\ \bibinfo {author} {\bibfnamefont {D.}~\bibnamefont
  {Aloise}},\ }in\ \href {\doibase 10.1109/AIKE.2019.00044} {\emph {\bibinfo
  {booktitle} {2019 {{IEEE Second International Conference}} on {{Artificial
  Intelligence}} and {{Knowledge Engineering}} ({{AIKE}})}}}\ (\bibinfo {year}
  {2019})\ pp.\ \bibinfo {pages} {211--214}\BibitemShut {NoStop}%
\bibitem [{\citenamefont {Schubert}\ \emph {et~al.}(2017)\citenamefont
  {Schubert}, \citenamefont {Sander}, \citenamefont {Ester}, \citenamefont
  {Kriegel},\ and\ \citenamefont {Xu}}]{schubertDBSCANRevisitedRevisited2017}%
  \BibitemOpen
  \bibfield  {author} {\bibinfo {author} {\bibfnamefont {E.}~\bibnamefont
  {Schubert}}, \bibinfo {author} {\bibfnamefont {J.}~\bibnamefont {Sander}},
  \bibinfo {author} {\bibfnamefont {M.}~\bibnamefont {Ester}}, \bibinfo
  {author} {\bibfnamefont {H.~P.}\ \bibnamefont {Kriegel}}, \ and\ \bibinfo
  {author} {\bibfnamefont {X.}~\bibnamefont {Xu}},\ }\href {\doibase
  10.1145/3068335} {\bibfield  {journal} {\bibinfo  {journal} {ACM Transactions
  on Database Systems}\ }\textbf {\bibinfo {volume} {42}},\ \bibinfo {pages}
  {19:1} (\bibinfo {year} {2017})}\BibitemShut {NoStop}%
\bibitem [{\citenamefont {{van der Maaten}}\ and\ \citenamefont
  {Hinton}(2008)}]{vandermaatenVisualizingHighDimensionalData2008}%
  \BibitemOpen
  \bibfield  {author} {\bibinfo {author} {\bibfnamefont {L.}~\bibnamefont {{van
  der Maaten}}}\ and\ \bibinfo {author} {\bibfnamefont {G.}~\bibnamefont
  {Hinton}},\ }\href@noop {} {\bibfield  {journal} {\bibinfo  {journal}
  {Journal of Machine Learning Research}\ }\textbf {\bibinfo {volume} {9}},\
  \bibinfo {pages} {2579} (\bibinfo {year} {2008})}\BibitemShut {NoStop}%
\bibitem [{\citenamefont {Cortes}\ and\ \citenamefont
  {Vapnik}(1995)}]{cortesSupportvectorNetworks1995}%
  \BibitemOpen
  \bibfield  {author} {\bibinfo {author} {\bibfnamefont {C.}~\bibnamefont
  {Cortes}}\ and\ \bibinfo {author} {\bibfnamefont {V.}~\bibnamefont
  {Vapnik}},\ }\href {\doibase 10.1007/BF00994018} {\bibfield  {journal}
  {\bibinfo  {journal} {Machine Learning}\ }\textbf {\bibinfo {volume} {20}},\
  \bibinfo {pages} {273} (\bibinfo {year} {1995})}\BibitemShut {NoStop}%
\bibitem [{\citenamefont {Breiman}(2001)}]{breimanRandomForests2001}%
  \BibitemOpen
  \bibfield  {author} {\bibinfo {author} {\bibfnamefont {L.}~\bibnamefont
  {Breiman}},\ }\href {\doibase 10.1023/A:1010933404324} {\bibfield  {journal}
  {\bibinfo  {journal} {Machine Learning}\ }\textbf {\bibinfo {volume} {45}},\
  \bibinfo {pages} {5} (\bibinfo {year} {2001})}\BibitemShut {NoStop}%
\bibitem [{\citenamefont {Pedregosa}\ \emph {et~al.}(2011)\citenamefont
  {Pedregosa}, \citenamefont {Varoquaux}, \citenamefont {Gramfort},
  \citenamefont {Michel}, \citenamefont {Thirion}, \citenamefont {Grisel},
  \citenamefont {Blondel}, \citenamefont {Prettenhofer}, \citenamefont {Weiss},
  \citenamefont {Dubourg}, \citenamefont {Vanderplas}, \citenamefont {Passos},
  \citenamefont {Cournapeau}, \citenamefont {Brucher}, \citenamefont {Perrot},\
  and\ \citenamefont {Duchesnay}}]{pedregosaScikitlearnMachineLearning2011}%
  \BibitemOpen
  \bibfield  {author} {\bibinfo {author} {\bibfnamefont {F.}~\bibnamefont
  {Pedregosa}}, \bibinfo {author} {\bibfnamefont {G.}~\bibnamefont
  {Varoquaux}}, \bibinfo {author} {\bibfnamefont {A.}~\bibnamefont {Gramfort}},
  \bibinfo {author} {\bibfnamefont {V.}~\bibnamefont {Michel}}, \bibinfo
  {author} {\bibfnamefont {B.}~\bibnamefont {Thirion}}, \bibinfo {author}
  {\bibfnamefont {O.}~\bibnamefont {Grisel}}, \bibinfo {author} {\bibfnamefont
  {M.}~\bibnamefont {Blondel}}, \bibinfo {author} {\bibfnamefont
  {P.}~\bibnamefont {Prettenhofer}}, \bibinfo {author} {\bibfnamefont
  {R.}~\bibnamefont {Weiss}}, \bibinfo {author} {\bibfnamefont
  {V.}~\bibnamefont {Dubourg}}, \bibinfo {author} {\bibfnamefont
  {J.}~\bibnamefont {Vanderplas}}, \bibinfo {author} {\bibfnamefont
  {A.}~\bibnamefont {Passos}}, \bibinfo {author} {\bibfnamefont
  {D.}~\bibnamefont {Cournapeau}}, \bibinfo {author} {\bibfnamefont
  {M.}~\bibnamefont {Brucher}}, \bibinfo {author} {\bibfnamefont
  {M.}~\bibnamefont {Perrot}}, \ and\ \bibinfo {author} {\bibfnamefont
  {{\'E}.}~\bibnamefont {Duchesnay}},\ }\href@noop {} {\bibfield  {journal}
  {\bibinfo  {journal} {Journal of Machine Learning Research}\ }\textbf
  {\bibinfo {volume} {12}},\ \bibinfo {pages} {2825} (\bibinfo {year}
  {2011})}\BibitemShut {NoStop}%
\bibitem [{\citenamefont {Hegde}(2021)}]{hegdeSampleefficientDeepLearning2021}%
  \BibitemOpen
  \bibfield  {author} {\bibinfo {author} {\bibfnamefont {R.}~\bibnamefont
  {Hegde}},\ }\href {\doibase 10.1364/OSAC.420977} {\bibfield  {journal}
  {\bibinfo  {journal} {OSA Continuum}\ }\textbf {\bibinfo {volume} {4}},\
  \bibinfo {pages} {1019} (\bibinfo {year} {2021})}\BibitemShut {NoStop}%
\bibitem [{\citenamefont {Majorel}\ \emph {et~al.}(2022)\citenamefont
  {Majorel}, \citenamefont {Girard}, \citenamefont {Arbouet}, \citenamefont
  {Muskens},\ and\ \citenamefont {Wiecha}}]{majorelDeepLearningEnabled2022}%
  \BibitemOpen
  \bibfield  {author} {\bibinfo {author} {\bibfnamefont {C.}~\bibnamefont
  {Majorel}}, \bibinfo {author} {\bibfnamefont {C.}~\bibnamefont {Girard}},
  \bibinfo {author} {\bibfnamefont {A.}~\bibnamefont {Arbouet}}, \bibinfo
  {author} {\bibfnamefont {O.~L.}\ \bibnamefont {Muskens}}, \ and\ \bibinfo
  {author} {\bibfnamefont {P.~R.}\ \bibnamefont {Wiecha}},\ }\href {\doibase
  10.1021/acsphotonics.1c01556} {\bibfield  {journal} {\bibinfo  {journal} {ACS
  Photonics}\ }\textbf {\bibinfo {volume} {9}},\ \bibinfo {pages} {575}
  (\bibinfo {year} {2022})},\ \Eprint {http://arxiv.org/abs/2110.02109}
  {arxiv:2110.02109} \BibitemShut {NoStop}%
\bibitem [{\citenamefont {Liu}\ \emph {et~al.}(2020{\natexlab{a}})\citenamefont
  {Liu}, \citenamefont {Raju}, \citenamefont {Zhu},\ and\ \citenamefont
  {Cai}}]{liuHybridStrategyDiscovery2020}%
  \BibitemOpen
  \bibfield  {author} {\bibinfo {author} {\bibfnamefont {Z.}~\bibnamefont
  {Liu}}, \bibinfo {author} {\bibfnamefont {L.}~\bibnamefont {Raju}}, \bibinfo
  {author} {\bibfnamefont {D.}~\bibnamefont {Zhu}}, \ and\ \bibinfo {author}
  {\bibfnamefont {W.}~\bibnamefont {Cai}},\ }\href {\doibase
  10.1109/JETCAS.2020.2970080} {\bibfield  {journal} {\bibinfo  {journal} {IEEE
  Journal on Emerging and Selected Topics in Circuits and Systems}\ }\textbf
  {\bibinfo {volume} {10}},\ \bibinfo {pages} {126} (\bibinfo {year}
  {2020}{\natexlab{a}})},\ \Eprint {http://arxiv.org/abs/1902.02293}
  {arxiv:1902.02293} \BibitemShut {NoStop}%
\bibitem [{\citenamefont {Kingma}\ and\ \citenamefont
  {Welling}(2022)}]{kingmaAutoEncodingVariationalBayes2022}%
  \BibitemOpen
  \bibfield  {author} {\bibinfo {author} {\bibfnamefont {D.~P.}\ \bibnamefont
  {Kingma}}\ and\ \bibinfo {author} {\bibfnamefont {M.}~\bibnamefont
  {Welling}},\ }\href {\doibase 10.48550/arXiv.1312.6114} {\enquote {\bibinfo
  {title} {Auto-{{Encoding Variational Bayes}}},}\ } (\bibinfo {year} {2022}),\
  \Eprint {http://arxiv.org/abs/1312.6114} {arxiv:1312.6114 [cs, stat]}
  \BibitemShut {NoStop}%
\bibitem [{\citenamefont {Rozi{\`e}re}\ \emph {et~al.}(2021)\citenamefont
  {Rozi{\`e}re}, \citenamefont {Riviere}, \citenamefont {Teytaud},
  \citenamefont {Rapin}, \citenamefont {LeCun},\ and\ \citenamefont
  {Couprie}}]{roziereInspirationalAdversarialImage2021}%
  \BibitemOpen
  \bibfield  {author} {\bibinfo {author} {\bibfnamefont {B.}~\bibnamefont
  {Rozi{\`e}re}}, \bibinfo {author} {\bibfnamefont {M.}~\bibnamefont
  {Riviere}}, \bibinfo {author} {\bibfnamefont {O.}~\bibnamefont {Teytaud}},
  \bibinfo {author} {\bibfnamefont {J.}~\bibnamefont {Rapin}}, \bibinfo
  {author} {\bibfnamefont {Y.}~\bibnamefont {LeCun}}, \ and\ \bibinfo {author}
  {\bibfnamefont {C.}~\bibnamefont {Couprie}},\ }\href {\doibase
  10.48550/arXiv.1906.11661} {\enquote {\bibinfo {title} {Inspirational
  {{Adversarial Image Generation}}},}\ } (\bibinfo {year} {2021}),\ \Eprint
  {http://arxiv.org/abs/1906.11661} {arxiv:1906.11661 [cs, stat]} \BibitemShut
  {NoStop}%
\bibitem [{\citenamefont {Kullback}\ and\ \citenamefont
  {Leibler}(1951)}]{kullbackInformationSufficiency1951}%
  \BibitemOpen
  \bibfield  {author} {\bibinfo {author} {\bibfnamefont {S.}~\bibnamefont
  {Kullback}}\ and\ \bibinfo {author} {\bibfnamefont {R.~A.}\ \bibnamefont
  {Leibler}},\ }\href {\doibase 10.1214/aoms/1177729694} {\bibfield  {journal}
  {\bibinfo  {journal} {The Annals of Mathematical Statistics}\ }\textbf
  {\bibinfo {volume} {22}},\ \bibinfo {pages} {79} (\bibinfo {year}
  {1951})}\BibitemShut {NoStop}%
\bibitem [{\citenamefont {Wiecha}\ \emph
  {et~al.}(2017{\natexlab{b}})\citenamefont {Wiecha}, \citenamefont {Arbouet},
  \citenamefont {Girard}, \citenamefont {Lecestre}, \citenamefont {Larrieu},\
  and\ \citenamefont
  {Paillard}}]{wiechaEvolutionaryMultiobjectiveOptimization2017}%
  \BibitemOpen
  \bibfield  {author} {\bibinfo {author} {\bibfnamefont {P.~R.}\ \bibnamefont
  {Wiecha}}, \bibinfo {author} {\bibfnamefont {A.}~\bibnamefont {Arbouet}},
  \bibinfo {author} {\bibfnamefont {C.}~\bibnamefont {Girard}}, \bibinfo
  {author} {\bibfnamefont {A.}~\bibnamefont {Lecestre}}, \bibinfo {author}
  {\bibfnamefont {G.}~\bibnamefont {Larrieu}}, \ and\ \bibinfo {author}
  {\bibfnamefont {V.}~\bibnamefont {Paillard}},\ }\href {\doibase
  10.1038/nnano.2016.224} {\bibfield  {journal} {\bibinfo  {journal} {Nature
  Nanotechnology}\ }\textbf {\bibinfo {volume} {12}},\ \bibinfo {pages} {163}
  (\bibinfo {year} {2017}{\natexlab{b}})}\BibitemShut {NoStop}%
\bibitem [{\citenamefont {Wiecha}\ \emph
  {et~al.}(2019{\natexlab{b}})\citenamefont {Wiecha}, \citenamefont {Majorel},
  \citenamefont {Girard}, \citenamefont {Cuche}, \citenamefont {Paillard},
  \citenamefont {Muskens}, \citenamefont {Arbouet},\ and\ \citenamefont
  {Arbouet}}]{wiechaDesignPlasmonicDirectional2019}%
  \BibitemOpen
  \bibfield  {author} {\bibinfo {author} {\bibfnamefont {P.~R.}\ \bibnamefont
  {Wiecha}}, \bibinfo {author} {\bibfnamefont {C.}~\bibnamefont {Majorel}},
  \bibinfo {author} {\bibfnamefont {C.}~\bibnamefont {Girard}}, \bibinfo
  {author} {\bibfnamefont {A.}~\bibnamefont {Cuche}}, \bibinfo {author}
  {\bibfnamefont {V.}~\bibnamefont {Paillard}}, \bibinfo {author}
  {\bibfnamefont {O.~L.}\ \bibnamefont {Muskens}}, \bibinfo {author}
  {\bibfnamefont {A.}~\bibnamefont {Arbouet}}, \ and\ \bibinfo {author}
  {\bibfnamefont {A.}~\bibnamefont {Arbouet}},\ }\href {\doibase
  10.1364/OE.27.029069} {\bibfield  {journal} {\bibinfo  {journal} {Optics
  Express}\ }\textbf {\bibinfo {volume} {27}},\ \bibinfo {pages} {29069}
  (\bibinfo {year} {2019}{\natexlab{b}})}\BibitemShut {NoStop}%
\bibitem [{\citenamefont {Liu}\ \emph {et~al.}(2020{\natexlab{b}})\citenamefont
  {Liu}, \citenamefont {Moreau}, \citenamefont {Preuss}, \citenamefont {Rapin},
  \citenamefont {Roziere}, \citenamefont {Teytaud},\ and\ \citenamefont
  {Teytaud}}]{liuVersatileBlackboxOptimization2020}%
  \BibitemOpen
  \bibfield  {author} {\bibinfo {author} {\bibfnamefont {J.}~\bibnamefont
  {Liu}}, \bibinfo {author} {\bibfnamefont {A.}~\bibnamefont {Moreau}},
  \bibinfo {author} {\bibfnamefont {M.}~\bibnamefont {Preuss}}, \bibinfo
  {author} {\bibfnamefont {J.}~\bibnamefont {Rapin}}, \bibinfo {author}
  {\bibfnamefont {B.}~\bibnamefont {Roziere}}, \bibinfo {author} {\bibfnamefont
  {F.}~\bibnamefont {Teytaud}}, \ and\ \bibinfo {author} {\bibfnamefont
  {O.}~\bibnamefont {Teytaud}},\ }in\ \href {\doibase 10.1145/3377930.3389838}
  {\emph {\bibinfo {booktitle} {Proceedings of the 2020 {{Genetic}} and
  {{Evolutionary Computation Conference}}}}},\ \bibinfo {series and number}
  {{{GECCO}} '20}\ (\bibinfo  {publisher} {{Association for Computing
  Machinery}},\ \bibinfo {address} {{New York, NY, USA}},\ \bibinfo {year}
  {2020})\ pp.\ \bibinfo {pages} {620--628}\BibitemShut {NoStop}%
\bibitem [{\citenamefont {Barry}\ \emph {et~al.}(2020)\citenamefont {Barry},
  \citenamefont {Berthier}, \citenamefont {Wilts}, \citenamefont {Cambourieux},
  \citenamefont {Bennet}, \citenamefont {Poll{\`e}s}, \citenamefont {Teytaud},
  \citenamefont {Centeno}, \citenamefont {Biais},\ and\ \citenamefont
  {Moreau}}]{barryEvolutionaryAlgorithmsConverge2020}%
  \BibitemOpen
  \bibfield  {author} {\bibinfo {author} {\bibfnamefont {M.~A.}\ \bibnamefont
  {Barry}}, \bibinfo {author} {\bibfnamefont {V.}~\bibnamefont {Berthier}},
  \bibinfo {author} {\bibfnamefont {B.~D.}\ \bibnamefont {Wilts}}, \bibinfo
  {author} {\bibfnamefont {M.-C.}\ \bibnamefont {Cambourieux}}, \bibinfo
  {author} {\bibfnamefont {P.}~\bibnamefont {Bennet}}, \bibinfo {author}
  {\bibfnamefont {R.}~\bibnamefont {Poll{\`e}s}}, \bibinfo {author}
  {\bibfnamefont {O.}~\bibnamefont {Teytaud}}, \bibinfo {author} {\bibfnamefont
  {E.}~\bibnamefont {Centeno}}, \bibinfo {author} {\bibfnamefont
  {N.}~\bibnamefont {Biais}}, \ and\ \bibinfo {author} {\bibfnamefont
  {A.}~\bibnamefont {Moreau}},\ }\href {\doibase 10.1038/s41598-020-68719-3}
  {\bibfield  {journal} {\bibinfo  {journal} {Scientific Reports}\ }\textbf
  {\bibinfo {volume} {10}},\ \bibinfo {pages} {12024} (\bibinfo {year}
  {2020})}\BibitemShut {NoStop}%
\bibitem [{\citenamefont {Br{\^u}l{\'e}}\ \emph {et~al.}(2022)\citenamefont
  {Br{\^u}l{\'e}}, \citenamefont {Wiecha}, \citenamefont {Cuche}, \citenamefont
  {Paillard},\ and\ \citenamefont
  {Des~Francs}}]{bruleMagneticElectricPurcell2022}%
  \BibitemOpen
  \bibfield  {author} {\bibinfo {author} {\bibfnamefont {Y.}~\bibnamefont
  {Br{\^u}l{\'e}}}, \bibinfo {author} {\bibfnamefont {P.}~\bibnamefont
  {Wiecha}}, \bibinfo {author} {\bibfnamefont {A.}~\bibnamefont {Cuche}},
  \bibinfo {author} {\bibfnamefont {V.}~\bibnamefont {Paillard}}, \ and\
  \bibinfo {author} {\bibfnamefont {G.~C.}\ \bibnamefont {Des~Francs}},\
  }\href@noop {} {\bibfield  {journal} {\bibinfo  {journal} {Optics Express}\
  }\textbf {\bibinfo {volume} {30}},\ \bibinfo {pages} {20360} (\bibinfo {year}
  {2022})}\BibitemShut {NoStop}%
\bibitem [{\citenamefont {Jing}\ \emph {et~al.}(2023)\citenamefont {Jing},
  \citenamefont {Chu}, \citenamefont {Huang}, \citenamefont {Luo},
  \citenamefont {Wang},\ and\ \citenamefont {Lai}}]{jingDeepNeuralNetwork2023}%
  \BibitemOpen
  \bibfield  {author} {\bibinfo {author} {\bibfnamefont {Y.}~\bibnamefont
  {Jing}}, \bibinfo {author} {\bibfnamefont {H.}~\bibnamefont {Chu}}, \bibinfo
  {author} {\bibfnamefont {B.}~\bibnamefont {Huang}}, \bibinfo {author}
  {\bibfnamefont {J.}~\bibnamefont {Luo}}, \bibinfo {author} {\bibfnamefont
  {W.}~\bibnamefont {Wang}}, \ and\ \bibinfo {author} {\bibfnamefont
  {Y.}~\bibnamefont {Lai}},\ }\href {\doibase 10.1515/nanoph-2022-0770}
  {\bibfield  {journal} {\bibinfo  {journal} {Nanophotonics}\ } (\bibinfo
  {year} {2023}),\ 10.1515/nanoph-2022-0770}\BibitemShut {NoStop}%
\bibitem [{\citenamefont {Augenstein}\ \emph {et~al.}(2023)\citenamefont
  {Augenstein}, \citenamefont {Rep{\"a}n},\ and\ \citenamefont
  {Rockstuhl}}]{augensteinNeuralOperatorbasedSurrogate2023}%
  \BibitemOpen
  \bibfield  {author} {\bibinfo {author} {\bibfnamefont {Y.}~\bibnamefont
  {Augenstein}}, \bibinfo {author} {\bibfnamefont {T.}~\bibnamefont
  {Rep{\"a}n}}, \ and\ \bibinfo {author} {\bibfnamefont {C.}~\bibnamefont
  {Rockstuhl}},\ }\href {\doibase 10.1021/acsphotonics.3c00156} {\bibfield
  {journal} {\bibinfo  {journal} {ACS Photonics}\ }\textbf {\bibinfo {volume}
  {10}},\ \bibinfo {pages} {1547} (\bibinfo {year} {2023})},\ \Eprint
  {http://arxiv.org/abs/2302.01934} {arxiv:2302.01934 [physics]} \BibitemShut
  {NoStop}%
\bibitem [{\citenamefont {Sitzmann}\ \emph {et~al.}(2020)\citenamefont
  {Sitzmann}, \citenamefont {Martel}, \citenamefont {Bergman}, \citenamefont
  {Lindell},\ and\ \citenamefont
  {Wetzstein}}]{sitzmannImplicitNeuralRepresentations2020}%
  \BibitemOpen
  \bibfield  {author} {\bibinfo {author} {\bibfnamefont {V.}~\bibnamefont
  {Sitzmann}}, \bibinfo {author} {\bibfnamefont {J.}~\bibnamefont {Martel}},
  \bibinfo {author} {\bibfnamefont {A.}~\bibnamefont {Bergman}}, \bibinfo
  {author} {\bibfnamefont {D.}~\bibnamefont {Lindell}}, \ and\ \bibinfo
  {author} {\bibfnamefont {G.}~\bibnamefont {Wetzstein}},\ }in\ \href@noop {}
  {\emph {\bibinfo {booktitle} {Advances in {{Neural Information Processing
  Systems}}}}},\ Vol.~\bibinfo {volume} {33}\ (\bibinfo  {publisher} {{Curran
  Associates, Inc.}},\ \bibinfo {year} {2020})\ pp.\ \bibinfo {pages}
  {7462--7473}\BibitemShut {NoStop}%
\bibitem [{\citenamefont {Ma}\ \emph {et~al.}(2022)\citenamefont {Ma},
  \citenamefont {Tobah}, \citenamefont {Wang},\ and\ \citenamefont
  {Guo}}]{maBenchmarkingDeepLearningbased2022}%
  \BibitemOpen
  \bibfield  {author} {\bibinfo {author} {\bibfnamefont {T.}~\bibnamefont
  {Ma}}, \bibinfo {author} {\bibfnamefont {M.}~\bibnamefont {Tobah}}, \bibinfo
  {author} {\bibfnamefont {H.}~\bibnamefont {Wang}}, \ and\ \bibinfo {author}
  {\bibfnamefont {L.~J.}\ \bibnamefont {Guo}},\ }\href {\doibase
  10.29026/oes.2022.210012} {\bibfield  {journal} {\bibinfo  {journal}
  {Opto-Electronic Science}\ }\textbf {\bibinfo {volume} {1}},\ \bibinfo
  {pages} {210012} (\bibinfo {year} {Fri Jan 07 15:19:58 CST
  2022})}\BibitemShut {NoStop}%
\bibitem [{\citenamefont {{I. Higgins}}\ \emph {et~al.}(2017)\citenamefont {{I.
  Higgins}}, \citenamefont {{L. Matthey}}, \citenamefont {{A. Pal}},
  \citenamefont {{C. Burgess}}, \citenamefont {{X. Glorot}}, \citenamefont {{M.
  Botvinick}}, \citenamefont {{S. Mohamed}},\ and\ \citenamefont {{A.
  Lerchner}}}]{i.higginsVVAELearningBasic2017}%
  \BibitemOpen
  \bibfield  {author} {\bibinfo {author} {\bibnamefont {{I. Higgins}}},
  \bibinfo {author} {\bibnamefont {{L. Matthey}}}, \bibinfo {author}
  {\bibnamefont {{A. Pal}}}, \bibinfo {author} {\bibnamefont {{C. Burgess}}},
  \bibinfo {author} {\bibnamefont {{X. Glorot}}}, \bibinfo {author}
  {\bibnamefont {{M. Botvinick}}}, \bibinfo {author} {\bibnamefont {{S.
  Mohamed}}}, \ and\ \bibinfo {author} {\bibnamefont {{A. Lerchner}}},\
  }\href@noop {} {\bibfield  {journal} {\bibinfo  {journal} {ICLR conference}\
  } (\bibinfo {year} {2017})}\BibitemShut {NoStop}%
\bibitem [{\citenamefont {Silver}\ \emph {et~al.}(2017)\citenamefont {Silver},
  \citenamefont {Hubert}, \citenamefont {Schrittwieser}, \citenamefont
  {Antonoglou}, \citenamefont {Lai}, \citenamefont {Guez}, \citenamefont
  {Lanctot}, \citenamefont {Sifre}, \citenamefont {Kumaran}, \citenamefont
  {Graepel}, \citenamefont {Lillicrap}, \citenamefont {Simonyan},\ and\
  \citenamefont {Hassabis}}]{silverMasteringChessShogi2017}%
  \BibitemOpen
  \bibfield  {author} {\bibinfo {author} {\bibfnamefont {D.}~\bibnamefont
  {Silver}}, \bibinfo {author} {\bibfnamefont {T.}~\bibnamefont {Hubert}},
  \bibinfo {author} {\bibfnamefont {J.}~\bibnamefont {Schrittwieser}}, \bibinfo
  {author} {\bibfnamefont {I.}~\bibnamefont {Antonoglou}}, \bibinfo {author}
  {\bibfnamefont {M.}~\bibnamefont {Lai}}, \bibinfo {author} {\bibfnamefont
  {A.}~\bibnamefont {Guez}}, \bibinfo {author} {\bibfnamefont {M.}~\bibnamefont
  {Lanctot}}, \bibinfo {author} {\bibfnamefont {L.}~\bibnamefont {Sifre}},
  \bibinfo {author} {\bibfnamefont {D.}~\bibnamefont {Kumaran}}, \bibinfo
  {author} {\bibfnamefont {T.}~\bibnamefont {Graepel}}, \bibinfo {author}
  {\bibfnamefont {T.}~\bibnamefont {Lillicrap}}, \bibinfo {author}
  {\bibfnamefont {K.}~\bibnamefont {Simonyan}}, \ and\ \bibinfo {author}
  {\bibfnamefont {D.}~\bibnamefont {Hassabis}},\ }\href {\doibase
  10.48550/arXiv.1712.01815} {\enquote {\bibinfo {title} {Mastering {{Chess}}
  and {{Shogi}} by {{Self-Play}} with a {{General Reinforcement Learning
  Algorithm}}},}\ } (\bibinfo {year} {2017}),\ \Eprint
  {http://arxiv.org/abs/1712.01815} {arxiv:1712.01815 [cs]} \BibitemShut
  {NoStop}%
\bibitem [{\citenamefont {Zoph}\ and\ \citenamefont
  {Le}(2017)}]{zophNeuralArchitectureSearch2017}%
  \BibitemOpen
  \bibfield  {author} {\bibinfo {author} {\bibfnamefont {B.}~\bibnamefont
  {Zoph}}\ and\ \bibinfo {author} {\bibfnamefont {Q.~V.}\ \bibnamefont {Le}},\
  }\href {\doibase 10.48550/arXiv.1611.01578} {\enquote {\bibinfo {title}
  {Neural {{Architecture Search}} with {{Reinforcement Learning}}},}\ }
  (\bibinfo {year} {2017}),\ \Eprint {http://arxiv.org/abs/1611.01578}
  {arxiv:1611.01578 [cs]} \BibitemShut {NoStop}%
\bibitem [{\citenamefont {Wang}\ \emph
  {et~al.}(2020{\natexlab{b}})\citenamefont {Wang}, \citenamefont {Zheng},
  \citenamefont {Ji},\ and\ \citenamefont
  {Guo}}]{wangAutomatedMultilayerOptical2020}%
  \BibitemOpen
  \bibfield  {author} {\bibinfo {author} {\bibfnamefont {H.}~\bibnamefont
  {Wang}}, \bibinfo {author} {\bibfnamefont {Z.}~\bibnamefont {Zheng}},
  \bibinfo {author} {\bibfnamefont {C.}~\bibnamefont {Ji}}, \ and\ \bibinfo
  {author} {\bibfnamefont {L.~J.}\ \bibnamefont {Guo}},\ }\href {\doibase
  10.1088/2632-2153/abc327} {\bibfield  {journal} {\bibinfo  {journal} {Machine
  Learning: Science and Technology}\ }\textbf {\bibinfo {volume} {2}},\
  \bibinfo {pages} {025013} (\bibinfo {year} {2020}{\natexlab{b}})}\BibitemShut
  {NoStop}%
\bibitem [{\citenamefont {Real}\ \emph {et~al.}(2017)\citenamefont {Real},
  \citenamefont {Moore}, \citenamefont {Selle}, \citenamefont {Saxena},
  \citenamefont {Suematsu}, \citenamefont {Tan}, \citenamefont {Le},\ and\
  \citenamefont {Kurakin}}]{realLargeScaleEvolutionImage2017}%
  \BibitemOpen
  \bibfield  {author} {\bibinfo {author} {\bibfnamefont {E.}~\bibnamefont
  {Real}}, \bibinfo {author} {\bibfnamefont {S.}~\bibnamefont {Moore}},
  \bibinfo {author} {\bibfnamefont {A.}~\bibnamefont {Selle}}, \bibinfo
  {author} {\bibfnamefont {S.}~\bibnamefont {Saxena}}, \bibinfo {author}
  {\bibfnamefont {Y.~L.}\ \bibnamefont {Suematsu}}, \bibinfo {author}
  {\bibfnamefont {J.}~\bibnamefont {Tan}}, \bibinfo {author} {\bibfnamefont
  {Q.}~\bibnamefont {Le}}, \ and\ \bibinfo {author} {\bibfnamefont
  {A.}~\bibnamefont {Kurakin}},\ }\href {\doibase 10.48550/arXiv.1703.01041}
  {\enquote {\bibinfo {title} {Large-{{Scale Evolution}} of {{Image
  Classifiers}}},}\ } (\bibinfo {year} {2017}),\ \Eprint
  {http://arxiv.org/abs/1703.01041} {arxiv:1703.01041 [cs]} \BibitemShut
  {NoStop}%
\bibitem [{\citenamefont {Cheng}\ \emph {et~al.}(2023)\citenamefont {Cheng},
  \citenamefont {Kahng}, \citenamefont {Kundu}, \citenamefont {Wang},\ and\
  \citenamefont {Wang}}]{chengAssessmentReinforcementLearning2023}%
  \BibitemOpen
  \bibfield  {author} {\bibinfo {author} {\bibfnamefont {C.-K.}\ \bibnamefont
  {Cheng}}, \bibinfo {author} {\bibfnamefont {A.~B.}\ \bibnamefont {Kahng}},
  \bibinfo {author} {\bibfnamefont {S.}~\bibnamefont {Kundu}}, \bibinfo
  {author} {\bibfnamefont {Y.}~\bibnamefont {Wang}}, \ and\ \bibinfo {author}
  {\bibfnamefont {Z.}~\bibnamefont {Wang}},\ }in\ \href {\doibase
  10.1145/3569052.3578926} {\emph {\bibinfo {booktitle} {Proceedings of the
  2023 {{International Symposium}} on {{Physical Design}}}}},\ \bibinfo {series
  and number} {{{ISPD}} '23}\ (\bibinfo  {publisher} {{Association for
  Computing Machinery}},\ \bibinfo {address} {{New York, NY, USA}},\ \bibinfo
  {year} {2023})\ pp.\ \bibinfo {pages} {158--166}\BibitemShut {NoStop}%
\bibitem [{\citenamefont {Markov}(2023)}]{markovFalseDawnReevaluating2023}%
  \BibitemOpen
  \bibfield  {author} {\bibinfo {author} {\bibfnamefont {I.~L.}\ \bibnamefont
  {Markov}},\ }\href {\doibase 10.48550/arXiv.2306.09633} {\enquote {\bibinfo
  {title} {The {{False Dawn}}: {{Reevaluating Google}}'s {{Reinforcement
  Learning}} for {{Chip Macro Placement}}},}\ } (\bibinfo {year} {2023}),\
  \Eprint {http://arxiv.org/abs/2306.09633} {arxiv:2306.09633 [cs]}
  \BibitemShut {NoStop}%
\bibitem [{\citenamefont {Goodfellow}\ \emph {et~al.}(2014)\citenamefont
  {Goodfellow}, \citenamefont {{Pouget-Abadie}}, \citenamefont {Mirza},
  \citenamefont {Xu}, \citenamefont {{Warde-Farley}}, \citenamefont {Ozair},
  \citenamefont {Courville},\ and\ \citenamefont
  {Bengio}}]{goodfellowGenerativeAdversarialNetworks2014}%
  \BibitemOpen
  \bibfield  {author} {\bibinfo {author} {\bibfnamefont {I.~J.}\ \bibnamefont
  {Goodfellow}}, \bibinfo {author} {\bibfnamefont {J.}~\bibnamefont
  {{Pouget-Abadie}}}, \bibinfo {author} {\bibfnamefont {M.}~\bibnamefont
  {Mirza}}, \bibinfo {author} {\bibfnamefont {B.}~\bibnamefont {Xu}}, \bibinfo
  {author} {\bibfnamefont {D.}~\bibnamefont {{Warde-Farley}}}, \bibinfo
  {author} {\bibfnamefont {S.}~\bibnamefont {Ozair}}, \bibinfo {author}
  {\bibfnamefont {A.}~\bibnamefont {Courville}}, \ and\ \bibinfo {author}
  {\bibfnamefont {Y.}~\bibnamefont {Bengio}},\ }\href@noop {} {\bibfield
  {journal} {\bibinfo  {journal} {arXiv:1406.2661 [cs, stat]}\ } (\bibinfo
  {year} {2014})},\ \Eprint {http://arxiv.org/abs/1406.2661} {arxiv:1406.2661
  [cs, stat]} \BibitemShut {NoStop}%
\bibitem [{\citenamefont {Arjovsky}\ \emph {et~al.}(2017)\citenamefont
  {Arjovsky}, \citenamefont {Chintala},\ and\ \citenamefont
  {Bottou}}]{arjovskyWassersteinGAN2017}%
  \BibitemOpen
  \bibfield  {author} {\bibinfo {author} {\bibfnamefont {M.}~\bibnamefont
  {Arjovsky}}, \bibinfo {author} {\bibfnamefont {S.}~\bibnamefont {Chintala}},
  \ and\ \bibinfo {author} {\bibfnamefont {L.}~\bibnamefont {Bottou}},\ }\href
  {\doibase 10.48550/arXiv.1701.07875} {\enquote {\bibinfo {title} {Wasserstein
  {{GAN}}},}\ } (\bibinfo {year} {2017}),\ \Eprint
  {http://arxiv.org/abs/1701.07875} {arxiv:1701.07875 [cs, stat]} \BibitemShut
  {NoStop}%
\bibitem [{\citenamefont {Gulrajani}\ \emph {et~al.}(2017)\citenamefont
  {Gulrajani}, \citenamefont {Ahmed}, \citenamefont {Arjovsky}, \citenamefont
  {Dumoulin},\ and\ \citenamefont
  {Courville}}]{gulrajaniImprovedTrainingWasserstein2017}%
  \BibitemOpen
  \bibfield  {author} {\bibinfo {author} {\bibfnamefont {I.}~\bibnamefont
  {Gulrajani}}, \bibinfo {author} {\bibfnamefont {F.}~\bibnamefont {Ahmed}},
  \bibinfo {author} {\bibfnamefont {M.}~\bibnamefont {Arjovsky}}, \bibinfo
  {author} {\bibfnamefont {V.}~\bibnamefont {Dumoulin}}, \ and\ \bibinfo
  {author} {\bibfnamefont {A.}~\bibnamefont {Courville}},\ }\href {\doibase
  10.48550/arXiv.1704.00028} {\enquote {\bibinfo {title} {Improved {{Training}}
  of {{Wasserstein GANs}}},}\ } (\bibinfo {year} {2017}),\ \Eprint
  {http://arxiv.org/abs/1704.00028} {arxiv:1704.00028 [cs, stat]} \BibitemShut
  {NoStop}%
\bibitem [{\citenamefont {Karras}\ \emph {et~al.}(2019)\citenamefont {Karras},
  \citenamefont {Laine},\ and\ \citenamefont
  {Aila}}]{karrasStyleBasedGeneratorArchitecture2019}%
  \BibitemOpen
  \bibfield  {author} {\bibinfo {author} {\bibfnamefont {T.}~\bibnamefont
  {Karras}}, \bibinfo {author} {\bibfnamefont {S.}~\bibnamefont {Laine}}, \
  and\ \bibinfo {author} {\bibfnamefont {T.}~\bibnamefont {Aila}},\ }\href
  {\doibase 10.48550/arXiv.1812.04948} {\enquote {\bibinfo {title} {A
  {{Style-Based Generator Architecture}} for {{Generative Adversarial
  Networks}}},}\ } (\bibinfo {year} {2019}),\ \Eprint
  {http://arxiv.org/abs/1812.04948} {arxiv:1812.04948 [cs, stat]} \BibitemShut
  {NoStop}%
\bibitem [{\citenamefont {Song}\ \emph {et~al.}(2021)\citenamefont {Song},
  \citenamefont {{Sohl-Dickstein}}, \citenamefont {Kingma}, \citenamefont
  {Kumar}, \citenamefont {Ermon},\ and\ \citenamefont
  {Poole}}]{songScoreBasedGenerativeModeling2021}%
  \BibitemOpen
  \bibfield  {author} {\bibinfo {author} {\bibfnamefont {Y.}~\bibnamefont
  {Song}}, \bibinfo {author} {\bibfnamefont {J.}~\bibnamefont
  {{Sohl-Dickstein}}}, \bibinfo {author} {\bibfnamefont {D.~P.}\ \bibnamefont
  {Kingma}}, \bibinfo {author} {\bibfnamefont {A.}~\bibnamefont {Kumar}},
  \bibinfo {author} {\bibfnamefont {S.}~\bibnamefont {Ermon}}, \ and\ \bibinfo
  {author} {\bibfnamefont {B.}~\bibnamefont {Poole}},\ }\href {\doibase
  10.48550/arXiv.2011.13456} {\enquote {\bibinfo {title} {Score-{{Based
  Generative Modeling}} through {{Stochastic Differential Equations}}},}\ }
  (\bibinfo {year} {2021}),\ \Eprint {http://arxiv.org/abs/2011.13456}
  {arxiv:2011.13456 [cs, stat]} \BibitemShut {NoStop}%
\bibitem [{\citenamefont {{Sohl-Dickstein}}\ \emph {et~al.}(2015)\citenamefont
  {{Sohl-Dickstein}}, \citenamefont {Weiss}, \citenamefont {Maheswaranathan},\
  and\ \citenamefont {Ganguli}}]{sohl-dicksteinDeepUnsupervisedLearning2015}%
  \BibitemOpen
  \bibfield  {author} {\bibinfo {author} {\bibfnamefont {J.}~\bibnamefont
  {{Sohl-Dickstein}}}, \bibinfo {author} {\bibfnamefont {E.}~\bibnamefont
  {Weiss}}, \bibinfo {author} {\bibfnamefont {N.}~\bibnamefont
  {Maheswaranathan}}, \ and\ \bibinfo {author} {\bibfnamefont {S.}~\bibnamefont
  {Ganguli}},\ }in\ \href@noop {} {\emph {\bibinfo {booktitle} {Proceedings of
  the 32nd {{International Conference}} on {{Machine Learning}}}}}\ (\bibinfo
  {publisher} {{PMLR}},\ \bibinfo {year} {2015})\ pp.\ \bibinfo {pages}
  {2256--2265}\BibitemShut {NoStop}%
\bibitem [{\citenamefont {Croitoru}\ \emph {et~al.}(2023)\citenamefont
  {Croitoru}, \citenamefont {Hondru}, \citenamefont {Ionescu},\ and\
  \citenamefont {Shah}}]{croitoruDiffusionModelsVision2023}%
  \BibitemOpen
  \bibfield  {author} {\bibinfo {author} {\bibfnamefont {F.-A.}\ \bibnamefont
  {Croitoru}}, \bibinfo {author} {\bibfnamefont {V.}~\bibnamefont {Hondru}},
  \bibinfo {author} {\bibfnamefont {R.~T.}\ \bibnamefont {Ionescu}}, \ and\
  \bibinfo {author} {\bibfnamefont {M.}~\bibnamefont {Shah}},\ }\href {\doibase
  10.1109/TPAMI.2023.3261988} {\bibfield  {journal} {\bibinfo  {journal} {IEEE
  Transactions on Pattern Analysis and Machine Intelligence}\ }\textbf
  {\bibinfo {volume} {45}},\ \bibinfo {pages} {10850} (\bibinfo {year}
  {2023})},\ \Eprint {http://arxiv.org/abs/2209.04747} {arxiv:2209.04747 [cs]}
  \BibitemShut {NoStop}%
\bibitem [{\citenamefont {Chang}\ \emph {et~al.}(2023)\citenamefont {Chang},
  \citenamefont {Koulieris},\ and\ \citenamefont
  {Shum}}]{changDesignFundamentalsDiffusion2023}%
  \BibitemOpen
  \bibfield  {author} {\bibinfo {author} {\bibfnamefont {Z.}~\bibnamefont
  {Chang}}, \bibinfo {author} {\bibfnamefont {G.~A.}\ \bibnamefont
  {Koulieris}}, \ and\ \bibinfo {author} {\bibfnamefont {H.~P.~H.}\
  \bibnamefont {Shum}},\ }\href {\doibase 10.48550/arXiv.2306.04542} {\enquote
  {\bibinfo {title} {On the {{Design Fundamentals}} of {{Diffusion Models}}:
  {{A Survey}}},}\ } (\bibinfo {year} {2023}),\ \Eprint
  {http://arxiv.org/abs/2306.04542} {arxiv:2306.04542 [cs]} \BibitemShut
  {NoStop}%
\bibitem [{\citenamefont {Zhang}\ \emph {et~al.}(2023)\citenamefont {Zhang},
  \citenamefont {Yang}, \citenamefont {Qin}, \citenamefont {Feng},
  \citenamefont {Feng},\ and\ \citenamefont
  {Li}}]{zhangDiffusionProbabilisticModel2023}%
  \BibitemOpen
  \bibfield  {author} {\bibinfo {author} {\bibfnamefont {Z.}~\bibnamefont
  {Zhang}}, \bibinfo {author} {\bibfnamefont {C.}~\bibnamefont {Yang}},
  \bibinfo {author} {\bibfnamefont {Y.}~\bibnamefont {Qin}}, \bibinfo {author}
  {\bibfnamefont {H.}~\bibnamefont {Feng}}, \bibinfo {author} {\bibfnamefont
  {J.}~\bibnamefont {Feng}}, \ and\ \bibinfo {author} {\bibfnamefont
  {H.}~\bibnamefont {Li}},\ }\href {\doibase 10.1515/nanoph-2023-0292}
  {\bibfield  {journal} {\bibinfo  {journal} {Nanophotonics}\ } (\bibinfo
  {year} {2023}),\ 10.1515/nanoph-2023-0292}\BibitemShut {NoStop}%
\bibitem [{\citenamefont {Behrmann}\ \emph {et~al.}(2019)\citenamefont
  {Behrmann}, \citenamefont {Grathwohl}, \citenamefont {Chen}, \citenamefont
  {Duvenaud},\ and\ \citenamefont
  {Jacobsen}}]{behrmannInvertibleResidualNetworks2019}%
  \BibitemOpen
  \bibfield  {author} {\bibinfo {author} {\bibfnamefont {J.}~\bibnamefont
  {Behrmann}}, \bibinfo {author} {\bibfnamefont {W.}~\bibnamefont {Grathwohl}},
  \bibinfo {author} {\bibfnamefont {R.~T.~Q.}\ \bibnamefont {Chen}}, \bibinfo
  {author} {\bibfnamefont {D.}~\bibnamefont {Duvenaud}}, \ and\ \bibinfo
  {author} {\bibfnamefont {J.-H.}\ \bibnamefont {Jacobsen}},\ }\href {\doibase
  10.48550/arXiv.1811.00995} {\enquote {\bibinfo {title} {Invertible {{Residual
  Networks}}},}\ } (\bibinfo {year} {2019}),\ \Eprint
  {http://arxiv.org/abs/1811.00995} {arxiv:1811.00995 [cs, stat]} \BibitemShut
  {NoStop}%
\bibitem [{\citenamefont {Ardizzone}\ \emph {et~al.}(2018)\citenamefont
  {Ardizzone}, \citenamefont {Kruse}, \citenamefont {Wirkert}, \citenamefont
  {Rahner}, \citenamefont {Pellegrini}, \citenamefont {Klessen}, \citenamefont
  {{Maier-Hein}}, \citenamefont {Rother},\ and\ \citenamefont
  {K{\"o}the}}]{ardizzoneAnalyzingInverseProblems2018}%
  \BibitemOpen
  \bibfield  {author} {\bibinfo {author} {\bibfnamefont {L.}~\bibnamefont
  {Ardizzone}}, \bibinfo {author} {\bibfnamefont {J.}~\bibnamefont {Kruse}},
  \bibinfo {author} {\bibfnamefont {S.}~\bibnamefont {Wirkert}}, \bibinfo
  {author} {\bibfnamefont {D.}~\bibnamefont {Rahner}}, \bibinfo {author}
  {\bibfnamefont {E.~W.}\ \bibnamefont {Pellegrini}}, \bibinfo {author}
  {\bibfnamefont {R.~S.}\ \bibnamefont {Klessen}}, \bibinfo {author}
  {\bibfnamefont {L.}~\bibnamefont {{Maier-Hein}}}, \bibinfo {author}
  {\bibfnamefont {C.}~\bibnamefont {Rother}}, \ and\ \bibinfo {author}
  {\bibfnamefont {U.}~\bibnamefont {K{\"o}the}},\ }\href@noop {} {\bibfield
  {journal} {\bibinfo  {journal} {arXiv:1808.04730 [cs, stat]}\ } (\bibinfo
  {year} {2018})},\ \Eprint {http://arxiv.org/abs/1808.04730} {arxiv:1808.04730
  [cs, stat]} \BibitemShut {NoStop}%
\bibitem [{\citenamefont {Raissi}\ \emph {et~al.}(2019)\citenamefont {Raissi},
  \citenamefont {Perdikaris},\ and\ \citenamefont
  {Karniadakis}}]{raissiPhysicsinformedNeuralNetworks2019}%
  \BibitemOpen
  \bibfield  {author} {\bibinfo {author} {\bibfnamefont {M.}~\bibnamefont
  {Raissi}}, \bibinfo {author} {\bibfnamefont {P.}~\bibnamefont {Perdikaris}},
  \ and\ \bibinfo {author} {\bibfnamefont {G.~E.}\ \bibnamefont
  {Karniadakis}},\ }\href {\doibase 10.1016/j.jcp.2018.10.045} {\bibfield
  {journal} {\bibinfo  {journal} {Journal of Computational Physics}\ }\textbf
  {\bibinfo {volume} {378}},\ \bibinfo {pages} {686} (\bibinfo {year}
  {2019})}\BibitemShut {NoStop}%
\bibitem [{\citenamefont {Grossmann}\ \emph {et~al.}(2023)\citenamefont
  {Grossmann}, \citenamefont {Komorowska}, \citenamefont {Latz},\ and\
  \citenamefont {Sch{\"o}nlieb}}]{grossmannCanPhysicsInformedNeural2023}%
  \BibitemOpen
  \bibfield  {author} {\bibinfo {author} {\bibfnamefont {T.~G.}\ \bibnamefont
  {Grossmann}}, \bibinfo {author} {\bibfnamefont {U.~J.}\ \bibnamefont
  {Komorowska}}, \bibinfo {author} {\bibfnamefont {J.}~\bibnamefont {Latz}}, \
  and\ \bibinfo {author} {\bibfnamefont {C.-B.}\ \bibnamefont
  {Sch{\"o}nlieb}},\ }\href {\doibase 10.48550/arXiv.2302.04107} {\enquote
  {\bibinfo {title} {Can {{Physics-Informed Neural Networks}} beat the {{Finite
  Element Method}}?}}\ } (\bibinfo {year} {2023}),\ \Eprint
  {http://arxiv.org/abs/2302.04107} {arxiv:2302.04107 [cs, math]} \BibitemShut
  {NoStop}%
\bibitem [{\citenamefont {Chen}\ \emph {et~al.}(2020)\citenamefont {Chen},
  \citenamefont {Lu}, \citenamefont {Karniadakis},\ and\ \citenamefont
  {Negro}}]{chenPhysicsinformedNeuralNetworks2020}%
  \BibitemOpen
  \bibfield  {author} {\bibinfo {author} {\bibfnamefont {Y.}~\bibnamefont
  {Chen}}, \bibinfo {author} {\bibfnamefont {L.}~\bibnamefont {Lu}}, \bibinfo
  {author} {\bibfnamefont {G.~E.}\ \bibnamefont {Karniadakis}}, \ and\ \bibinfo
  {author} {\bibfnamefont {L.~D.}\ \bibnamefont {Negro}},\ }\href {\doibase
  10.1364/OE.384875} {\bibfield  {journal} {\bibinfo  {journal} {Optics
  Express}\ }\textbf {\bibinfo {volume} {28}},\ \bibinfo {pages} {11618}
  (\bibinfo {year} {2020})},\ \Eprint {http://arxiv.org/abs/1912.01085}
  {arxiv:1912.01085} \BibitemShut {NoStop}%
\bibitem [{\citenamefont {Fang}\ and\ \citenamefont
  {Zhan}(2020)}]{fangDeepPhysicalInformed2020}%
  \BibitemOpen
  \bibfield  {author} {\bibinfo {author} {\bibfnamefont {Z.}~\bibnamefont
  {Fang}}\ and\ \bibinfo {author} {\bibfnamefont {J.}~\bibnamefont {Zhan}},\
  }\href {\doibase 10.1109/ACCESS.2019.2963375} {\bibfield  {journal} {\bibinfo
   {journal} {IEEE Access}\ }\textbf {\bibinfo {volume} {8}},\ \bibinfo {pages}
  {24506} (\bibinfo {year} {2020})}\BibitemShut {NoStop}%
\bibitem [{\citenamefont {Lu}\ \emph {et~al.}(2021)\citenamefont {Lu},
  \citenamefont {Pestourie}, \citenamefont {Yao}, \citenamefont {Wang},
  \citenamefont {Verdugo},\ and\ \citenamefont
  {Johnson}}]{luPhysicsinformedNeuralNetworks2021}%
  \BibitemOpen
  \bibfield  {author} {\bibinfo {author} {\bibfnamefont {L.}~\bibnamefont
  {Lu}}, \bibinfo {author} {\bibfnamefont {R.}~\bibnamefont {Pestourie}},
  \bibinfo {author} {\bibfnamefont {W.}~\bibnamefont {Yao}}, \bibinfo {author}
  {\bibfnamefont {Z.}~\bibnamefont {Wang}}, \bibinfo {author} {\bibfnamefont
  {F.}~\bibnamefont {Verdugo}}, \ and\ \bibinfo {author} {\bibfnamefont
  {S.~G.}\ \bibnamefont {Johnson}},\ }\href {\doibase 10.1137/21M1397908}
  {\bibfield  {journal} {\bibinfo  {journal} {SIAM Journal on Scientific
  Computing}\ }\textbf {\bibinfo {volume} {43}},\ \bibinfo {pages} {B1105}
  (\bibinfo {year} {2021})}\BibitemShut {NoStop}%
\bibitem [{\citenamefont {Klocek}\ \emph {et~al.}(2019)\citenamefont {Klocek},
  \citenamefont {Maziarka}, \citenamefont {Wo{\l}czyk}, \citenamefont {Tabor},
  \citenamefont {Nowak},\ and\ \citenamefont
  {{\'S}mieja}}]{klocekHypernetworkFunctionalImage2019}%
  \BibitemOpen
  \bibfield  {author} {\bibinfo {author} {\bibfnamefont {S.}~\bibnamefont
  {Klocek}}, \bibinfo {author} {\bibfnamefont {{\L}.}~\bibnamefont {Maziarka}},
  \bibinfo {author} {\bibfnamefont {M.}~\bibnamefont {Wo{\l}czyk}}, \bibinfo
  {author} {\bibfnamefont {J.}~\bibnamefont {Tabor}}, \bibinfo {author}
  {\bibfnamefont {J.}~\bibnamefont {Nowak}}, \ and\ \bibinfo {author}
  {\bibfnamefont {M.}~\bibnamefont {{\'S}mieja}},\ }in\ \href {\doibase
  10.1007/978-3-030-30493-5_48} {\emph {\bibinfo {booktitle} {Artificial
  {{Neural Networks}} and {{Machine Learning}} \textendash{} {{ICANN}} 2019:
  {{Workshop}} and {{Special Sessions}}}}},\ \bibinfo {series and number}
  {Lecture {{Notes}} in {{Computer Science}}},\ \bibinfo {editor} {edited by\
  \bibinfo {editor} {\bibfnamefont {I.~V.}\ \bibnamefont {Tetko}}, \bibinfo
  {editor} {\bibfnamefont {V.}~\bibnamefont {K{\r{u}}rkov{\'a}}}, \bibinfo
  {editor} {\bibfnamefont {P.}~\bibnamefont {Karpov}}, \ and\ \bibinfo {editor}
  {\bibfnamefont {F.}~\bibnamefont {Theis}}}\ (\bibinfo  {publisher} {{Springer
  International Publishing}},\ \bibinfo {address} {{Cham}},\ \bibinfo {year}
  {2019})\ pp.\ \bibinfo {pages} {496--510}\BibitemShut {NoStop}%
\bibitem [{\citenamefont {Pestourie}\ \emph {et~al.}(2020)\citenamefont
  {Pestourie}, \citenamefont {Mroueh}, \citenamefont {Nguyen}, \citenamefont
  {Das},\ and\ \citenamefont {Johnson}}]{pestourieActiveLearningDeep2020}%
  \BibitemOpen
  \bibfield  {author} {\bibinfo {author} {\bibfnamefont {R.}~\bibnamefont
  {Pestourie}}, \bibinfo {author} {\bibfnamefont {Y.}~\bibnamefont {Mroueh}},
  \bibinfo {author} {\bibfnamefont {T.~V.}\ \bibnamefont {Nguyen}}, \bibinfo
  {author} {\bibfnamefont {P.}~\bibnamefont {Das}}, \ and\ \bibinfo {author}
  {\bibfnamefont {S.~G.}\ \bibnamefont {Johnson}},\ }\href@noop {} {\bibfield
  {journal} {\bibinfo  {journal} {arXiv:2008.12649 [physics]}\ } (\bibinfo
  {year} {2020})},\ \Eprint {http://arxiv.org/abs/2008.12649} {arxiv:2008.12649
  [physics]} \BibitemShut {NoStop}%
\bibitem [{\citenamefont {{Blanchard-Dionne}}\ and\ \citenamefont
  {Martin}(2021)}]{blanchard-dionneSuccessiveTrainingGenerative2021}%
  \BibitemOpen
  \bibfield  {author} {\bibinfo {author} {\bibfnamefont {A.-P.}\ \bibnamefont
  {{Blanchard-Dionne}}}\ and\ \bibinfo {author} {\bibfnamefont {O.~J.~F.}\
  \bibnamefont {Martin}},\ }\href {\doibase 10.1364/OSAC.413394} {\bibfield
  {journal} {\bibinfo  {journal} {OSA Continuum}\ }\textbf {\bibinfo {volume}
  {4}},\ \bibinfo {pages} {87} (\bibinfo {year} {2021})},\ \Eprint
  {http://arxiv.org/abs/2005.08832} {arxiv:2005.08832} \BibitemShut {NoStop}%
\bibitem [{\citenamefont {Yao}\ \emph {et~al.}(2007)\citenamefont {Yao},
  \citenamefont {Rosasco},\ and\ \citenamefont
  {Caponnetto}}]{yaoEarlyStoppingGradient2007}%
  \BibitemOpen
  \bibfield  {author} {\bibinfo {author} {\bibfnamefont {Y.}~\bibnamefont
  {Yao}}, \bibinfo {author} {\bibfnamefont {L.}~\bibnamefont {Rosasco}}, \ and\
  \bibinfo {author} {\bibfnamefont {A.}~\bibnamefont {Caponnetto}},\ }\href
  {\doibase 10.1007/s00365-006-0663-2} {\bibfield  {journal} {\bibinfo
  {journal} {Constructive Approximation}\ }\textbf {\bibinfo {volume} {26}},\
  \bibinfo {pages} {289} (\bibinfo {year} {2007})}\BibitemShut {NoStop}%
\bibitem [{\citenamefont {Wang}\ \emph {et~al.}(2019)\citenamefont {Wang},
  \citenamefont {Fan}, \citenamefont {Luo}, \citenamefont {Cao}, \citenamefont
  {Wu}, \citenamefont {Zhang}, \citenamefont {Heller},\ and\ \citenamefont
  {You}}]{wangMassiveComputationalAcceleration2019}%
  \BibitemOpen
  \bibfield  {author} {\bibinfo {author} {\bibfnamefont {S.}~\bibnamefont
  {Wang}}, \bibinfo {author} {\bibfnamefont {K.}~\bibnamefont {Fan}}, \bibinfo
  {author} {\bibfnamefont {N.}~\bibnamefont {Luo}}, \bibinfo {author}
  {\bibfnamefont {Y.}~\bibnamefont {Cao}}, \bibinfo {author} {\bibfnamefont
  {F.}~\bibnamefont {Wu}}, \bibinfo {author} {\bibfnamefont {C.}~\bibnamefont
  {Zhang}}, \bibinfo {author} {\bibfnamefont {K.~A.}\ \bibnamefont {Heller}}, \
  and\ \bibinfo {author} {\bibfnamefont {L.}~\bibnamefont {You}},\ }\href
  {\doibase 10.1038/s41467-019-12342-y} {\bibfield  {journal} {\bibinfo
  {journal} {Nature Communications}\ }\textbf {\bibinfo {volume} {10}},\
  \bibinfo {pages} {4354} (\bibinfo {year} {2019})}\BibitemShut {NoStop}%
\bibitem [{\citenamefont {Eigen}\ \emph {et~al.}(2014)\citenamefont {Eigen},
  \citenamefont {Ranzato},\ and\ \citenamefont
  {Sutskever}}]{eigenLearningFactoredRepresentations2014}%
  \BibitemOpen
  \bibfield  {author} {\bibinfo {author} {\bibfnamefont {D.}~\bibnamefont
  {Eigen}}, \bibinfo {author} {\bibfnamefont {M.}~\bibnamefont {Ranzato}}, \
  and\ \bibinfo {author} {\bibfnamefont {I.}~\bibnamefont {Sutskever}},\ }\href
  {\doibase 10.48550/arXiv.1312.4314} {\enquote {\bibinfo {title} {Learning
  {{Factored Representations}} in a {{Deep Mixture}} of {{Experts}}},}\ }
  (\bibinfo {year} {2014}),\ \Eprint {http://arxiv.org/abs/1312.4314}
  {arxiv:1312.4314 [cs]} \BibitemShut {NoStop}%
\bibitem [{\citenamefont {Shazeer}\ \emph {et~al.}(2017)\citenamefont
  {Shazeer}, \citenamefont {Mirhoseini}, \citenamefont {Maziarz}, \citenamefont
  {Davis}, \citenamefont {Le}, \citenamefont {Hinton},\ and\ \citenamefont
  {Dean}}]{shazeerOutrageouslyLargeNeural2017}%
  \BibitemOpen
  \bibfield  {author} {\bibinfo {author} {\bibfnamefont {N.}~\bibnamefont
  {Shazeer}}, \bibinfo {author} {\bibfnamefont {A.}~\bibnamefont {Mirhoseini}},
  \bibinfo {author} {\bibfnamefont {K.}~\bibnamefont {Maziarz}}, \bibinfo
  {author} {\bibfnamefont {A.}~\bibnamefont {Davis}}, \bibinfo {author}
  {\bibfnamefont {Q.}~\bibnamefont {Le}}, \bibinfo {author} {\bibfnamefont
  {G.}~\bibnamefont {Hinton}}, \ and\ \bibinfo {author} {\bibfnamefont
  {J.}~\bibnamefont {Dean}},\ }\href {\doibase 10.48550/arXiv.1701.06538}
  {\enquote {\bibinfo {title} {Outrageously {{Large Neural Networks}}: {{The
  Sparsely-Gated Mixture-of-Experts Layer}}},}\ } (\bibinfo {year} {2017}),\
  \Eprint {http://arxiv.org/abs/1701.06538} {arxiv:1701.06538 [cs, stat]}
  \BibitemShut {NoStop}%
\bibitem [{\citenamefont {Moreau}(2023)}]{moreauPyMoosh2023}%
  \BibitemOpen
  \bibfield  {author} {\bibinfo {author} {\bibfnamefont {A.}~\bibnamefont
  {Moreau}},\ }\href@noop {} {\enquote {\bibinfo {title} {{{PyMoosh}}},}\
  }\bibinfo {howpublished} {https://github.com/AnMoreau/PyMoosh} (\bibinfo
  {year} {2023})\BibitemShut {NoStop}%
\bibitem [{\citenamefont {Defrance}\ \emph {et~al.}(2016)\citenamefont
  {Defrance}, \citenamefont {Lema{\^i}tre}, \citenamefont {Ajib}, \citenamefont
  {Benedicto}, \citenamefont {Mallet}, \citenamefont {Poll{\`e}s},
  \citenamefont {Plumey}, \citenamefont {Mihailovic}, \citenamefont {Centeno},
  \citenamefont {Cirac{\`i}}, \citenamefont {Smith},\ and\ \citenamefont
  {Moreau}}]{defranceMooshNumericalSwiss2016}%
  \BibitemOpen
  \bibfield  {author} {\bibinfo {author} {\bibfnamefont {J.}~\bibnamefont
  {Defrance}}, \bibinfo {author} {\bibfnamefont {C.}~\bibnamefont
  {Lema{\^i}tre}}, \bibinfo {author} {\bibfnamefont {R.}~\bibnamefont {Ajib}},
  \bibinfo {author} {\bibfnamefont {J.}~\bibnamefont {Benedicto}}, \bibinfo
  {author} {\bibfnamefont {E.}~\bibnamefont {Mallet}}, \bibinfo {author}
  {\bibfnamefont {R.}~\bibnamefont {Poll{\`e}s}}, \bibinfo {author}
  {\bibfnamefont {J.-P.}\ \bibnamefont {Plumey}}, \bibinfo {author}
  {\bibfnamefont {M.}~\bibnamefont {Mihailovic}}, \bibinfo {author}
  {\bibfnamefont {E.}~\bibnamefont {Centeno}}, \bibinfo {author} {\bibfnamefont
  {C.}~\bibnamefont {Cirac{\`i}}}, \bibinfo {author} {\bibfnamefont
  {D.}~\bibnamefont {Smith}}, \ and\ \bibinfo {author} {\bibfnamefont
  {A.}~\bibnamefont {Moreau}},\ }\href {\doibase 10.5334/jors.100} {\bibfield
  {journal} {\bibinfo  {journal} {Journal of Open Research Software}\ }\textbf
  {\bibinfo {volume} {4}},\ \bibinfo {pages} {e13} (\bibinfo {year}
  {2016})}\BibitemShut {NoStop}%
\bibitem [{\citenamefont {Bennet}\ \emph {et~al.}(2021)\citenamefont {Bennet},
  \citenamefont {Doerr}, \citenamefont {Moreau}, \citenamefont {Rapin},
  \citenamefont {Teytaud},\ and\ \citenamefont
  {Teytaud}}]{bennetNevergradBlackboxOptimization2021}%
  \BibitemOpen
  \bibfield  {author} {\bibinfo {author} {\bibfnamefont {P.}~\bibnamefont
  {Bennet}}, \bibinfo {author} {\bibfnamefont {C.}~\bibnamefont {Doerr}},
  \bibinfo {author} {\bibfnamefont {A.}~\bibnamefont {Moreau}}, \bibinfo
  {author} {\bibfnamefont {J.}~\bibnamefont {Rapin}}, \bibinfo {author}
  {\bibfnamefont {F.}~\bibnamefont {Teytaud}}, \ and\ \bibinfo {author}
  {\bibfnamefont {O.}~\bibnamefont {Teytaud}},\ }\href@noop {} {\bibfield
  {journal} {\bibinfo  {journal} {ACM SIGEVOlution}\ }\textbf {\bibinfo
  {volume} {14}},\ \bibinfo {pages} {8} (\bibinfo {year} {2021})}\BibitemShut
  {NoStop}%
\bibitem [{\citenamefont {Wiecha}(2018)}]{wiechaPyGDMPythonToolkit2018}%
  \BibitemOpen
  \bibfield  {author} {\bibinfo {author} {\bibfnamefont {P.~R.}\ \bibnamefont
  {Wiecha}},\ }\href {\doibase 10.1016/j.cpc.2018.06.017} {\bibfield  {journal}
  {\bibinfo  {journal} {Computer Physics Communications}\ }\textbf {\bibinfo
  {volume} {233}},\ \bibinfo {pages} {167} (\bibinfo {year}
  {2018})}\BibitemShut {NoStop}%
\bibitem [{\citenamefont {Wiecha}\ \emph {et~al.}(2022)\citenamefont {Wiecha},
  \citenamefont {Majorel}, \citenamefont {Arbouet}, \citenamefont {Patoux},
  \citenamefont {Br{\^u}l{\'e}}, \citenamefont {des Francs},\ and\
  \citenamefont {Girard}}]{wiechaPyGDMNewFunctionalities2022}%
  \BibitemOpen
  \bibfield  {author} {\bibinfo {author} {\bibfnamefont {P.~R.}\ \bibnamefont
  {Wiecha}}, \bibinfo {author} {\bibfnamefont {C.}~\bibnamefont {Majorel}},
  \bibinfo {author} {\bibfnamefont {A.}~\bibnamefont {Arbouet}}, \bibinfo
  {author} {\bibfnamefont {A.}~\bibnamefont {Patoux}}, \bibinfo {author}
  {\bibfnamefont {Y.}~\bibnamefont {Br{\^u}l{\'e}}}, \bibinfo {author}
  {\bibfnamefont {G.~C.}\ \bibnamefont {des Francs}}, \ and\ \bibinfo {author}
  {\bibfnamefont {C.}~\bibnamefont {Girard}},\ }\href {\doibase
  10.1016/j.cpc.2021.108142} {\bibfield  {journal} {\bibinfo  {journal}
  {Computer Physics Communications}\ }\textbf {\bibinfo {volume} {270}},\
  \bibinfo {pages} {108142} (\bibinfo {year} {2022})},\ \Eprint
  {http://arxiv.org/abs/2105.04587} {arxiv:2105.04587} \BibitemShut {NoStop}%
\bibitem [{\citenamefont
  {Wiecha}(2023{\natexlab{b}})}]{wiechaDeepLearningNanophotonic2023}%
  \BibitemOpen
  \bibfield  {author} {\bibinfo {author} {\bibfnamefont {P.~R.}\ \bibnamefont
  {Wiecha}},\ }\href {\doibase 10.48550/arXiv.2310.08618} {\enquote {\bibinfo
  {title} {Deep learning for nano-photonic materials -- {{The}} solution to
  everything!?}}\ } (\bibinfo {year} {2023}{\natexlab{b}}),\ \Eprint
  {http://arxiv.org/abs/2310.08618} {arxiv:2310.08618 [physics]} \BibitemShut
  {NoStop}%
\end{thebibliography}%

\end{document}